\documentclass[journal]{IEEEtran}

\ifCLASSINFOpdf
\else
\fi
\usepackage{cite}
\usepackage{booktabs}
\usepackage{amsmath,amssymb,mathrsfs}
\usepackage{graphicx}
\usepackage{caption2}
\usepackage{subfigure}
\usepackage{bbding}
\usepackage{makecell}
\usepackage{longtable}
\usepackage{rotating}
\usepackage{multirow}

\usepackage{enumitem}
\usepackage{array}
\usepackage{color}
\usepackage{bm}
\usepackage{pifont}
\usepackage[colorlinks]{hyperref}
\usepackage[table]{xcolor}
\usepackage{multicol,lipsum}


\begin{document}
\title{A Comprehensive Survey on Community Detection with Deep Learning}

\author{Xing Su,
        Shan Xue,
        Fanzhen Liu,
        Jia Wu,~\IEEEmembership{Senior Member,~IEEE,}
        Jian Yang, 
        Chuan Zhou,
        Wenbin Hu,\\
        Cecile Paris, 
        Surya Nepal,
        Di Jin,
        Quan Z. Sheng, 
        and Philip S. Yu,~\IEEEmembership{Fellow,~IEEE}
\thanks{ X. Su, S. Xue, F. Liu, J. Wu, J. Yang, Q. Z. Sheng are with School of Computing, Macquarie University, Sydney,  Australia. 
E-mail: \{xing.su2, fanzhen.liu\}@students.mq.edu.au, \{emma.xue, jia.wu, jian.yang, michael.sheng\}@mq.edu.au}
\thanks{ C. Zhou is with  Academy of Mathematics and Systems Science, Chinese Academy of Sciences, Beijing, China. 
E-mail: zhouchuan@amss.ac.cn.}
\thanks{ W. Hu is with School of Computer Science, Wuhan University, Wuhan, China. E-mail: hwb@whu.edu.cn.}
\thanks{ S. Xue, C. Paris, S. Nepal are with CSIRO Data61, Sydney, Australia.
E-mail: \{emma.xue, surya.nepal, cecile.paris\}@data61.csiro.au}
\thanks{ D. Jin is with School of Computer Science and Technology,
Tianjin University, China.  E-mail: jindi@tju.edu.cn.}
\thanks{P.S. Yu is with Department of Computer Science, University of Illinois at Chicago, Chicago, USA. Email: psyu@uic.edu.}}

\markboth{IEEE Transactions on Neural Networks and Learning Systems,~Vol.XX, No.XX, 2021}%
{Su \MakeLowercase{\textit{et al.}}:A Comprehensive Survey on Community Detection with Deep Learning}

\maketitle

\begin{abstract}
A community reveals the features and connections of its members that are different from those in other communities in a network. Detecting communities is of great significance in network analysis. Despite the classical spectral clustering and statistical inference methods, we notice a significant development of deep learning techniques for community detection in recent years with their advantages in handling high dimensional network data. Hence, a comprehensive overview of community detection's latest progress through deep learning is timely to academics and practitioners. This survey devises and proposes a new taxonomy covering different state-of-the-art methods, including deep learning-based models upon deep neural networks, deep nonnegative matrix factorization and deep sparse filtering. The main category, \textit{i.e.}, deep neural networks, is further divided into convolutional networks, graph attention networks, generative adversarial networks and autoencoders. The survey also summarizes the popular benchmark data sets, evaluation metrics, and open-source implementations to address experimentation settings. We then discuss the practical applications of community detection in various domains and point to implementation scenarios. Finally, we outline future directions by suggesting challenging topics in this fast-growing deep learning field.
\end{abstract}


\maketitle
\IEEEdisplaynontitleabstractindextext
\IEEEpeerreviewmaketitle

\section{Introduction}\label{sec-introduction}

Communities have been investigated as early as the 1920s in sociology and social anthropology \cite{rice1927identification}. However, it is only after the 21st century that advanced scientific tools were primarily developed upon real-world data \cite{newman2004fast}. Since 2002 \cite{girvan2002community}, Girvan and Newman opened a new direction with graph partition. In the past ten years, researchers from computer science have extensively studied community detection \cite{Liu_2020} by utilizing network topological structures \cite{kim2015community,xie2013overlapping,javed2018community,xin2017deep} and semantic information \cite{CHUNAEV2020100286,jin2019graph} for both static and dynamic networks \cite{10.1145/3172867,8004509,cai2016survey} and small and large networks \cite{cui2014local,li2017most}. Graph-based approaches are developed increasingly to detect communities in environments with complex data structures \cite{leskovec2012learning,nicolini2017community}. Network dynamics and community impacts can be analyzed in details within community detection, such as rumor spread, virus outbreak, and tumour evolution.

\begin{figure}[!t]
\centering
\subfigure[graph]{\includegraphics[width=0.19\textwidth]{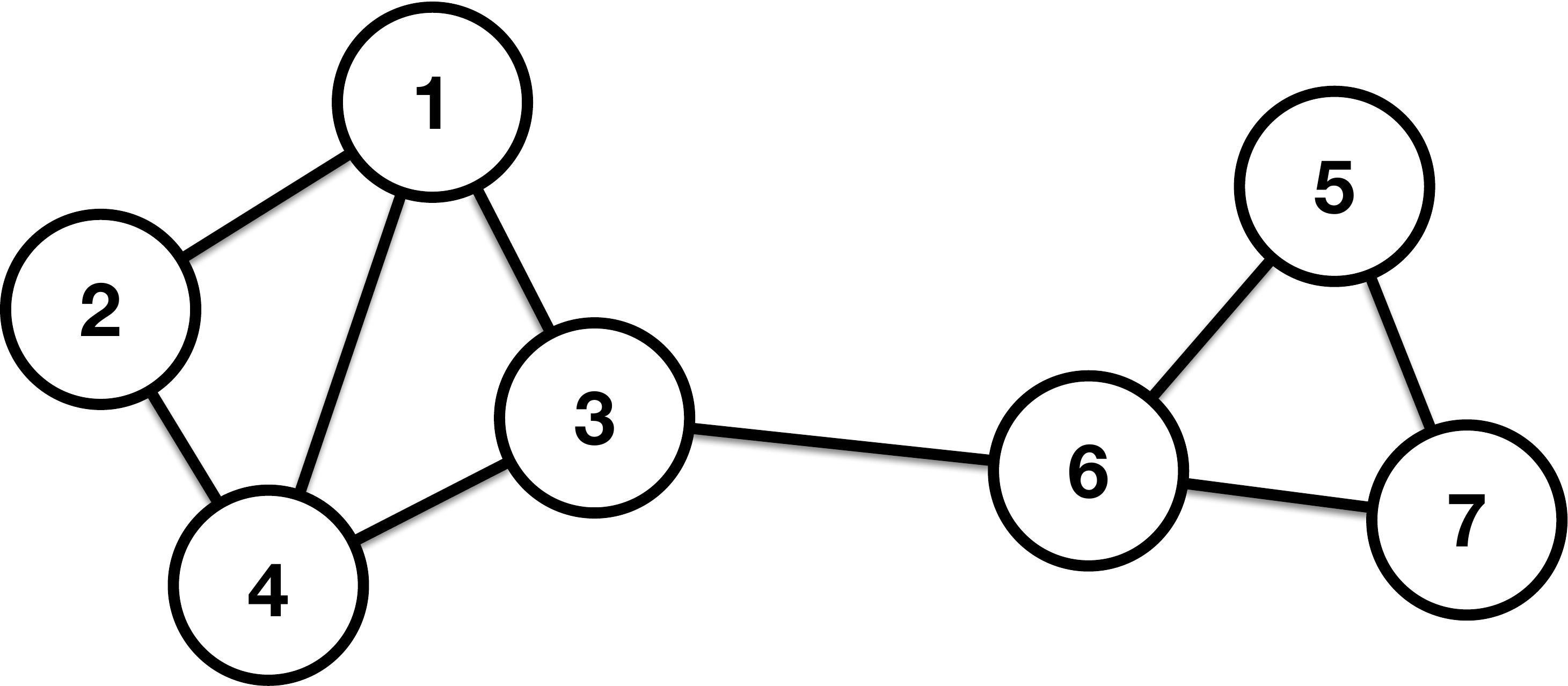}} \hfill
\subfigure[communities]{\includegraphics[width=0.27\textwidth]{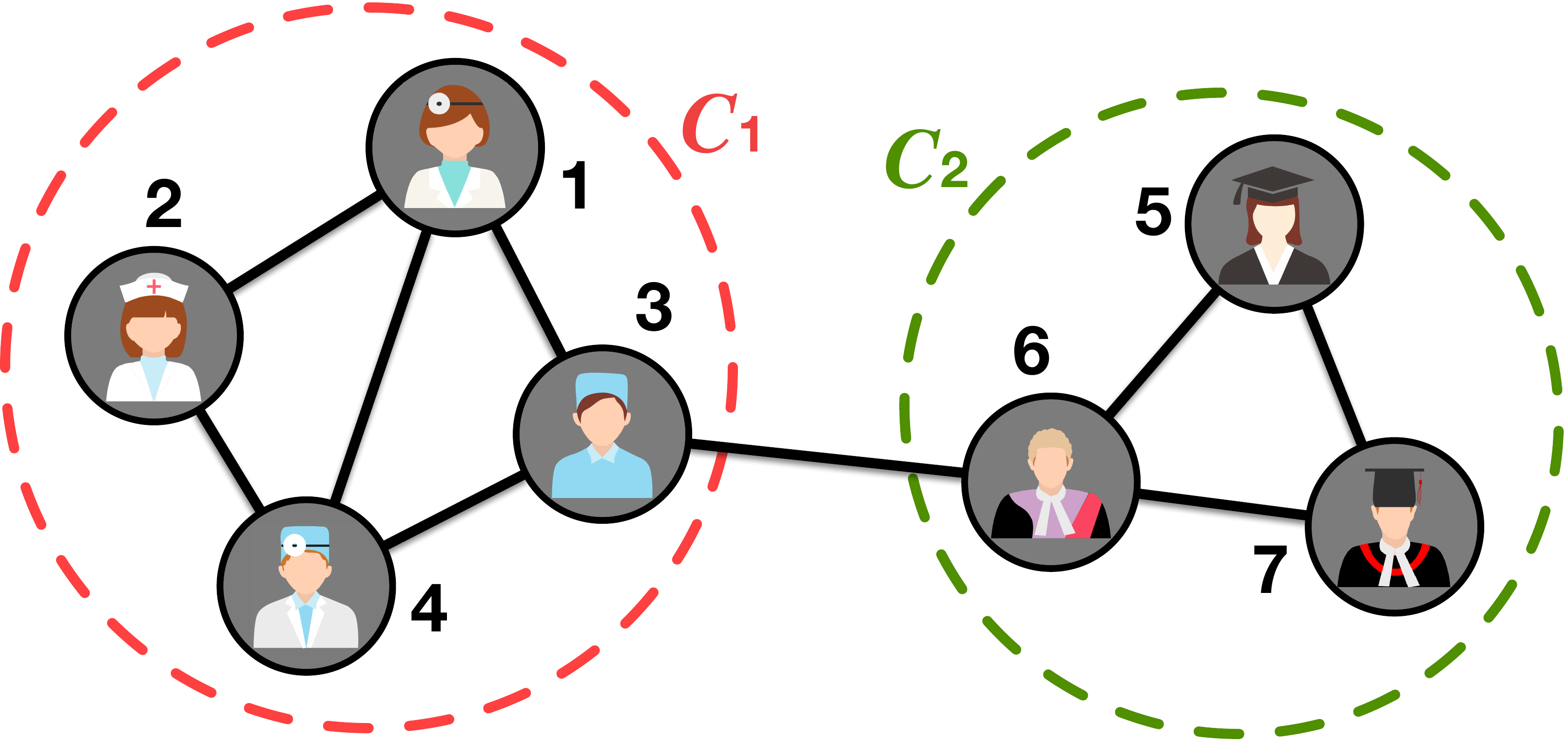}} 
\caption{\footnotesize (a) An illustration of a graph where nodes denote users in a social network. (b) An illustration of two communities ($C_1$ and $C_2$) based on the prediction of users' occupations. The detection utilizes users' closeness in online activities (topology) and account profiles (attributes). }
\label{fig-osn}
\end{figure}

Community detection is a research area with increasing practical significance. As the saying goes, \textit{Birds of a feather flock together} \cite{mcpherson2001birds}. Based on the theory of \textit{Six Degrees of Separation}, any person in the world can know anyone else through six acquaintances \cite{watts1998collective}. Indeed, our world is a vast network formed by a series of communities. For example, in social networks (Fig. \ref{fig-osn}), platform sponsors promote products to targeted users in detected communities   \cite{xie2012towards,yang2013community}. Community detection in citation networks \cite{chen2010community} determines the importance, interconnectedness, evolution of research topics and identifies research trends. In metabolic networks \cite{ravasz2002hierarchical,guimera2005functional} and Protein-Protein Interaction (PPI) networks \cite{chen2006detecting}, community detection reveals the complexities of metabolisms and proteins with similar biological functions. Similarly, community detection in brain networks \cite{sporns2016modular,nicolini2017community} reflects the functional and anatomical segregation of brain regions.

Many traditional techniques, such as spectral clustering \cite{amini2013pseudo,de2014laplacian} and statistical inference \cite{holland1983stochastic,karrer2011stochastic,airoldi2008mixed}, are applied to small networks and simple cases. However, real-world networks in rich nonlinear information make traditional models less applicable to practical applications, including complex topology and high dimensional features. Their computational costs are expensive. The powerful techniques of \textbf{deep learning} offer flexible solutions with good community detection performance to: (1) learn nonlinear network properties, such as relations denoting edges between nodes, (2) represent lower-dimensional network embeddings preserving the complicated network structure, and (3) achieve better community detection from various information. Therefore, deep learning for community detection is a new trend that demands a timely comprehensive survey\footnote{This paper is an extended vision of our published survey \cite{Liu_2020} in IJCAI-20, which is the first published work on the review of community detection approaches with deep learning.}. 

To the best of our knowledge, this paper is the first comprehensive survey focusing on deep learning contributions in community detection. In discovering inherent patterns and functions \cite{fortunato2010community,FORTUNATO20161}, existing surveys mainly focused on community detection on \textit{specific techniques} \cite{JMLR:v18:16-480,GARZA2019122058,chin2017semi,cai2016survey,8004509,jin2021survey}, \textit{different network types} \cite{10.1145/3172867,malliaros2013clustering,kim2015community}, \textit{community types} \cite{xie2013overlapping,javed2018community}, and \textit{application scenarios} \cite{bedi2016community,CHUNAEV2020100286}. The surveys on \textit{specific techniques} are summarized but not limited to partial detection based on probabilistic graphical models \cite{JMLR:v18:16-480,jin2021survey}, Label Propagation Algorithm (LPAs) \cite{GARZA2019122058,chin2017semi}, and evolutionary computation for single- and multi-objective optimizations \cite{cai2016survey,8004509}. In terms of different \textit{network types}, researchers provide overviews on dynamic networks \cite{10.1145/3172867}, directed networks \cite{malliaros2013clustering} and multi-layer networks \cite{kim2015community}. Moreover, detection techniques are reviewed over disjoint and overlapping \cite{xie2013overlapping,javed2018community} \textit{community types}. Regarding \textit{application scenarios}, the focus has been on techniques on social networks \cite{bedi2016community,CHUNAEV2020100286}.

\begin{figure*}[!t]
\centering
\includegraphics[width=0.75\textwidth]{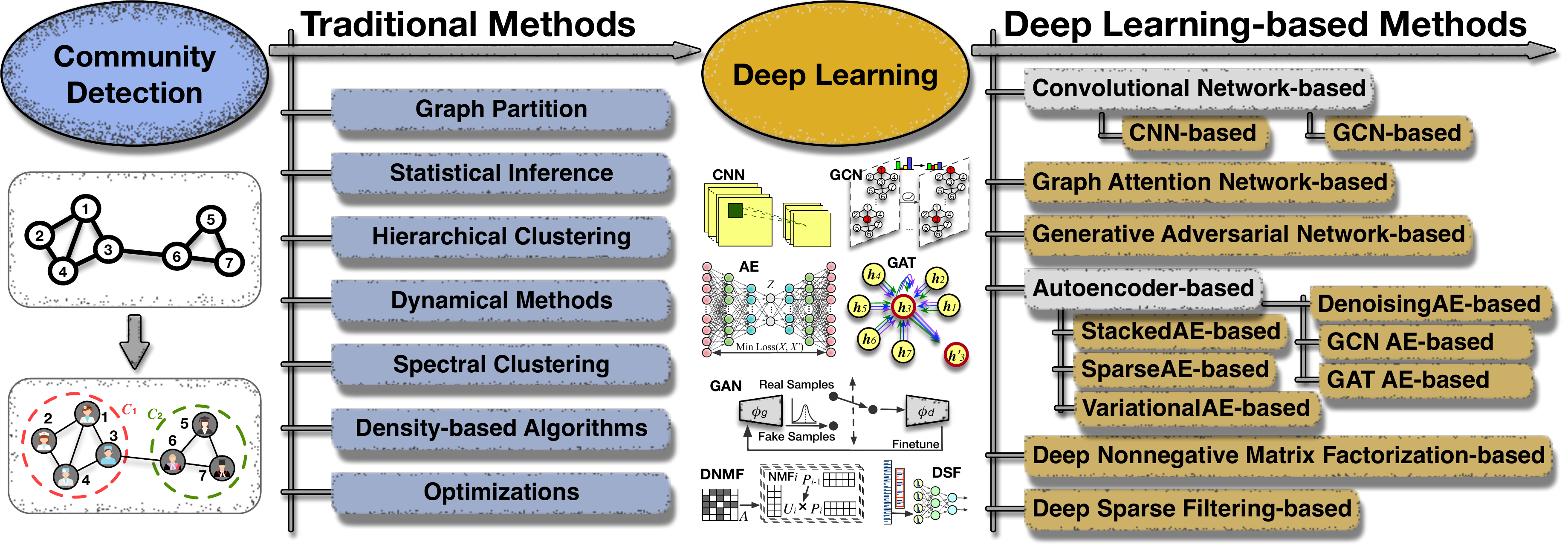}
\caption{\footnotesize Traditional community detection methods and the taxonomy of deep learning-based methods.}
\label{fig-taxonomy}
\end{figure*}

Observing the past, current and future trends, this paper aims to support researchers and practitioners to understand the community detection field with respect to:

\begin{itemize}
\item \textbf{Systematic Taxonomy and Comprehensive Review.} We propose a new systematic taxonomy for this survey (see Fig. \ref{fig-taxonomy}). For each category, we review, summarize and compare the representative works. We also briefly introduce community detection applications in the real world. These scenarios provide horizons for future community detection research and practices.

\item \textbf{Abundant Resources and High-impact References.} The survey collects open resources, including benchmark data sets, evaluation metrics and technique implementations. Publications in the latest high-impact international conferences and high-quality peer-reviewed journals cover data mining, artificial intelligence, machine learning and knowledge discovery.

\item \textbf{Future Directions.} As deep learning is a new research area, we discuss current limitations, critical challenges and open opportunities for future directions.
\end{itemize}

The rest of the article is organized as follows: Section \ref{sec-preliminaries} defines essential notations, concepts, input and output of deep learning approaches. Section \ref{sec-traditional-alg} overviews community detection development. Section \ref{sec-categorization} introduces a deep learning taxonomy. Sections \ref{sec-model-cn}--\ref{sec-model-filtering} summarize comprehensive reviews on each category in the taxonomy. Section \ref{sec-experimentation} and Section \ref{sec-application} organize the popular implementation resources and real-world applications. Lastly, Section \ref{sec-future} discusses current challenges, suggests future research directions before the conclusion in Section \ref{sec-conclusion}. The supporting materials can be found in APPENDIX \ref{sec-stdlcd} (core techniques of reviewed literature summarized in tables), APPENDIX \ref{sec-ddds}--\ref{sec-appcode} (resource descriptions of data sets, evaluation metrics and implementation projects) and APPENDIX \ref{sec-abbr} (abbreviations).

\section{Definitions and Preliminaries} \label{sec-preliminaries}

In this section, preliminaries include primary definitions, notations in TABLE \ref{tab-notations}, and general inputs and outputs of deep learning-based community detection models. 

\vspace{2mm}
\noindent
\textbf{\textit{Definition 1: Network.}} Given a basic network $\mathcal{G} = (V, E)$, where $V=\{v_{1}, \cdots, v_{n}\}$ is the node set with $E=\{e_{ij}\}_{i,j=1}^{n}$ representing the edge set among nodes. $N(v_i) = \{u\in V | (v_i, u)\in E\}$ defines the neighborhood of a node $v_i$. $\bm{A}=[a_{ij}]$ denotes an $n\times n$ dimensional adjacency matrix, where $a_{ij} = 1$ if $e_{ij}\in E$, otherwise $a_{ij} = 0$. If $a_{ij}\neq a_{ji}$, $\mathcal{G}$ is a \textit{directed network}, or else it is an \textit{undirected network}. If $a_{ij}$ is weighted by $w_{ij} \in \bm{W}$, $\mathcal{G} = (V,E,\bm{W})$ is a \textit{weighted network}, otherwise it is an \textit{unweighted network}. If $a_{ij}$'s value differs in $+1$ (positive) and $-1$ (negative), $\mathcal{G}$ is a \textit{signed network}. If node $v_i \in V$ is attributed by $\bm{x}_i \in \bm{X} \subseteq \mathbb{R}^{n\times d}$, $\mathcal{G} = (V,E,\bm{X})$ is an \textit{attributed network}, otherwise it is an \textit{unattributed network}. 

\begin{table}[!t]
\centering \footnotesize
\renewcommand\arraystretch{1}
\setlength{\tabcolsep}{3mm}
\caption{Notations and descriptions used in this paper.}
\rowcolors{2}{white}{gray!30}
\begin{tabular}{l|l}
\toprule[1pt]
\textbf{Notations} & \textbf{Descriptions }\\ \midrule
$\mathbb{R}$ & A data space \\ \hline
$\mathcal{G}$ & A graph \\
$V$, $E$, $\mathcal{C}$ & A set of nodes, edges, communities \\
$v_i$, $e_{ij}$, $C_k$ & The $i$-th node, edge of ($v_i,v_j$), $k$-th community \\
$N(v_i)$ & A neighborhood of $v_i$ \\
$\bm{A}$, $a_{ij}$ & An adjacency matrix, value \\
$\bm{X}$, $\bm{x}_i$ & A node attribute matrix, vector \\ 
$y_i$, $c_k$ & The node label of $v_i$, community label of $C_k$ \\
$y^k_i$ & The binary community label of $v_i$ in $C_k$ \\
$n$, $m$, $K$, $d$ & The number of nodes, edges, communities, attributes \\ \hline
$\mathcal{A}_{ij}$ & The anchor links between graphs $(\mathcal{G}_i,\mathcal{G}_j)$  \\
$\mathcal{V}$, $\mathcal{E}$, $\mathcal{X}$ & A set of heterogeneous nodes, edges, attributes \\ 
$E^{(r)}$ & $r$-th type edges in multiplex network \\
$\bm{A}(+,-)$ & The adjacency matrix of a signed network \\
$\hat{\bm{A}}$ & A reconstructed adjacency matrix\\
$\widetilde{\bm{X}}$ & The corrupted node attribute matrix \\ \hline
$\bm{D}$ & A degree matrix \\
$\bm{B}$, $b_{ij}$ & A modularity matrix, value \\
$\bm{S}$, $s_{ij}$ & A similarity matrix, value \\
$\bm{O}$, $o_{ij}$ & Node pairwise constraint matrix, value \\
$\bm{P}$, $p_{ij}$ & Community membership matrix, probability of ($v_i,C_j$) \\
$\bm{L}$, $\bm{M}$ & A Laplacian, Markov matrix \\
$\bm{Z}$, $\bm{z}$ & A latent variable matrix, vector \\
$\Theta$ & Trainable parameters \\
$\bm{W}^{(l)}$ & The weight matrix of the $l$-th layer in DNN \\
$\bm{H}^{(l)}$, $\bm{h}_i^{(l)}$ & The $l$-th layer representation matrix, vector \\ \hline
$\sigma(\cdot)$ & An activation function\\
$\mathcal{L}$ & A loss function \\
$\Omega$ & A sparsity penalty \\
$\vert \cdot \vert$  & The length of a set \\
$\|\cdot \|$ & The norm operator \\
$\phi_g$, $\phi_d$ & A generator, discriminator \\
$\phi_e$, $\phi_r$ & An encoder, decoder \\
\bottomrule[1pt]
\end{tabular}
\label{tab-notations}
\end{table}

\vspace{2mm}
\noindent
\textbf{\textit{Definition 2: Community.}} Given a set of communities $\mathcal{C} = \{C_1, C_2, \cdots, C_K\}$, each community $C_{k}$ is a partition of $\mathcal{G}$ which keeps regional structure and cluster properties. A node $v_i$ clustered into community $C_{k}$ should satisfy the condition that the internal node degree inside the community exceeds its external degree. Suppose $C_{k} \cap C_{k'} = \emptyset$, ($\forall k, k'$), $\mathcal{C}$ denotes \textit{disjoint communities}; otherwise \textit{overlapping communities}.

\begin{figure*}[!t]
\centering
\includegraphics[width=0.95\textwidth]{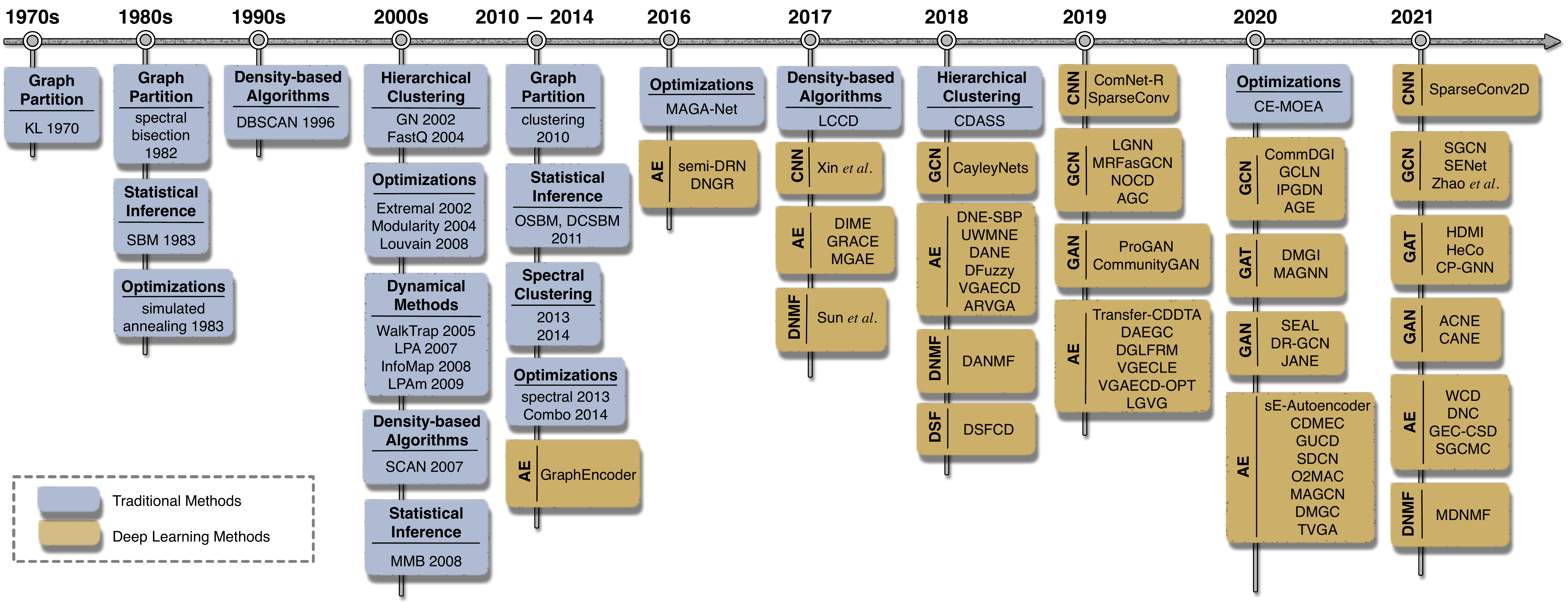}
\caption{\footnotesize A timeline of community detection development.}
\label{fig-timeline}
\end{figure*}

\vspace{2mm}
\noindent
\textbf{Community Detection Input.} Deep learning models take inputs as the network topology and network attributes. The topology formed by nodes and edges can be represented in matrices such as adjacency matrix $\bm{A}$, signed adjacency matrix $\bm{A}(+,-)$ and the measurement matrices like the modularity matrix $\bm{B}$. Network attributes denote additional information on the network entities such as node attributes $\bm{X}$.

\vspace{2mm}
\noindent
\textbf{Community Detection Output.} 
Community detection methods aim to output a set of communities which can be either disjoint or overlapping. TABLEs \ref{table-gcn}--\ref{table-autoencoder} in APPENDIX \ref{sec-stdlcd} indicate different output by different community detection methods. Disjoint communities represent, for example, student clubs that allow a student to join only one club. Overlapping communities describe, for instance, users who participate in several circles in the social network. The methods for overlapping communities can detect disjoint communities.

\section{A Development of Community Detection} \label{sec-traditional-alg}

Community detection has been significant in network analysis and data mining. Fig. \ref{fig-timeline} illustrates its development from traditional to deep learning in a timeline. Their respective categories are summarized in Fig. \ref{fig-taxonomy} in the left and right parts, respectively. Two categories of methods suggest the development changes. Traditional methods mainly explore communities from network structures. We briefly review these seven categories of methods in this section. Deep learning uncovers deep network information and models complex relationships from high-dimensional data to lower-dimensional vectors. These will be reviewed in the following sections.

\vspace{2mm}
\noindent
\textbf{Graph Partition.} These methods, well known as graph clustering \cite{fortunato2010community}, are employed in deep learning models. They partition a network into communities with a given number $K$. Kernighan-Lin \cite{kernighan1970efficient} is a representative heuristic algorithm. It initially divides a network into two arbitrary subgraphs and optimizes on nodes. Spectral bisection \cite{barnes1982algorithm} is another representative method applying spectrum Laplacian matrix.

\vspace{2mm}
\noindent
\textbf{Statistical Inference.} Stochastic Block Model (SBM) \cite{holland1983stochastic} is a widely applied generative model by assigning nodes into communities and controlling their probabilities of likelihood. The variants include Degree-Corrected SBM (DCSBM) \cite{karrer2011stochastic} and Mixed Membership SBM (MMB) \cite{airoldi2008mixed}.

\vspace{2mm}
\noindent
\textbf{Hierarchical Clustering.} This group of methods discover hierarchical community structures (\textit{i.e.}, dendrogram) in three ways: divisive, agglomerative and hybrid. The Girvan-Newman (GN) algorithm finds community structure in a divisive way by successively removing edges such that a new community occurs \cite{girvan2002community,newman2004finding}. Fast Modularity (Fast\textit{Q}) \cite{newman2004fast,clauset2004finding}, an agglomerative algorithm, gradually merges nodes, each of which is initially regarded as a community. Community Detection Algorithm based on Structural Similarity (CDASS) \cite{zarandi2018community} jointly applies divisive and agglomerative strategies in a hybrid way. 

\vspace{2mm}
\noindent
\textbf{Dynamical Methods.} Random walks are utilized to detect communities dynamically. For example, the random walk in WalkTrap \cite{pons2005computing} calculates node distances and the probability of community membership. Information Mapping (InfoMap) \cite{rosvall2008maps} applies the minimal-length encoding. Label Propagation Algorithm (LPA) \cite{raghavan2007near} identifies diffusion communities through an information propagation mechanism. 

\vspace{2mm}
\noindent
\textbf{Spectral Clustering.} The network spectra reflects the community structure. Spectral clustering \cite{amini2013pseudo} partitions the network on the normalized Laplacian matrix and the regularized adjacency matrix, and fits SBM in the pseudo-likelihood algorithm. On the spectra of normalized Laplacian matrices, Siemon \textit{et al.}\cite{de2014laplacian} integrated communities in macroscopic and microscopic neural brain networks to obtain clusters.

\vspace{2mm}
\noindent
\textbf{Density-based Algorithms.} Significant clustering algorithms include Density-Based Spatial Clustering of Applications with Noise (DBSCAN) \cite{ester1996density}, Structural Clustering Algorithm for Networks (SCAN) \cite{xu2007scan} and Locating Structural Centers for Community Detection (LCCD) \cite{wang2017locating}. They identify communities, hubs and outliers by measuring entities' density. 

\begin{figure*}[!t]
\centering
\includegraphics[width=0.9\textwidth]{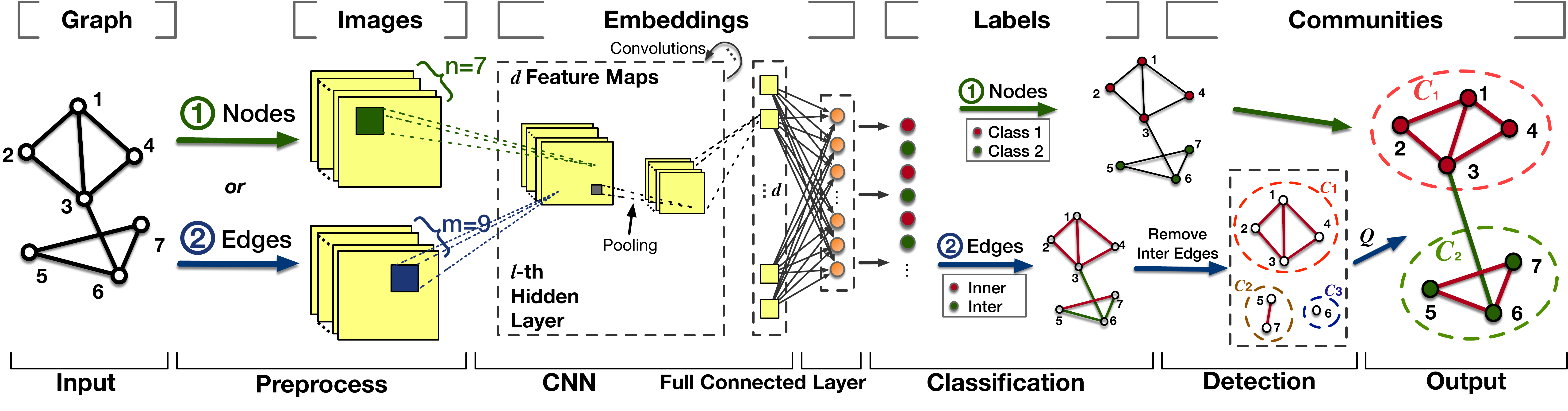}
\caption{\footnotesize A general framework for CNN-based community detection with details in Section \ref{sec-model-cnn}. As Convolutional Neural Network (CNN) input, the graph is preprocessed into the image data on nodes or edges. The $d$-dimensional latent features are convolutionally mapped within multiple CNN hidden layers and a final fully connected layer outputs representations of each node or each edge for classifications. Focusing on nodes, the flow \ding{172} predicts community labels in $k$ classes that nodes with same labels are clustered into a community. Focusing on edges, the flow \ding{173} predicts edge labels in two classes, \textit{i.e.}, inner and inter. The preliminary communities are formed by removing inter-community edges and optimized by merging with a measurement such as Modularity $Q$.} \label{fig-cnn}
\end{figure*}

\vspace{2mm}
\noindent
\textbf{Optimizations.} Community detection generally maximizes the likelihood. Modularity (\textit{Q}) \cite{newman2004finding} is the most classic optimization function following its variant Fast\textit{Q} \cite{newman2004fast,clauset2004finding}. Louvain \cite{blondel2008fast} is another well-known optimization algorithm that employs the node-moving strategy on optimized modularity. Moreover, the extensions of greedy optimizations include simulated annealing \cite{kirkpatrick1983optimization}, extremal optimization \cite{boettcher2002optimization} and spectral optimization \cite{newman2013spectral}. Effective in local and global searching \cite{liu2019evolutionary}, the evolutionary optimizations consist of single and multiple objectives. For example, Multi-Agent Genetic Algorithm (MAGA-Net) \cite{li2016multi} applies a single modularity function. Combo \cite{sobolevsky2014general} mixes Normalized Mutual Information (NMI) \cite{ana2003robust} and Conductance (CON) \cite{kannan2004clusterings}. \textit{Q}, NMI and CON estimate the network partition quality with details in APPENDIX \ref{sec-EM}.

\vspace{2mm}
\noindent
\textbf{Why community detection needs deep learning?} Capturing information from connections in a straightforward way may lead to sub-optimal community detection results. Deep learning models \cite{leCun2015deep} bring the following additional advantages to community detection. Thus, deep learning-based community detection has been a new emerging branch. Its general framework learns lower-dimensional vectors from high-dimensional data of complex structural relationships \cite{jin2019graph,park2020unsupervised,jane2020yang,yang2016modularity}. It, therefore, enables knowledge discoveries via the state-of-the-art machine learning and data mining techniques. The framework with representations can further embed non-structural features, such as node attributes, to increase the knowledge of community memberships \cite{zhang2019attributed,jing2021hdmi,gao2019progan}. Besides, groups of information from nodes \cite{park2020unsupervised}, edges \cite{shen2018deep}, neighborhoods \cite{yang2017graph} or multi-graphs \cite{fan2020one2multi} can be jointly recognized with special attentions in the deep learning process, leading to effective community detection results. With the deep learning ability to big data, more and more large-scale \cite{li2017most}, high-sparse \cite{sperli2019deep}, complex structural \cite{du2018galaxy,fu2020magnn}, dynamic \cite{wang2020evolutionary} networks in real-world scenarios can be explored. Along with the achievements in a relatively short time (in Fig. \ref{fig-timeline}), the development of community detection within the deep learning domain keeps facing a series of challenges. In this paper, we review accomplishments in Sections \ref{sec-model-cn}--\ref{sec-application}, and point out opportunities and challenges in Section \ref{sec-future}.

\section{A Taxonomy of Community Detection with Deep Learning} \label{sec-categorization}

This survey proposes a taxonomy for deep community detection methods according to the iconic characteristics of the employed deep learning models. The taxonomy summarizes six categories: convolutional networks, Graph Attention Network (GAT), Generative Adversarial Network (GAN), Autoencoder (AE), Deep Nonnegative Matrix Factorization (DNMF) and Deep Sparse Filtering (DSF). In this taxonomy, convolutional networks include Convolutional Neural Network (CNN) and Graph Convolutional Network (GCN). They both contribute convolutions in representing latent features for community detection. GATs are significant at special attentions to community signals. The adversarial training process between the input graph and fake samples in the GAN model is successfully employed by community detection. On a universal AE framework, subcategories consist of stacked AE, sparse AE, denoising AE, graph convolutional AE, graph attention AE and Variational AE (VAE). The taxonomy structure is shown in Fig. \ref{fig-taxonomy}. The following Sections \ref{sec-model-cn}--\ref{sec-model-filtering} overview each category of methods, respectively.

\section{Convolutional Network-based Community Detection} \label{sec-model-cn}

Community detection applying convolutional network models includes Convolutional Neural Networks (CNNs) and Graph Convolutional Networks (GCNs). CNNs \cite{lecun1995convolutional} are a particular class of feed-forward Deep Neural Network (DNN) proposed for grid-like topology data such as image data, where convolution layers reduce computational costs and pooling operators ensure CNN's robustness in feature representations. GCNs \cite{kipf2016semi} are proposed for graph-structured data based on CNNs and performed on the first-order approximation of spectral filters. The rule of GCNs' layer-wise propagation is: 
\begin{equation}
\small
\label{equ-gcn}
\bm{H}^{(l+1)} = \sigma (\tilde{\bm{D}}^{-\frac{1}{2}} \tilde{\bm{A}} \tilde{\bm{D}}^{-\frac{1}{2}} \bm{H}^{(l)} \bm{W}^{(l)} ), 
\end{equation}
where the latent representations of the $l$-th layer are preserved in the matrix $\bm{H}^{(l)}$ ($\bm{H}^{(0)}=\bm{X}$), through the activation function $\sigma(\cdot)$ with the layer-specific trainable weight matrix $\bm{W}^{(l)}$; $\tilde{\bm{A}}=\bm{A}+\bm{I}_n$ with $\bm{I}_n$ denoting the identity matrix; and $\tilde{\bm{D}}_{ii}=\sum_j \tilde{a}_{ij}$ where $\tilde{a}_{ij} \in \tilde{\bm{A}}$.

\subsection{CNN-based Community Detection} \label{sec-model-cnn}
The existing CNN-based community detection methods implement CNN models with strict data input limitations so that they need to preprocess for image-formatted and labeled data (Fig. \ref{fig-cnn}). Techniques below solve particular problems in community detection with summaries in TABLE \ref{table-cnn}.

The traditional community detection is an unsupervised learning task for deep learning models, incomplete topological structure affects its neighborhood analyzes and reduces the accuracy of community detection. However, networks in the real world have limited structural information. To this end, Xin \textit{et al.} \cite{xin2017deep} proposed the first community detection model based on supervised CNN for Topologically Incomplete Networks (TINs). The model has two CNN layers with max-pooling operators for network representation and a fully connected DNN layer for community detection. The CNN architecture gradually recovers intact latent features from the rudimental input. The convolutional layers represent each node's local features from different views. The last full connection layer $f$ updates communities for each node $v_i$:
\begin{equation}
\small
o_i^k = \sigma(b_k^f + \bm{W}_k^f\bm{h}^{(2)}_i), 
\end{equation}
where $\sigma$ indicates the \textit{sigmoid} function, $\bm{W}_k^f$ and $b_k^f$ are weights and bias of the $k$-th neuron $o_i^k$, and $\bm{h}^{(2)}_i$ is the node representation vector output by the previous two convolutional layers. The model performs back-propagation to minimize:
\begin{equation}
\small
\mathcal{L} = \frac{1}{2}\sum\nolimits_i\|\bm{o}_i - \bm{y}_i\|_2^2 = \frac{1}{2}\sum\nolimits_i\sum\nolimits_k(o_i^k - y^k_i)^2, 
\end{equation}
where $\bm{y}_i$ denotes a ground truth label vector of node $v_i$, in which $y^k_i\in\{0,1\}$ represents whether $v_i$ belongs to the $k$-th community or not. The experiments on this model achieve a community detection accuracy around 80\% in TINs with 10\% labeled nodes and the rest unlabeled, indicating that a high-order neighborhood representation in a range of multiple hops can improve the community detection accuracy. 

To deal with the high sparsity in large-scale social networks, the following two methods work on particular sparse matrices (\textit{i.e.}, non-zero elements of adjacency matrix) for efficient community detection. Sperl{\'\i} \cite{sperli2019deep} designed a sparse matrix convolution (SparseConv) into CNN. Santo \textit{et al.} \cite{de2021deep} further proposed a sparse CNN approach with a SparseConv2D operator so that the number of operations in the approach are significantly decreased.

Community Network Local Modularity R (ComNet-R) \cite{cai2020edge} is an edge-2-image model for community detection, classifying edges within and across communities through CNN. ComNet-R removes inter-community edges to prepare the disconnected preliminary communities. The optimization process is designed to merge communities based on the local modularity.

\begin{figure*}[!t]
\centering
\includegraphics[width=0.9\textwidth]{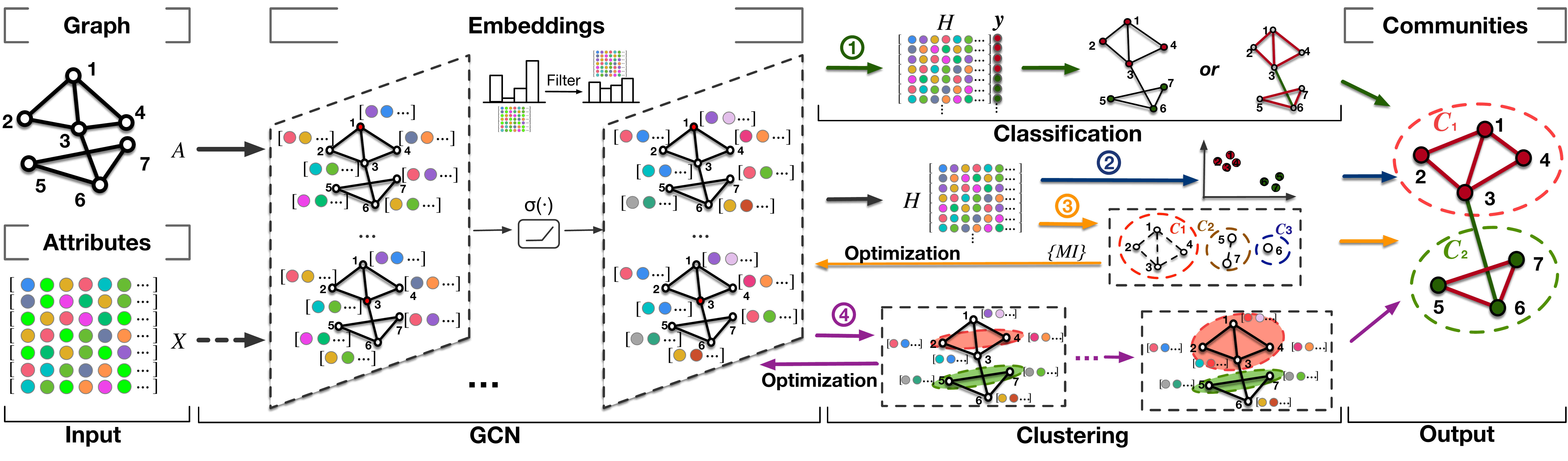}
\caption{\footnotesize A general framework for GCN-based community detection with details in Section \ref{sec-model-gcn}. It inputs a graph structure $\bm{A}$ and optional node attributes $\bm{X}$. Within multiple Graph Convolutional Network (GCN) layers, graph latent features are smoothed based on community detection requirements. The graph representation learning is activated by $\sigma(\cdot)$. Four community detection frameworks are illustrated in \ding{172}--\ding{175} applying either final node representations (\ding{172} and \ding{173}) or temporal representations in hidden layers (\ding{174} and \ding{175}). Given node labels, communities are detected based on node classification in \ding{172} while \ding{173} implements node clustering on embeddings $\bm{H}$ and can further optimized in \ding{174} with measurements, \textit{e.g.}, Mutual Information (MI), for best community affiliations. \ding{175} jointly optimizes clustering results and GCN representations that gradually detects each node into communities with convolutionally represented node embeddings.}
\label{fig-gcn}
\end{figure*}

\subsection{GCN-based Community Detection} \label{sec-model-gcn}

GCNs aggregate node neighborhood's information in deep graph convolutional layers to globally capture complex features for community detection (Fig. \ref{fig-gcn}). There are two classes of community detection methods based on GCNs: (1) supervised/semi-supervised community classification, and (2) community clustering with unsupervised network representation. Community classification methods are limited by a lack of labels in the real world. In comparison, network representations are more flexible to cluster communities through techniques such as matrix reconstructions and objective optimizations. TABLE \ref{table-gcn} in APPENDIX \ref{sec-stdlcd} compares the techniques.

GCNs employ a few traditional community detection methods as deep graph operators, such as SBMs for statistical inference, Laplacian matrix for spectrum analysis and probabilistic graphical models for belief propagation. For example, Line Graph Neural Network (LGNN) \cite{chen2019supervised} is a supervised community detection model, which improves SBMs with better community detection performance and reduces the computational cost. Integrating the non-backtracking operator with belief propagation's message-passing rules, LGNN learns node represented features in directed networks. The softmax function identifies conditional probabilities that a node $v_i$ belongs to the community $C_k$ ($o_{i,k} = p(y_i=c_k|\Theta,\mathcal{G}$), and minimizes the cross-entropy loss over all possible permutations $S_{\mathcal{C}}$ of community labels:
\begin{equation}
\small
\mathcal{L}(\Theta)=\min_{\pi \in S_{\mathcal{C}}}{-\sum\nolimits_i\log{o_{i, \pi(y_{i})}}}.
\end{equation}

Since GCN is not initially designed for the community detection task, community structures are not the focus in learning node embeddings. To meet this gap, a semi-supervised GCN community detection model, named MRFasGCN, characterizes hidden communities through extending the network-specific Markov Random Field as a new convolutional layer (eMRF) which makes MRFasGCN community oriented and performs a smooth refinement to the coarse results from GCN. To enable unsupervised community detection, the GCN-based framework, denoted as SGCN \cite{wang2021unsupervised}, designs a local label sampling model to locate the structural centers for community detection. By integrating the label sampling model with GCN, SGCN encodes both network topology and node attributes in the training of each node's community membership without any prior label information. 

In terms of a probabilistic inference framework, detecting overlapping communities can be solved by a generative model inferring the community affiliations of nodes. For example, Neural Overlapping Community Detection (NOCD) \cite{shchur2019overlapping} combines the Bernoulli–Poisson (BP) probabilistic model and a two-layer GCN to learn community affiliation vectors by minimizing BP's negative log-likelihood. By setting a threshold to keep recognizing and removing weak affiliations, final communities are obtained.

Spectral GCNs represent all latent features from the node's neighborhood. The features of neighboring nodes converge to the same values by repeatedly operating Laplacian smoothing in deep GCN layers. However, these models lead to an over-smoothing problem in community detection. To reduce the negative impact, Graph Convolutional Ladder-shape Networks (GCLN) \cite{gcln2020hu} is designed as a new GCN architecture for unsupervised community detection ($k$-means), which is based on the U-Net in the CNN field. A contracting path and an expanding path are built symmetrically in GCLN. The contextual features captured from the contracting path fuse with the localized information learned in the expanding path. The layer-wise propagation follows Eq. (\ref{equ-gcn}).

Since different types of connections are generally treated as plain edges, GCNs represent each type of connection and aggregate them, leading to redundant representations. Independence Promoted Graph Disentangled Network (IPGDN) \cite{ipgdn2020liu} distinguishes the neighborhood into different parts and automatically discovers the nuances of a graph's independent latent features to reduce the difficulty in detecting communities. IPGDN is supported by Hilbert-Schmidt Independence Criterion (HSIC) regularization \cite{gretton2005measuring} in neighborhood routings.

For attributed graphs, community detection by GCNs relies on both structural information and attributed features, where neighboring nodes and the nodes with similar features are likely to cluster in the same community. Therefore, graph convolutions multiply the above two graph signals and need to filter out high-frequency noises smoothly. To this end, a low-pass filter is embedded in Adaptive Graph Convolution (AGC) \cite{zhang2019attributed} with spectral clustering through:
\begin{equation}
\small
p(\lambda_q) = (1 - \frac{1}{2} \lambda_q)^k,
\end{equation}
where the frequency response function of $\mathcal{G}$ denotes $p(\Lambda)=\text{diag}\left(p\left(\lambda_1\right),\cdots,p\left(\lambda_n\right)\right)$ decreasing and nonnegative on all eigenvalues. AGC convolutionally selects suitable neighborhood hop sizes in $k$ and represents graph features by the $k$-order graph convolution as:
\begin{equation}
\small
\bar{\bm{X}} = (I - \frac{1}{2}\bm{L}_s)^k\bm{X}, 
\end{equation}
where $\bm{L}_s$ denotes the symmetrically normalized graph Laplacian on $\lambda_q$. The filter smooths the node embedding $\bar{\bm{X}}$ adjusting to $k$ that higher $k$ leads to better filtering performance.

Adaptive Graph Encoder (AGE) \cite{cui2020adaptive} is another smoothing filter model scalable to community detection. AGE adaptively performs node similarity measurement ($\bm{S} = \left[s_{ij}\right]$) and $t$-stacked Laplacian smoothing filters ($\bar{\bm{X}} = (\bm{I} - \gamma \bm{L})^t \bm{X}$):
\begin{equation}
\small
\mathcal{L} = \sum\nolimits_{(v_i,v_j)\in V'} -s'_{ij}\log(s_{ij}) - (1-s'_{ij})\log(1 - s_{ij}),  
\end{equation}
where $V'$ denotes balanced training set over positive (similar) and negative (dissimilar) samples, $s'_{ij}$ is the ranked binary similarity label on node pairs $(v_i,v_j)$.

A few works make significant contributions to GCN's filters. For example, Graph Convolutional Neural Networks with Cayley Polynomials (CayleyNets) \cite{levie2018cayleynets}, in a spectral graph convolution architecture, proposes an effective Cayley filter for high-order approximation on community detection. It specializes in narrow-band filtering since low frequencies contain extensive community information for community detection-aimed representations. Cooperating with the Cayley filter, CayleyNets involves a mean pooling in spectral convolutional layers and a semi-supervised softmax classifier on nodes for community membership prediction. 

To capture the global cluster structure for community detection, the Spectral Embedding Network for attributed graph clustering (SENet) \cite{zhang2021spectral} introduces the loss of spectral clustering into a three-layer GCN's output layer by minimizing: 
\begin{equation}
\small
\mathcal{L} = -\operatorname{tr}((\bm{H}^{(3)})^{\top} \bm{D}^{-\frac{1}{2}} \bm{K} \bm{D}^{-\frac{1}{2}} \bm{H}^{(3)}), 
\end{equation}
where $\bm{K}$ is a kernel matrix encoding typologies and attributes. 

Community Deep Graph Infomax (CommDGI) \cite{zhang2020commdgi} jointly optimizes graph representations and clustering through Mutual Information (MI) on nodes and communities, and measures graph modularity for maximization. It applies contrastive training to obtain better representations, $k$-means for node clustering and targets cluster centers. Zhao \textit{et al.} \cite{ijcai2021-473} proposed a graph debiased contrastive learning which simultaneously performs representations and clustering so that the clustering results and discriminative representations are both improved.

\begin{figure}[!t]
\centering
\includegraphics[width=0.48\textwidth]{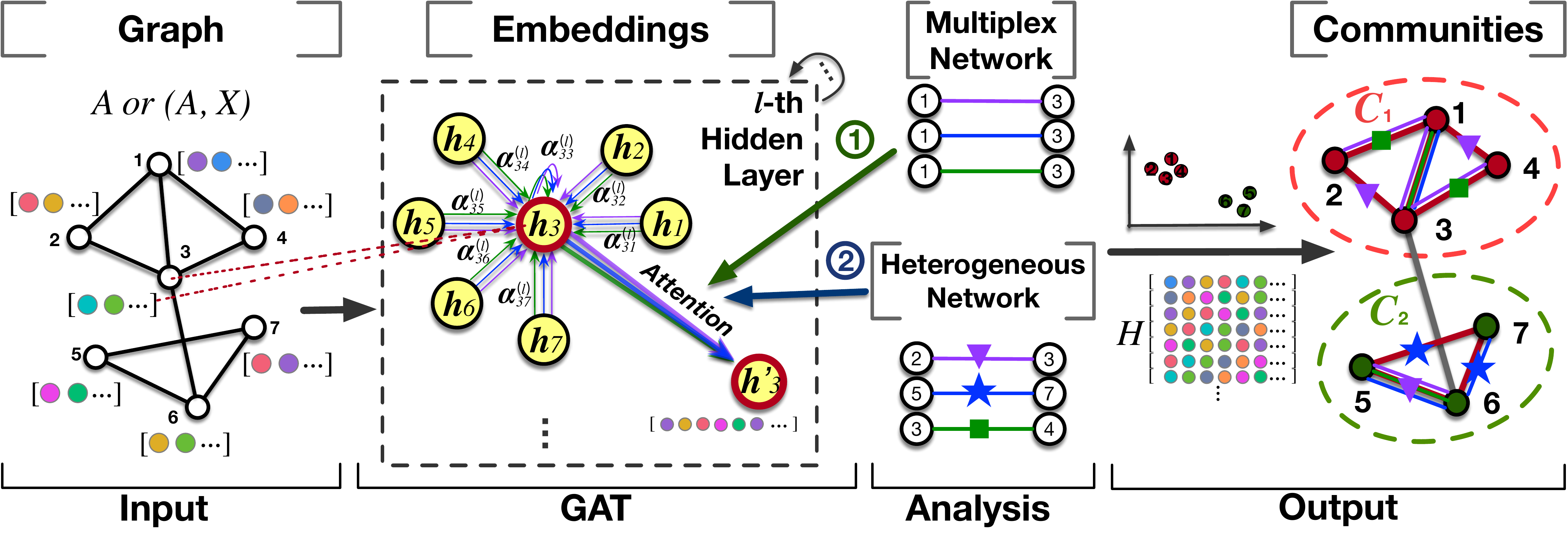}
\caption{\footnotesize A general framework for GAT-based community detection with details in Section \ref{sec-model-gat}. Graph Attention Network (GAT) assigns attention coefficients $\bm{\alpha}^{(l)}_{ij}$: \{green, blue, purple\} in each neighborhood $N(v_i)$ in $l$-th hidden layer. The represented vector $\bm{h}'_i$ aggregates available information: \ding{172} differently colored edges between the same node pairs in multiplex networks, or \ding{173} metapaths or context paths in a heterogeneous network. The GAT embeddings $\bm{H}$ are analyzed to cluster communities.} \label{fig-gat}
\end{figure}

\section{Graph Attention Network-based Community Detection} \label{sec-model-gat}

Graph Attention Networks (GATs) \cite{velivckovic2018graph} aggregate nodes' features in neighborhoods by trainable weights with attentions:
\begin{equation}
\small
\bm{h}_{i}^{(l+1)}=\sigma\left(\sum\nolimits_{j \in N(v_i)} \alpha_{i j}^{(l+1)} \bm{W}^{(l+1)} \bm{h}_{j}^{(l)}\right),
\end{equation}
where $\bm{h}_{i}^{(l)}$ represents the node $v_i$'s representation in the $l$-th layer ($\bm{h}_{i}^{(0)} = \bm{x}_i$) and $\alpha_{i j}^{(l)}$ is the attention coefficient between $v_i$ and $v_j \in N(v_i)$. In community detection (Fig. \ref{fig-gat}), the attention mechanism contributes to adaptively learning the importance of each node in a neighborhood. Correlations between similar nodes are computed close to the ground truth of community membership, which is inherited in GAT's representations. These attentions filter spatial relations or integrate metapaths. Therefore, communities in attributed, multiplex, heterogeneous networks can be easily detected (TABLE \ref{table-gat}).

The relations need special attention in deep community detection models. For example, citations and co-subject relationships are both significant in clustering papers into research topics. The multiplex network is constructed by multiple relation types reflected on edges (\textit{i.e.}, $E^{(r)}$). Each type of edges is grouped in a layer $r$. Thus, GATs can pay attentions to relation types. Deep Graph Infomax for attributed Multiplex network embedding (DMGI) \cite{park2020unsupervised} independently embeds each relation type and computes network embeddings in maximizing the globally shared features to detect communities. Its contrastive learning conducts between the original network and a corrupted network on each layer through the discriminator. A consensus regularization is subsequently applied with the attention coefficient to integrate final embeddings. The benefit of applying the attention mechanism is to preprocess the less significant relations by weakening their coefficient, especially when various types of relations are presented.

Despite the extrinsic supervision signal sharing globally in multiple relations, DMGI cannot capture the intrinsic signal into node embeddings where node attributes are considered. To fill the gap, High-order Deep Multiplex Infomax (HDMI) \cite{jing2021hdmi} is developed to capture both signals and their interactions. HDMI further designs a fusion module for the node embedding combination over multiplex network layers based on the semantic attention \cite{you2016image} on community memberships. 

Heterogeneous Information Network (HIN) involves diverse types of nodes $\mathcal{V}$ and edges $\mathcal{E}$. Deep community detection over heterogeneous networks represents heterogeneous typologies and attributes $\mathcal{X}$. Metapath Aggregated Graph Neural Network (MAGNN) \cite{fu2020magnn} offers a superior community detection solution by multi-informative semantic metapaths which distinguish heterogeneous structures in graph attention layers. MAGNN generates node attributes from semantic information. Since heterogeneous nodes and edges exist in intra- and inter-metapaths, MAGNN utilizes the attention mechanism in both of their embeddings by aggregating semantic variances over nodes and metapaths. Thus, MAGNN reduces the high heterogeneity and embeds richer topological and semantic information to enhance community detection results. Moreover, HeCo \cite{wang2021self} embeds network schema and metapaths from HINs. A modified contrastive learning is employed to guide two parts of embeddings. Sequentially, the attention mechanism acts on node importance, network heterogeneity and semantic features, where community detection benefits.

The above two works both use metapaths to facilitate community detection, but defining meaningful metapaths requires lots of domain knowledge. Thus, Context Path-based Graph Neural Network (CP-GNN) \cite{luo2021detecting} is built to learn node embeddings by context paths where attention mechanisms are exploited to discriminate the importance of different relationships. The non-predefined context paths are significant in capturing the high-order relationship for community detection. 

\begin{figure*}[!t]
\centering
\includegraphics[width=0.7\textwidth]{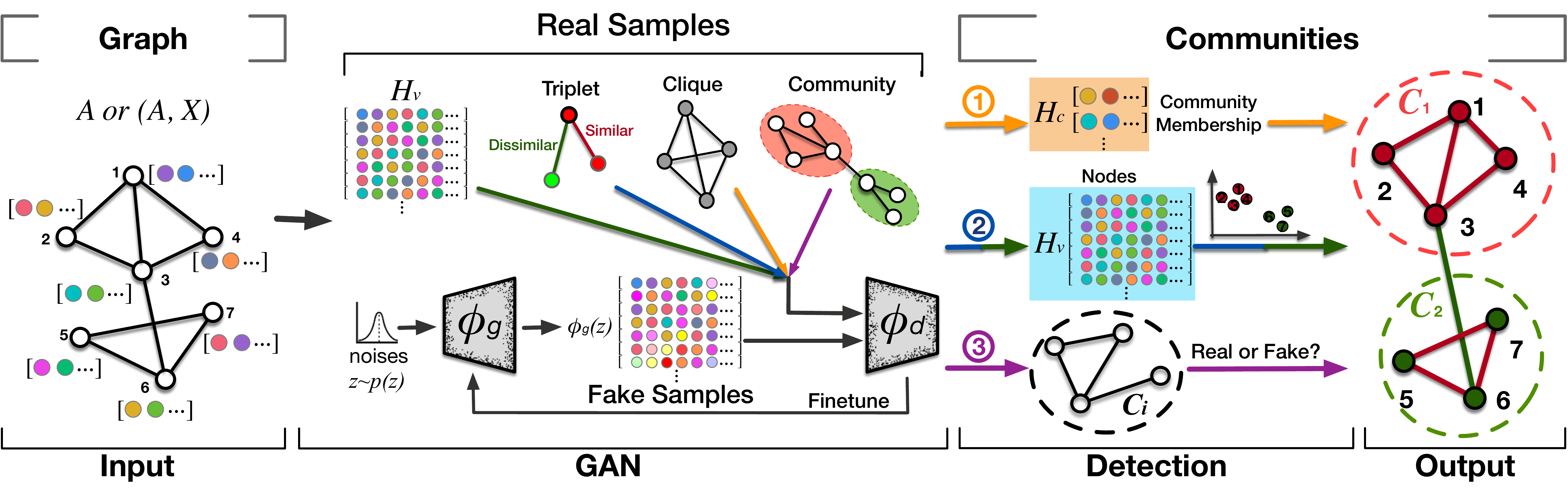}
\caption{\footnotesize A general framework for GAN-based community detection with details in Section \ref{sec-model-gan}. Generative Adversarial Network (GAN) produces fake samples $\phi_{g}(\bm{z})$ by the generator $\phi_g$ to fool the discriminator $\phi_d$. GAN's real samples can be node embeddings $\bm{H}_v$, local topology (\textit{e.g.,} triplet, clique) or communities. Thus, real and fake samples competitively finetune community features: The flow \ding{172} obtains community membership via clique-level GAN. The flow \ding{173} detects communities based on competitive node-level representations from $\bm{H}_v$ or triplets. The flow \ding{174} directly discriminates communities through the discriminator.} 
\label{fig-gan}
\end{figure*}

\section{Generative Adversarial Network-based Community Detection} 
\label{sec-model-gan}

Adversarial training is effective in generative models and improves discriminative ability. However, it needs to solve the overfitting problem when applied to community detection (see Fig. \ref{fig-gan} and TABLE \ref{table-gan}). Generative Adversarial Networks (GANs) \cite{goodfellow2014generative} competitively trains a generator $\phi_g$ and a discriminator $\phi_d$ in the adversarial framework. $\phi_d(\bm{x})$ represents the probability of input data, while $\phi_g(\bm{z})$ learns the generator's distribution $p_g$ via input noise variables $p_{\bm{z}}(\bm{z})$. The generator fools the discriminator by generating fake samples. Its objective function is defined as: 
\begin{equation}
\small
\begin{aligned}
\min_{\phi_g} \max_{\phi_d} & \mathbb{E}_{\bm{x} \sim p_{data}(\bm{x})} [\log \phi_d(\bm{x})] \\
& +\mathbb{E}_{\bm{z} \sim p_{\bm{z} }(\bm{z} )} [\log (1-\phi_d(\phi_g(\bm{z} )))]. 
\end{aligned}
\end{equation}

Seed Expansion with generative Adversarial Learning (SEAL) \cite{zhang2020seal} generates seed-aware communities from selected seed nodes by Graph Pointer Network with incremental updates (iGPN). It consists of four components in the community level, \textit{i.e.}, generator, discriminator, seed selector and locator. The discriminator adopts Graph Isomorphism Networks (GINs) to modify generated communities with the ground truth of community labels. The locator is designed to provide regularization signals to the generator, so that irrelevant nodes in community detection can be eliminated.

To imbalanced communities, Dual-Regularized Graph Convolutional Networks (DR-GCN) \cite{ijcai2020-398} utilizes a conditional GAN into the dual-regularized GCN model, \textit{i.e.}, a latent distribution alignment regularization and a class-conditioned adversarial regularization. The first
regularization $\mathcal{L}=(1-\alpha)\mathcal{L}_{gcn}+\alpha \mathcal{L}_{dist}$ balances the communities by minimizing the Kullback-Leibler (KL) divergence between majority and minority community classes $\mathcal{L}_{dist}$ guided with a standard GCN training $\mathcal{L}_{gcn}$. The second regularization is designed to distinguish communities on labeled node representations:
\begin{equation}
\small
\begin{aligned}
\min_{\phi_{g}, \mathcal{L}} &\max_{\phi_{d}} \mathcal{L}(\phi_{d}, \phi_{g}) = \mathbb{E}_{v_{i} \sim p_{data}(v_{i})} \log \phi_{d}(v_{i} |y_{i}) \\
& +\mathbb{E}_{\bm{z} \sim p_{\bm{z}}(\bm{z})} [\log(1-\phi_{d}(\phi_{g}(\bm{z} |y_{i}))) + \mathcal{L}_{reg}], 
\end{aligned}
\end{equation}
where $\mathcal{L}_{reg} = \sum_{u \in N(v_{i})} \| \bm{h}_{v'_{i}} - \bm{h}_{u} \|_2$ forces the generated fake node $v'_{i}$ to reconstruct the respective neighborhood relations (as $u \sim v_{i}$), $y_{i}$ is the ground truth label of node $v_{i}$. 

Instead of generating only one kind of fake samples, Jointly Adversarial Network Embedding (JANE) \cite{jane2020yang} employs two network information of topology and node attributes to capture semantic variations from adversarial groups of real and fake samples. Specifically, JANE represents embeddings $\bm{H}$ through a multi-head self-attention encoder $\phi_e$, where Gaussian noises are added for fake features for competition over the generator $\phi_g$ and the discriminator $\phi_d$:
\begin{equation}
\small
\begin{aligned}
\min_{\phi_g,\phi_e} \max_{\phi_d} & {\mathcal{L} (\phi_d, \phi_e, \phi_g)}
= \\
&\mathbb{E}_{(\bm{a}, \bm{x}) \sim p_{\bm{AX}}}[\underbrace{\mathbb{E}_{\bm{h} \sim p_{\phi_e}(\cdot |\bm{a}, \bm{x})}[\log \phi_d(\bm{h}, \bm{a}, \bm{x})]}_{\log \phi_d(\phi_e(\bm{a}, \bm{x}), \bm{a}, \bm{x})}] \\
+& \mathbb{E}_{\bm{h} \sim p_{\bm{H}}}[\underbrace{\mathbb{E}_{(\bm{a}, \bm{x}) \sim p_{\phi_g}(\cdot |\bm{h})}[\log (1-\phi_d(\bm{h}, \bm{a}, \bm{x}))]}_{\log (1-\phi_d(\bm{h}, \phi_g(\bm{h})))}], 
\end{aligned}
\end{equation}
where $p_{\bm{AX}}$ denotes the joint distribution of the topology $\bm{A}$ ($\bm{a} \subseteq \bm{A}$) and sampled node attributes $\bm{X}$ ($\bm{x} \subseteq \bm{X}$).

\begin{figure*}[!t]
\centering \footnotesize
\subfigure[\footnotesize A general framework of AE-based community detection]{\includegraphics[width=0.8\textwidth]{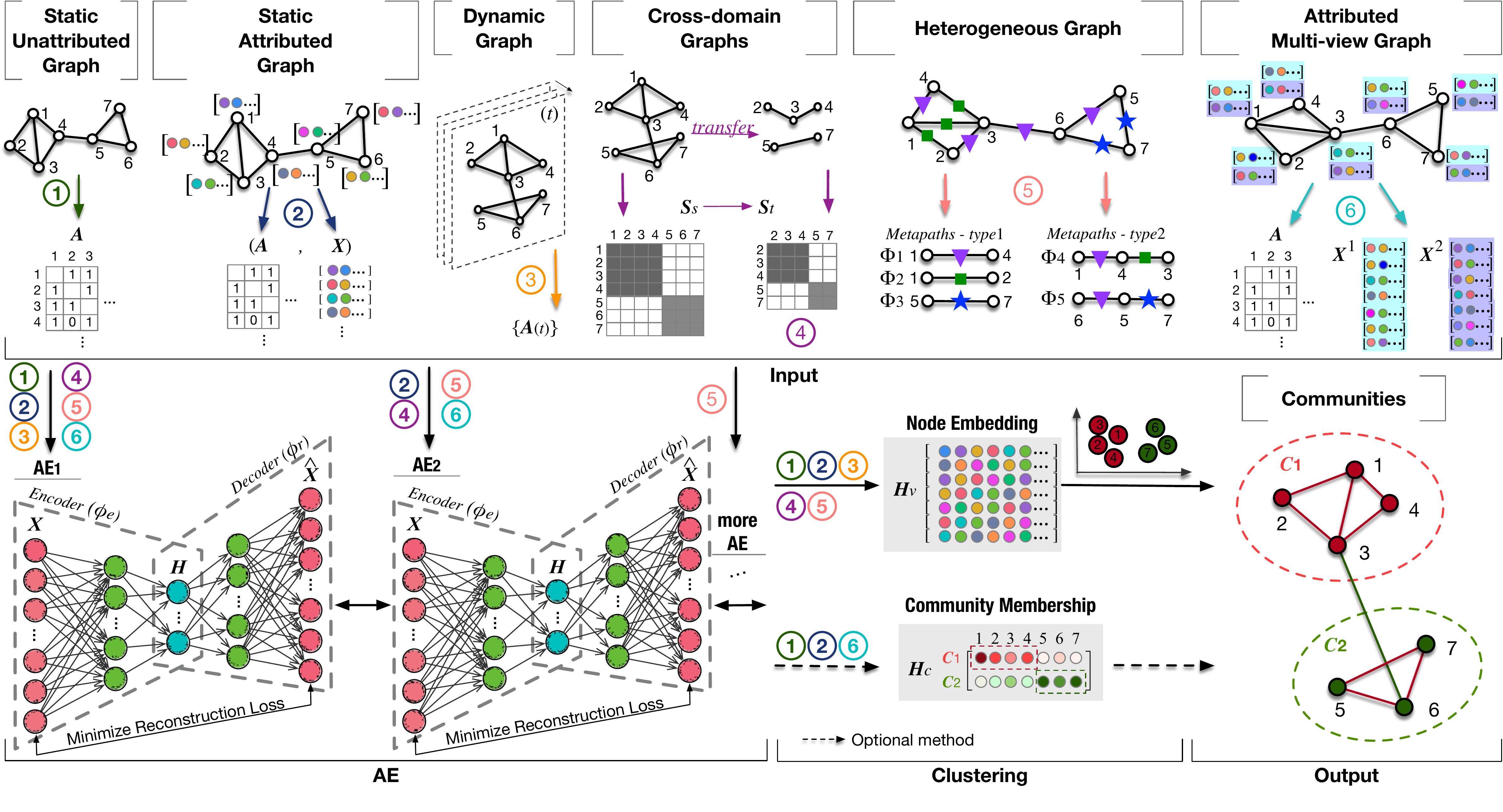}}
\subfigure[\footnotesize static unattributed graph]{\includegraphics[width=0.3\textwidth]{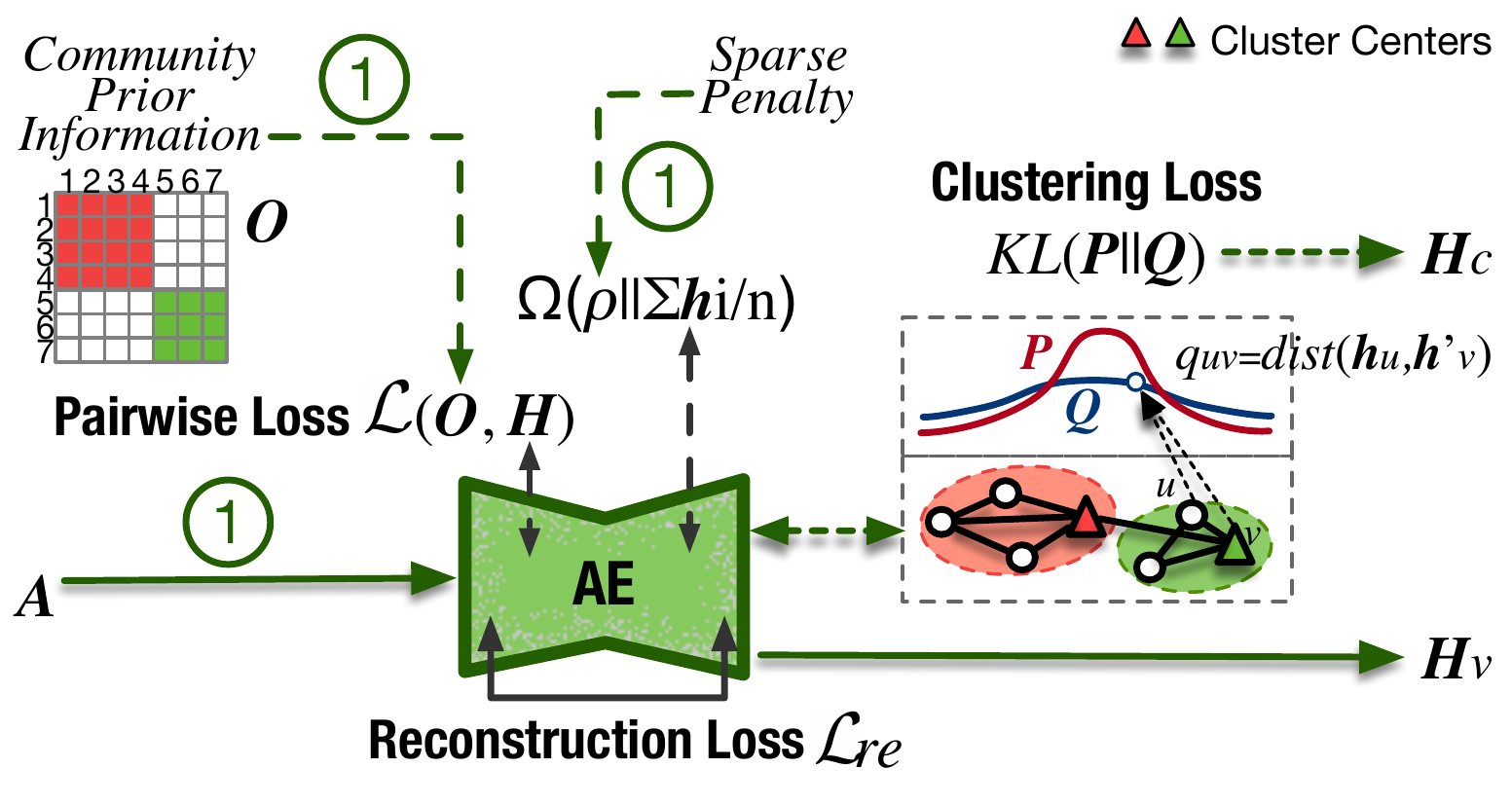}} 
\subfigure[\footnotesize static attributed graph]{\includegraphics[width=0.27\textwidth]{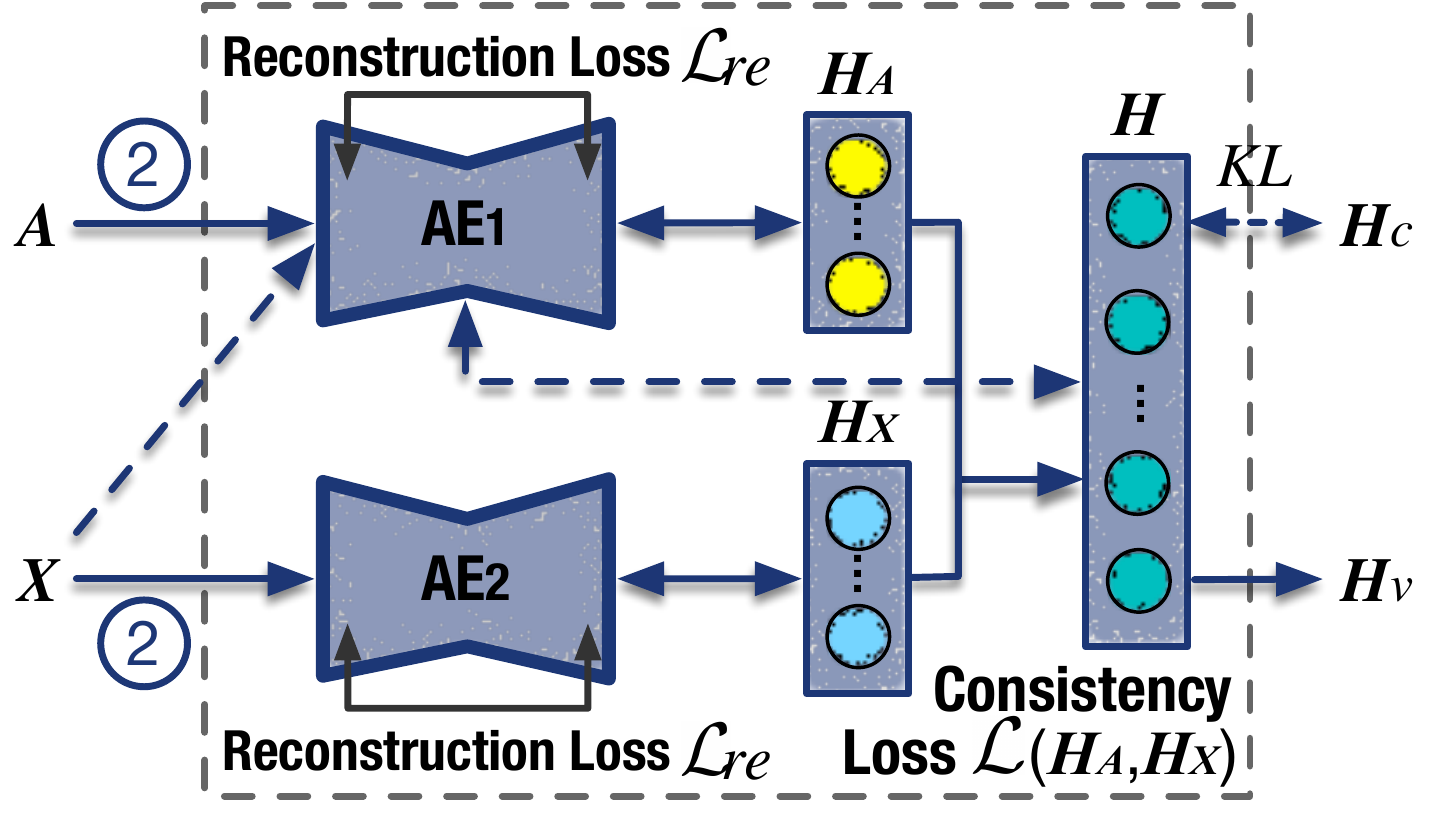}} 
\subfigure[\footnotesize dynamic graph]{\includegraphics[width=0.26\textwidth]{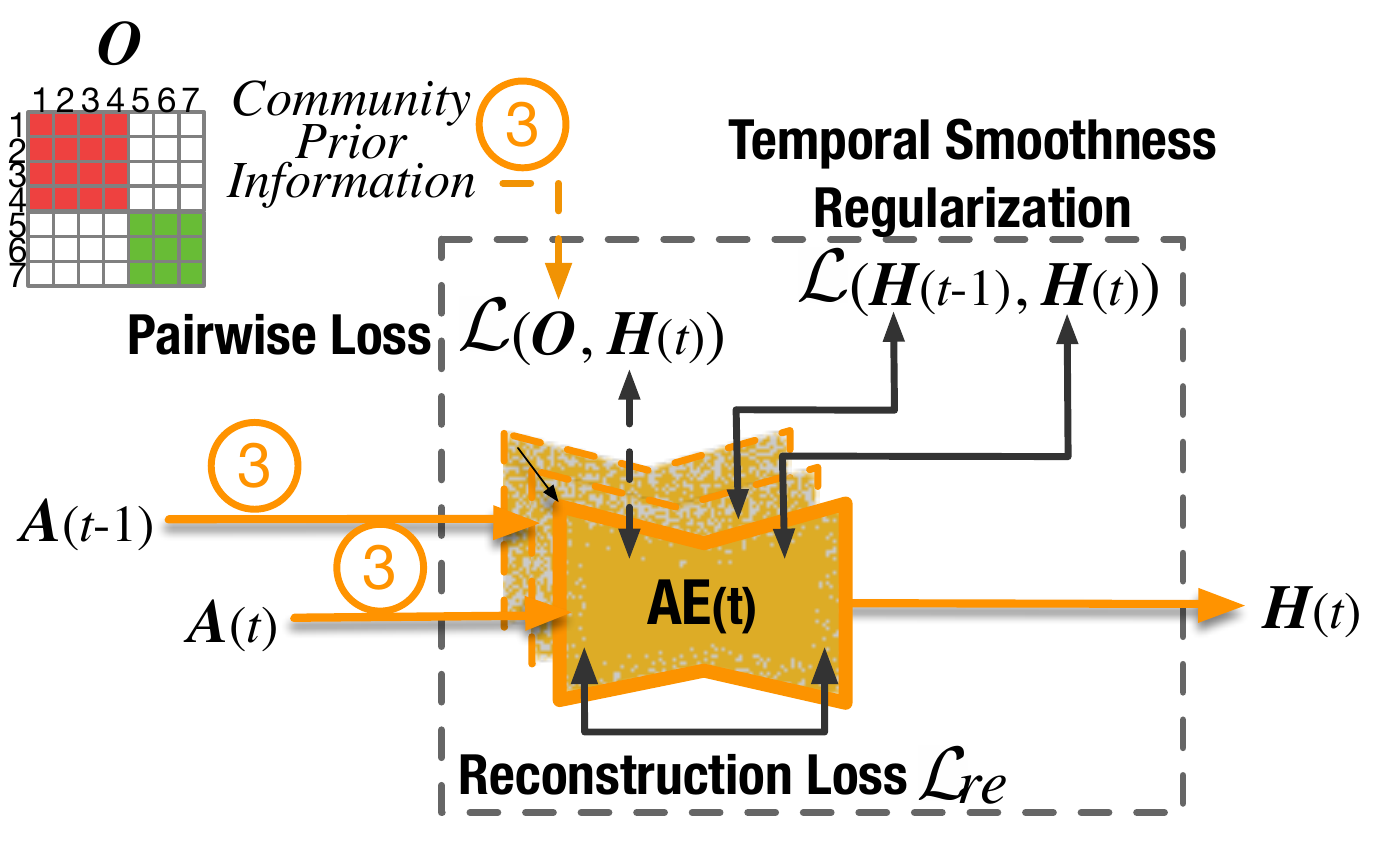}} 
\subfigure[\footnotesize cross-domain graph]{\includegraphics[width=0.2\textwidth]{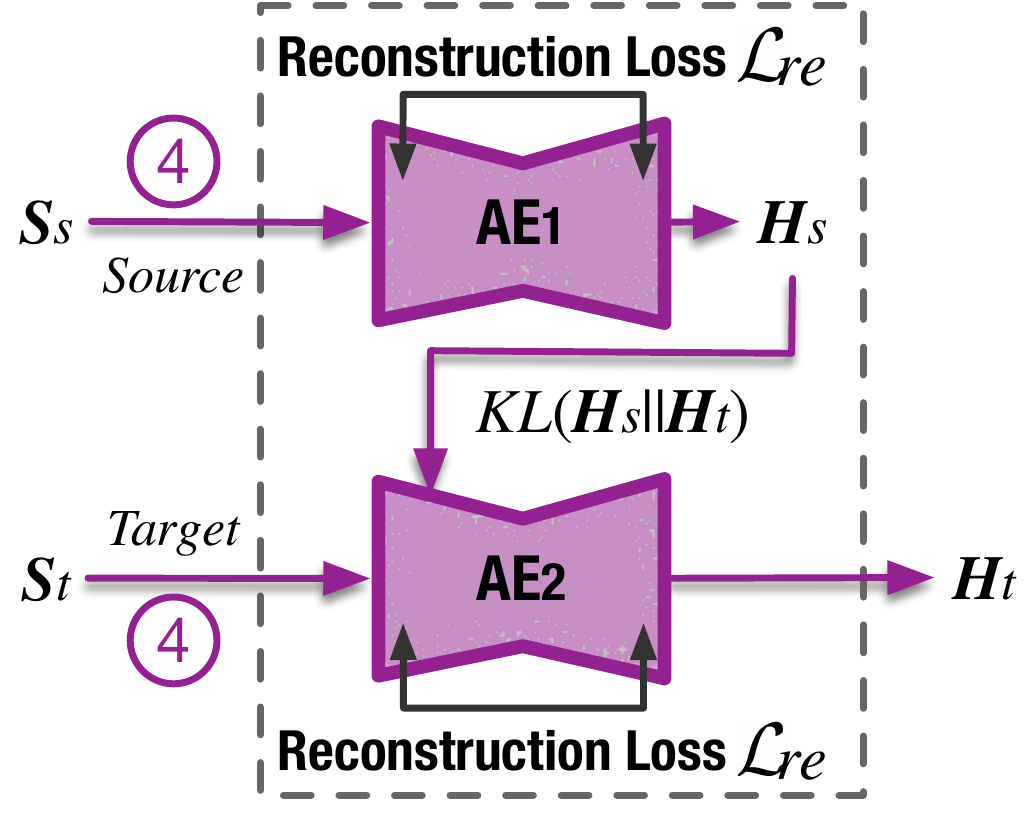}} 
\subfigure[\footnotesize heterogeneous graph]{\includegraphics[width=0.33\textwidth]{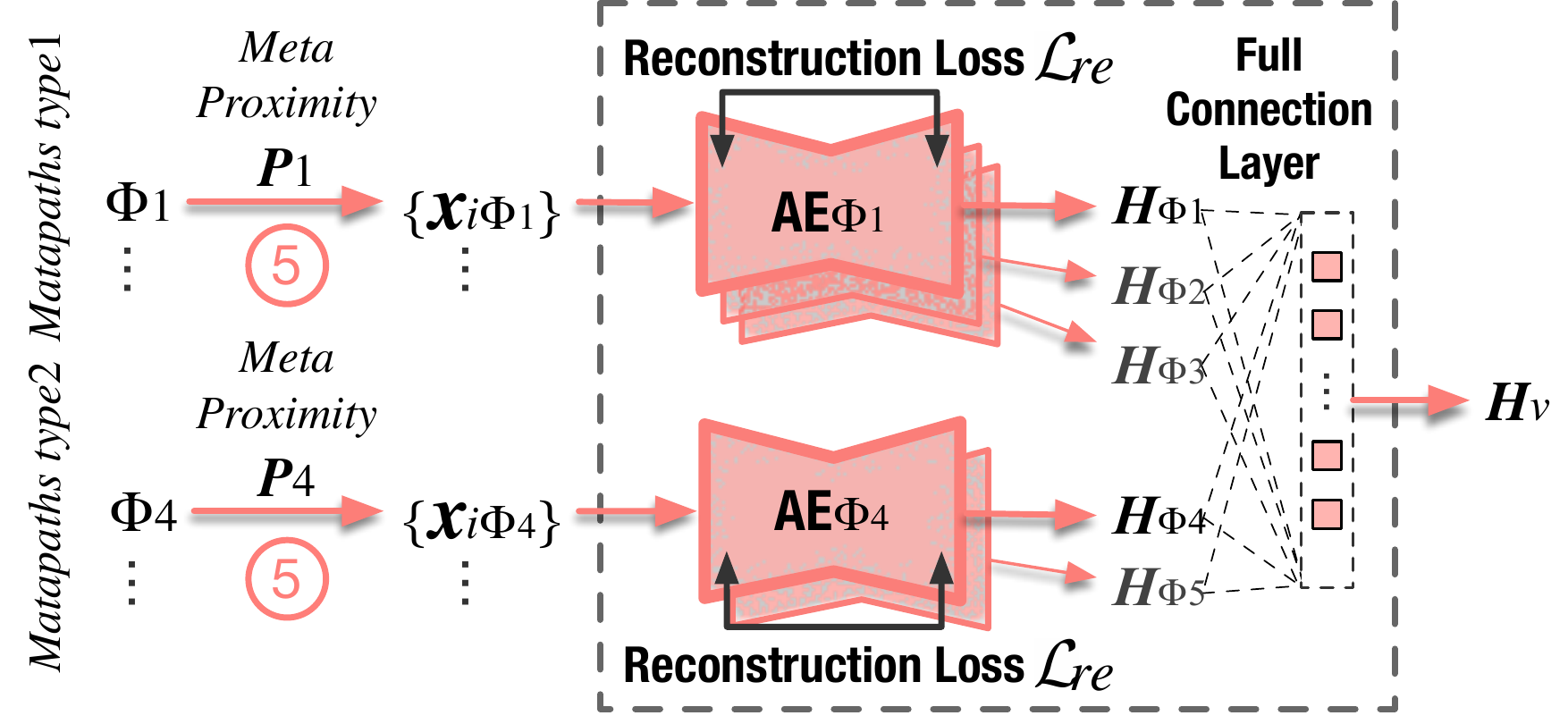}}
\subfigure[\footnotesize attributed multi-view graph]{\includegraphics[width=0.41\textwidth]{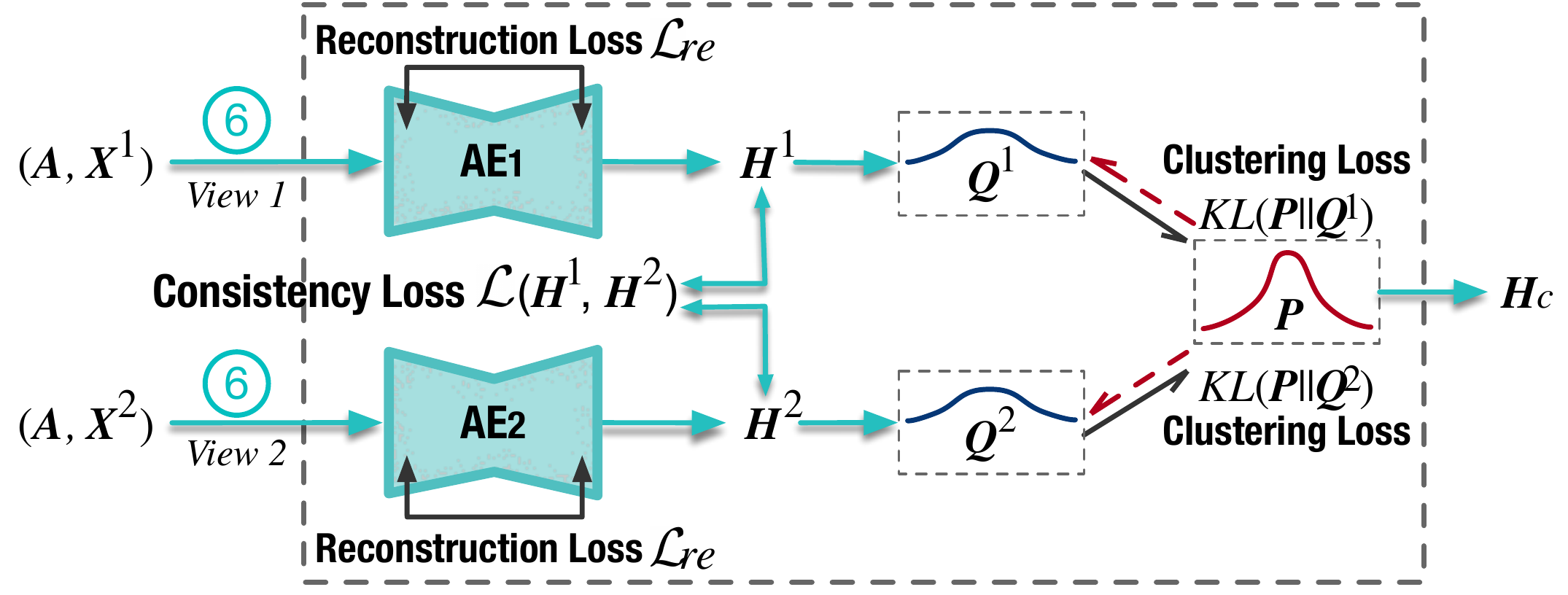}} 
\caption{\footnotesize A general framework for AE-based community detection consists of (a) the framework of the input layer, AE and clustering layers and the output layer and (b)-(g) AE-based community detection processes for each type of graph: (a) The input can be one of the graphs. They have topology and attributes information which are represented by the adjacency matrix $\bm{A}$ (\ding{172}-\ding{174}, \ding{177}), node attribute vectors $\bm{X}$ (\ding{173}, \ding{177}), node similarity matrix $\bm{S}$ (\ding{175}) and metapaths $\Phi$ (\ding{176}). More than one AE can be included that each AE embeds a part of input. Two optional work flows representively output node embeddings $\bm{H}_v$ and community membership matrix $\bm{H}_c$ for community detection results (\textit{i.e.}, the AE output layer). (b) Inputting typologies $\bm{A}$, the AE outputs node embeddings $\bm{H}_{v}$ by minimizing the reconstruction loss $\mathcal{L}_{re}$ (solid arrows) or outputs the community membership matrix $\bm{H}_{c}$ by further minimizing the clustering loss over $\bm{P}$ and $\bm{Q}$ distributions which measures node's probability in a community (dashed arrows). The community prior information for $\mathcal{L}(\bm{O},\bm{H})$ (dashed arrows) or the sparse penalty for $\Omega(\rho\|\sum{\bm{h}_{i}/n})$ (dashed arrows) optionally guides AE learning upon hidden layer embeddings $\bm{h}_{i} \in \bm{H}$. (c) Aiming at $\bm{H}_{v}$ or $\bm{H}_{c}$, typologies $\bm{A}$ and node attributes $\bm{X}$ are respectively represented by AE1 and AE2 (solid arrows). Their embeddings $\bm{H}_{\bm{A}}$ and $\bm{H}_{\bm{X}}$ are aggregated by minimizing the consistency loss. Otherwise (dashed arrows), an AE embeds all input over $(\bm{A},\bm{X})$. (d) The $t$-th snapshot $\bm{A}_{(t)}$ in the dynamic graph is represented for the node embeddings $\bm{H}_{(t)}$. The temporal smoothness regularization optimizes $\bm{H}_{(t)}$ over the previous snapshot's embeddings. $\bm{O}$ is optionally employed. (e) Two AEs respectively represent similarity matrices from the source and target domains into $\bm{H}_{s}$ and $\bm{H}_{t}$ where $\bm{H}_{s}$ guides AE2's training and node embeddings of the target graph $\bm{H}_{t}$ is outputted. (f) Each metapath is represented in an AE with the input feature vectors $\{\bm{x}_{i\Phi}\}$ denoting the meta proximity $\bm{p}_{i} \in \bm{P}$ between node $v_{i}$ in the metapath $\Phi$. All metapaths' embeddings are aggregated for $\bm{H}_{v}$ in the fully connected layer in these stacked AEs. (g) Attributes in each view are input with $A$ into an AE where node embeddings in different views are optimized by the clustering loss.}
\label{fig-stacked-ae}
\end{figure*}

The proximity can capture underlying relationships within communities. However, sparsely connected networks in the real world do not provide enough edges. Attributes in networks cannot be measured by proximity. To address these limitations, Proximity Generative Adversarial Network (ProGAN) \cite{gao2019progan} encodes each node's proximity from a set of instantiated triplets, so that community relationships are discovered and preserved in a low-dimensional space.

Community Detection with Generative Adversarial Nets (CommunityGAN) \cite{jia2019communitygan} is proposed for overlapping communities, which obtains node representation by assigning each node-community pair a nonnegative factor. Its objective function optimizes through a motif-level generator $\phi_g(\cdot|v_i;\Theta_g)$ and discriminator $\phi_d(\cdot,\Theta_d)$: 
\begin{equation}
\small
\begin{aligned}
\min_{\Theta_g} \max_{\Theta_d} & {\mathcal{L}(\phi_g, \phi_d)} =
\sum_i (\mathbb{E}_{C' \sim p_{true} (\cdot |v_{i})} \left[ \log \phi_d (C'; \Theta_d) \right] \\
& + \mathbb{E}_{V' \sim \phi_g(V' |v_i ; \Theta_g)} \left[ \log (1-\phi_d(V'; \Theta_d) \right] ), 
\end{aligned}
\end{equation}
where $\Theta_g$ and $\Theta_d$ unifies all nonnegative representation vector of node $v_i$ in the generator and the discriminator, $V'\subseteq V$ denotes a node subset, $C'$ represents motifs (\textit{i.e.}, cliques), and conditional probability $p_{true}(C'| v_i)$ describes the preference distribution of $C'$ covering $v_i$ over all other motifs $C'\in \mathcal{C'}$. 

To capture the global structural information, Community-Aware Network Embedding (CANE) \cite{wang2021cane} integrates a community detection model, \textit{i.e.} Latent Dirichlet Allocation (LDA), into the adversarial node representation process: 
\begin{equation}
\small
\begin{aligned}
\min_{\Theta_g} & \max_{\Theta_d}  {\mathcal{L}(\phi_g, \phi_d)} = \\
\sum_i & (\mathbb{E}_{v \sim p_{true} (v |v_{i})} \left[ \log \phi_{d} (v, v_{i}; \Theta_d) + P_{\phi}(v |v_{i}) \right] \\
& + \mathbb{E}_{v \sim \phi_{g}(v |v_{i} ; \Theta_g)} \left[ \log (1-\phi_{d}(v, v_{i}; \Theta_d) - P_{\phi}(v |v_{i}) \right] ), 
\end{aligned}
\end{equation}
where $P_{\phi}(v |v_{i})$ indicates the community similarity between a node $v_{i}$ and the node sample $v$, $\phi_{g}(v |v_{i} ; \Theta_g)$ outputs the node connectivity distribution. In turn, the representations characterize the community property.

Network Embedding and overlapping Community detection with Adversarial learning (ACNE) \cite{chen2021self} further generates negative communities as fake samples to learn distinguished community representation. ACNE utilizes walking strategy with perception to assist sampling of overlapping communities for nodes and exploits the discriminator to jointly map the node and community membership embeddings, so as to preserve the correlation of node-community in representations.

\section{Autoencoder-based Community Detection} \label{sec-model-autoencoder}

Autoencoders (AEs) are most commonly used in community detection as its unsupervised setting, including stacked AE, sparse AE, denoising AE, convolutional AE and Variational AE (VAE). AEs can depict nonlinear, noisy real-world networks and represent communities from rich information through reconstruction. A general framework for AE \cite{hinton2006reducing} is formed by an encoder $\bm{H} = \phi_e(\bm{A},\bm{X})$ and decoder $\hat{\bm{A}} = \phi_r(\bm{H})$ or $\bm{\hat{X}} = \phi_r(\bm{H})$. The encoder $\phi_e$ maps a high-dimensional network structure $\bm{A}$ and possible attributes $\bm{X}$ into a low-dimensional latent feature space $\bm{H}$. The decoder $\phi_r$ reconstructs a decoded network from encoder's representations $\bm{H}$ that $\hat{\bm{A}}$ and $\bm{\hat{X}}$ inherits preferred information in $\bm{A}$ and $\bm{X}$. A loss function $\mathcal{L}(\bm{x}, \phi_r(\phi_e(\bm{x})))$ is designed to maximize the likelihood between source data $\bm{x}$ and decoded data $\phi_r(\phi_e(\bm{x}))$. Fig. \ref{fig-stacked-ae} shows how AEs are generally applied in community detection, and TABLE \ref{table-autoencoder} compares the techniques. 

\subsection{Stacked AE-based Community Detection}  \label{sec-stacked-ae}

Compared to a single AE, a group of stacked AEs developed in deep hidden layers can better embed high-dimensional community features with representing each encoder in the stack individually representing one type of input data. In this architecture, stacked AEs represent multi-level and dynamic information to flexibly support wide community detection implementations \cite{liu2019detecting}.

Semi-supervised Nonlinear Reconstruction Algorithm with DNN (semi-DRN) \cite{yang2016modularity} designs a stacked AE into a community detection model, where a modularity matrix learns the nonlinear node representations in AEs and $k$-means obtains final community structure. Given an edge between nodes $v_i$ and $v_j$ in the adjacency matrix $\bm{A}=[a_{ij}]$, the modularity \cite{newman2006modularity} $b_{ij} = a_{ij} - \frac{k_i k_j}{2m}$ in the modularity matrix $\bm{B}$ is optimized for maximization. Encoding the node pairwise similarities (community membership) based on node representations, a pairwise constraint matrix $\bm{O} = [o_{i,j}\in \{0,1\}]$ is, meanwhile, defined to provide a prior knowledge that nodes $v_i$ and $v_j$ belong to the same community ($o_{i,j} = 1$) or not ($o_{i,j} = 0$). Thus, semi-DRN is optimized by minimizing the loss function:
\begin{equation}
\small
\mathcal{L} = \mathcal{L}(\bm{B}, \bm{\hat{B}}) + \lambda \mathcal{L}(\bm{O},\bm{H}), 
\end{equation}
where $\bm{\hat{B}}$ represents the reconstructed $\bm{B}$ over stacked representations by a series of AEs, $\lambda$ denotes an adjusting weight between the AE reconstruction loss $\mathcal{L}(\bm{B}, \bm{\hat{B}})$ and the pairwise constraints $\mathcal{L}(\bm{O},\bm{H})$, and $\mathcal{L}(\bm{O},\bm{H})$ measures each pair of community membership $o_{ij}$ and latent representations $(\bm{h}_i,\bm{h}_j)$ within stacked AEs. 

Similarly, Deep Network Embedding with Structural Balance Preservation (DNE-SBP) \cite{shen2018deep} incorporates the adjusting weight on pairwise constraints for signed networks so that the stacked AE clusters the closest nodes distinguished by positive and negative connections. Unified Weight-free Multi-component Network Embedding (UWMNE) and its variant with Local Enhancement (WMCNE-LE) \cite{jin2018integrative} preserve community properties from network topology and semantic information and integrate the diverse information in deep AE from a local network structure perspective.

To discover $t$ time-varying community structure, Semi-supervised Evolutionary Autoencoder (sE-Autoencoder) \cite{wang2020evolutionary} is developed within an evolutionary clustering framework, assuming community structures at previous time steps successively guide the detection at the current time step. To this end, sE-Autoencoder adds a temporal smoothness regularization $\mathcal{L}(\bm{H}_{(t)}, \bm{H}_{(t-1)})$ for minimization \cite{yang2016modularity} in:
\begin{equation}
\small
\mathcal{L} = \mathcal{L}(\bm{S}_{(t)}, \bm{\hat{S}}_{(t)})+ \lambda \mathcal{L}(\bm{O}, \bm{H}_{(t)}) + (1-\lambda) \mathcal{L}(\bm{H}_{(t)}, \bm{H}_{(t-1)}),
\end{equation}
where reconstruction error $\mathcal{L}(\bm{S}_{(t)}, \bm{\hat{S}}_{(t)})$ minimizes the loss between the similarity matrix $\bm{S}_{(t)}$ and reconstructed matrix $\bm{\hat{S}}_{(t)}$ at the time step $t$, and the parameter $\lambda$ controls the node pairwise constraint $\mathcal{L}(\bm{O},\bm{H}_{(t)})$ and the temporal smoothness regularization over time $t$-th graph representation $\bm{H}_{(t)}$.

To attributed networks, Deep Attributed Network Embedding (DANE) \cite{gao2018deep} develops a two-branch AE framework: one branch maps the highly nonlinear network structure to the low-dimensional feature space, and the other collaboratively learns node attributes. As similar nodes are more likely to be clustered in the same community, DANE measures these similarities by a series of proximity regarding network topological and attribute information in the representation learning, where optimizations are applied on reconstruction losses at first-order proximity $\mathcal{L}_f$, higher-order proximity $\mathcal{L}_h$ and semantic proximity $\mathcal{L}_s$, and a negative log likelihood control $\mathcal{L}_c$ for a consistent and complementary representation.

To solve the problem of shifted distributions, imbalanced features or a lack of data, Transfer Learning-inspired Community Detection with Deep Transitive Autoencoder (Transfer-CDDTA) \cite{xie2019high} uses the transfer learning framework and modifies it to the unsupervised community detection task. On the source and target domains, Transfer-CDDTA balances stacked AEs via KL divergence while learning node embeddings. Aiming to map community information into one smooth feature space, CDDTA separates the input adjacency matrix $\bm{A}$ into a source domain $s$ and a target domain $t$ by similarity matrices $\bm{S}_s$ and $\bm{S}_t$ to keep node pairwise similarity values for each stacked AE. Transfer-CDDTA then incorporates domain-independent features into the following minimization: 
\begin{equation}
\small
\mathcal{L} = \mathcal{L}_s(\bm{S}_s, \bm{\hat{S}}_s) + \mathcal{L}_t(\bm{S}_t, \bm{\hat{S}}_t) + \alpha KL(\bm{H}_s,\bm{H}_t)+\beta\mathcal{L}(\Theta; \gamma),
\end{equation}
where $\alpha$, $\beta$, $\gamma$ are trade-off parameters, $\mathcal{L}_s$ and $\mathcal{L}_t$ denote reconstruction losses of source and target domains, $KL$ smooths the KL divergence on encoded features ($\bm{H}_s$, $\bm{H}_t$) across two domains, and $\mathcal{L}(\Theta)$ is a regularization term on trainable variables to reduce overfitting in the optimization. 

Deep alIgned autoencoder-based eMbEdding (DIME) \cite{zhang2017bl} stacks AEs for the multiple aligned structures in heterogeneous social networks. It employs metapaths to represent different relations (heterogeneous links denoting $\mathcal{A}_{ij}$ which represents anchor links between multiple aligned network $\mathcal{G}_i$ and $\mathcal{G}_j$) and various attribute information $\mathcal{X}=\{\bm{X}_i\}$. Correspondingly, a set of meta proximity measurements are developed for each meta path and embed the close nodes to a close area in the low-dimensional latent feature space. The relatively close region reflects communities going to be detected. 
	
\subsection{Sparse AE-based Community Detection} 
\label{sec-sparse-ae}

The sparsity commonly exists in real-world networks and causes computational difficulties in community detection algorithms. To solve these issues, sparse AEs \cite{ng2011sparse} introduce a sparsity penalty $\Omega(\bm{h})$ in hidden layers $\bm{h}$. The reconstruction loss function is as follows:
\begin{equation}
\small
\mathcal{L}(\bm{x}, \phi_r(\phi_e(\bm{x})))+\Omega(\bm{h}). 
\end{equation}

Autoencoder-based Graph Clustering Model (GraphEncoder) \cite{tian2014learning} is the first work that uses AE for graph clustering. It processes the sparsity by a sparsity term as a part of the following loss function: 
\begin{equation}
\small
\mathcal{L}(\Theta)=\sum\nolimits_{i}\|\bm{\hat{x}_{i}}-\bm{x}_{i}\|_{2}+\beta \Omega(\rho \| \frac{1}{n}\sum\nolimits_{i} \bm{h}_i),
\end{equation}
where the weight parameter $\beta$ controls the sparsity penalty $\Omega(\cdot\|\cdot)$ over a configuration value $\rho$ (a small constraint such as $0.01$) and the average of hidden layers' activation values. GraphEncoder improves the clustering efficiency of large-scale networks and proves that sparse networks can provide enough structural information for representations. A Weighted Community Detection model (WCD) \cite{li2021weighted} is further developed for weighted networks by sparse AE in second-order neighbors.

Deep Learning-based Fuzzy Clustering Model (DFuzzy) \cite{bhatia2018dfuzzy} is proposed for detecting overlapping communities in sparse large-scale networks within a parallel processing framework. DFuzzy introduces a stacked sparse AE targeting head nodes to evolve overlapping and disjoint communities based on modularity measurements. DFuzzy performs 63\% (\textit{Q}) and 34\% (CON) higher than non-deep learning baselines. 

Community Detection Method via Ensemble Clustering (CDMEC) \cite{xu2020stacked} combines sparse AEs with a transfer learning model to discover more valuable information from local network structures. To this end, CDMEC constructs four similarity matrices and employs transfer learning to share local information via AEs' parameters. A consensus matrix is applied to aggregate community detection results, individually produced by four similarity matrices and supported by $k$-means. The final communities are globally determined based on the factorization of the consensus matrix.

\subsection{Denoising AE-based Community Detection} 
\label{sec-denoise-ae}
The denoising process, subtracting noises within DNN layers, can effectively enhance models' robustness. Denoising AEs \cite{wang2017mgae} minimize the reconstruction loss between the input $\bm{x}$ and AE features upon corrupt $\widetilde{\bm{x}}$:
\begin{equation}
\small
\mathcal{L}(\bm{x}, \phi_r(\phi_e(\widetilde{\bm{x}}))). 
\label{eq-ae-denonising}
\end{equation}

Deep Neural Networks for Graph Representation (DNGR) \cite{10.5555/3015812.3015982} applies stacked denoising encoder to increase the robustness of capturing the local structural information when detecting communities. Specifically, it generates a probabilistic co-occurrence matrix and a shifted positive pointwise MI matrix by randomly walking over communities. A Deep Neural network-based Clustering-oriented network  network embedding (DNC) \cite{li2021dnc} is the extension of DNGR which joint learns node embeddings and cluster assignments.

To corrupted node attributes, Graph Clustering with dynamic Embedding (GRACE) \cite{yang2017graph} focuses on the dynamically changing inter-community activities via a self-training clustering guided by the influence propagation within neighborhoods.

Marginalized Graph AutoEncoder (MGAE) \cite{wang2017mgae} denoises both graph attributes and structures to improve community detection through the marginalization process. It obtains the corrupted attributes $\widetilde{\bm{X}}$ in $\eta$ times by randomly removing some attributes. The objective function in MGAE training is: 
\begin{equation}
\small
\mathcal{L} = \frac{1}{\eta}\sum\nolimits_{i=1}^{\eta}\|\bm{X}-\widetilde{\bm{D}}^{-\frac{1}{2}} \widetilde{\bm{A}} \widetilde{\bm{D}}^{-\frac{1}{2}} \widetilde{\bm{X}} \bm{W}\|_2+\lambda \mathcal{L}(\bm{W}), 
\end{equation}
where $\mathcal{L}(\bm{W})$ denotes a regularization term on AE's parameters $\bm{W}$ with a coeffcient $\lambda$. 

\subsection{Graph Convolutional AE-based Community Detection}
\label{sec-gcn-ae}

It is a great success to introduce GCNs into AEs: GCNs provide by neighborhood aggregation over community representations, and AEs alleviate the over-smoothing issue in GCNs to clarify community boundaries. For example, Structural Deep Clustering Network (SDCN) \cite{bo2020structural} designs a delivery operator to connect AE and GCN in DNN layers for the structure-aware representations. When SDCN integrates the structural information into deep clustering, it updates communities by applying a dual self-supervised optimization in backpropagation.

GCN-based approach for Unsupervised Community Detection (GUCD) \cite{ijcai2020-486} employs the semi-supervised MRF \cite{jin2019graph} as the encoder in the convolutional layer and proposes a community-centric dual decoder to identify communities in attributed networks. The dual decoder reconstructs the network topology and node attributes. 

One2Multi Graph Autoencoder for Multi-view Graph Clustering (O2MAC) \cite{fan2020one2multi} deals with the multi-view graph by dividing it into multiple one-view graphs and assigning each with an AE, named One2Multi. In the encoder, a GCN is applied to embed a set of view-separated graphs. Decoders respectively for one-view graphs select the most informative graph to represent the multi-view graph. O2MAC significantly captures shared features among multi-view graphs and improves clustering results through a self-training optimization. 

\subsection{Graph Attention AE-based Community Detection}
\label{sec-gat-ae}

Instead of integrating GCNs, this community detection category applies GATs into AEs, in which a GAT is employed as the encoder to rank the importance of nodes within a neighborhood. For example, Deep Attentional Embedded Graph Clustering (DAEGC) \cite{wang2019attributed} exploits high-order neighbors to cluster communities under self-training. Graph Embedding Clustering with Cluster-Specificity Distribution (GEC-CSD) \cite{xu2021graph} utilizes graph attention AE as a generator to learn the distinguished community representations in the framework of adversarial learning integrating self-training, where discriminator can make sure the diversity of cluster distributions.

Considering multi-view attributes in networks, Multi-View Attribute Graph Convolution Networks (MAGCN) \cite{magcn2020cheng} designs a two-pathway encoder: the first pathway encodes with a multi-view attribute GAT capable of denoising, and the second pathway develops a encoder to obtain consistent embeddings over multi-view attributes. Thus, noises and distribution variances are removed for community detection. Self-supervised Graph Convolutional network for Multi-view Clustering (SGCMC) \cite{xia2021self} is developed from MAGCN through sharing a coefficient matrix in different views. 

Deep Multi-Graph Clustering (DMGC) \cite{luo2020deep} introduces AEs to represent each graph with an attention coefficient that node embeddings of multiple graphs cluster cross-graph centroids to obtain communities on Cauthy distribution.

\subsection{VAE-based Community Detection}\label{sec-model-vae}

Variational AutoEncoder (VAE) is an extension of AEs based on variational inference (\textit{e.g.}, mean and covariance of features) \cite{kingma2013auto}. It is first introduced into the graph learning field by Variational Graph AutoEncoder (VGAE) \cite{kipf2016variational} assuming prior distribution and applying GCN as the encoder, through which the learned representation $\bm{Z}$ is a latent distribution. Community detection based on VAEs is activated by algorithms such as SBM, to fast inference community memberships in node representations \cite{mehta2019stochastic}. The inference process considers the uncertainty of the network \cite{chen2019variational, choong2018learning}, \textit{e.g.}, community contradiction between neighbors of boundary nodes connecting multiple communities. VAEs are also required to handle sparsity in detecting communities. Meanwhile, VAEs easily incorporate deeper nonlinear relationship information. For example, Triad (Variational) Graph Autoencoder (TGA/TVGA) \cite{shi2020effective} replaces VAE/VGAE's decoder with a new triad decoder, which describes a real-world existing triadic closure property in communities.

Variational Graph Embedding and Clustering with Laplacian Eigenmaps (VGECLE) \cite{chen2019variational} divides the graph representation into mean and covariance while generatively detecting communities, indicating each node's uncertainty of implicit relationships to its actual geographic position. With a Mixture-of-Gaussian prior and a Teacher-Student (T-S) as regularization, VGECLE aims to let the node $v_i$ (student) learn a distribution close to its neighbors' (teacher).

Deep Generative Latent Feature Relational Model (DGLFRM) \cite{mehta2019stochastic} and Ladder Gamma Variational Autoencoder for Graphs (LGVG) \cite{Sarkar2020LGVG} further capture the community membership strength on each node. DGLFRM models the sparse node embeddings by a Beta-Bernoulli process which can detect overlapping communities and infer the number of communities. LGVG is devised to learn the multi-layered and gamma-distributed embeddings so that it detects communities from fine-grained and coarse-grained granularities.

To capture the higher-order structural features of communities, Variational Graph Autoencoder for Community Detection (VGAECD) \cite{choong2018learning} employs a Gaussian Mixture Model (GMM) to generalize the network generation process via bringing in a parameter of community assignments. VGAECD defines the joint probability of generative process, it can be detailed as: \begin{equation}
\small
p(\bm{a}, \bm{z}, c) = p(\bm{a} |\bm{z}) p(\bm{z} |c) p(c),
\end{equation} 
where $p(c) = \pi_{c}$ denotes the prior probability of community $C_c$, $p(\bm{z} |c)$ is calculated from the Gaussain distribution corresponding to community $C_c$ with the mean $\mu_c$ and the standard deviation ${\sigma}_c$ from a GCN layer (\textit{i.e.}, $\bm{z} \sim \mathcal{N}(\mu_c, {\sigma}_c^2 \bm{I})$), and $\bm{a}$ with the posterior (reconstruction) probability $p(\bm{a} |\bm{z})$ is sampled from the multivariate Bernoulli distribution parametrized by $\mu_x$ that is learned from the decoder on $\bm{z}$. The variational distribution can be obtained by the overall loss function as: 
\begin{equation}
\small
\mathcal{L}_{\mathrm{ELBO}}(\bm{x}) = \mathbb{E}_{q(\bm{z} | \bm{x}, \bm{a})}[\log p(\bm{x},\bm{a} | \bm{z})] - KL[q(c | \bm{x}) \| p(c | \bm{z})],
\end{equation}
where the first term calculates reconstruction loss, and $KL$ divergence minimizes the clustering loss. The evidence lower bound (ELBO) is maximized to optimize the function. 

The training progress of VGAECD reveals a favoured treatment for minimizing the reconstruction loss over the community loss, which leads to a sub-optimal community detection result. To resolve it, Optimizing Variational Graph Autoencoder for Community Detection (VGAECD-OPT) \cite{chen2019variational} proposes a dual optimization to learn the community-aware latent representation, the ELBO maximization becomes:
\begin{equation}
\small
\mathcal{L}_{\mathrm{ELBO}}(\bm{a}) = \mathbb{E}_{q(\bm{z}, c | \bm{a})}[\log p(\bm{a} | \bm{z})] - KL[q(\bm{z}, c | \bm{a}) \| p(\bm{z}, c)].
\end{equation}

To the sparsity and noises in the real world, Adversarially Regularized Variational Graph AutoEncoder (ARVGA) \cite{pan2018adversarially} embeds the latent representations $\bm{Z}$ based on the prior distribution $p_{z}$ under the adversarial AE framework. $\bm{Z}$ is represented by a variational graph encoder aiming at a robust community detection. Meanwhile, a special discriminator $\phi_d$ is designed to decide adversarially on $p_{z}$ guided embeddings $\bm{Z}$ and encodings on real samples $\phi_g(\bm{X}, \bm{A})$:
\begin{equation}
\small
\begin{aligned}
\min_{\Theta_g}\max_{\Theta_d} & \mathbb{E}_{\bm{z} \sim p_{z}}[\log \phi_d(\bm{Z})]\\
& +\mathbb{E}_{\bm{x} \sim p(\bm{x})}[\log (1-\phi_d(\phi_g(\bm{X}, \bm{A})))].
\end{aligned} 
\end{equation}

\section{Deep Nonnegative Matrix Factorization-based Community Detection} \label{sec-model-nmf}

In the network embedding domain, Nonnegative Matrix Factorization (NMF)\cite{lee1999learning} is a particular technique factorizing an adjacency matrix $\bm{A}$ into two nonnegative matrices ($\bm{U}\in \mathbb{R}^{n\times k}$ and $\bm{P}\in \mathbb{R}^{k\times n}$) with the nonnegative constraints that $\bm{U}\geq 0$ and $\bm{P}\geq 0$. The matrix $\bm{U}$ corresponds to the mapping between the given network and the community membership space. Each column in the matrix $\bm{P} = [p_{ij}]$ indicates the a community membership for each node $(v_i,C_j)$ in the probability of $p_{ij}$. NMF is applicable for disjoint and overlapping community detection. While real-world networks contain complicated topology information, the traditional NMF cannot fully uncover them to detect communities. Deep NMF (DNMF) \cite{song2015hierarchical} stacks multiple layers of NMF $\{\bm{U}_1,\cdots,\bm{U}_p\}$ to capture pairwise node similarities in various levels/aspects, like deep layers in GCN and AE.

In the community detection field, Deep Autoencoder-like Nonnegative Matrix Fatorization (DANMF) \cite{ye2018deep} is the influential model in an unsupervised learning setting. In contrast with the conventional NMF-based community detection mapping simple community membership, DANMF employs the AE framework to make network reconstruction on hierarchical mappings. The learning objective of community membership $\bm{P}_p$ and the hierarchical mappings $\{\bm{U}_i\}^p_1$ are trained by combining reconstruction losses and a $\lambda$ weighted graph regularization:
\begin{equation}
\small
\begin{aligned}
\min_{\bm{P}_p,\bm{U}_i} & \mathcal{L}(\bm{P}_p,\bm{U}_i) = \|\bm{A} - \bm{U}_1\cdots \bm{U}_p \bm{P}_p\|_F^2 \\ 
&  +\|\bm{P}_p - \bm{U}_p^T\cdots \bm{U}_1^T \bm{A}\|_F^2 + \lambda tr(\bm{P}_p \bm{L} \bm{P}_p^T) \\
\text{s.t.} ~~ & \bm{P}_p \geq 0, \bm{U}_i \geq 0, \forall i=1,\cdots,p
\end{aligned},
\end{equation}
where $\|\cdot\|_F$ denotes the Frobenius norm, $\bm{L}$ represents the graph Laplacian matrix, and graph regularization focuses on the network topological similarity to cluster neighboring nodes. A further work \cite{sun2017non} adds a sparsity constraint into the above DNMF-based community detection.

Although DNMF provides a solution to map multiple factors in forming communities, the computational cost is relatively high on matrix factorizations. To address this issue, Modularized Deep Nonnegative Matrix Factorization (MDNMF) \cite{huang2020community} applies modularity (details in Eq. \eqref{eq-modularity}) directly into a basic multi-layer deep learning structure targeting community detection. The node membership in communities $\bm{P}_p$ can be reached by minimizing the objective function below:
\begin{equation}
\small
\begin{aligned}
\mathcal{L} = & \|\bm{A} - \bm{U}_1\cdots \bm{U}_p \bm{P}_p\|_F^2 + \alpha \|\bm{M'} - \bm{P}_{p}^{T} \bm{K}^{T}\|_F^2 \\
& - \beta tr(\bm{M'}^{T}\bm{BM'}) + \lambda tr(\bm{P}_p \bm{L} \bm{P}_p^T) \\
\textrm{s.t.}  ~~  & \quad \bm{P}_p \geq 0, \bm{U}_i \geq 0, \forall i=1,...,p
\end{aligned},
\end{equation}
where $\bm{B}$ and $\bm{M'}$ are modularity matrix and the corresponding community membership matrix, $\bm{K}$ is an extra nonnegative matrix combining modularity information so that DNMF can explore the hidden features of network topology.

\section{Deep Sparse Filtering-based Community Detection} \label{sec-model-filtering} 

Sparse Filtering (SF)\cite{ngiam2011sparse} is a two-layer learning model capable of handling high-dimensional graph data. The high sparse input $\bm{A}$ with many $0$ elements is represented into lower-dimensional feature vectors $\bm{h}_i$ with non-zero values. To explore deeper information such as community membership, Deep SF (DSF) stacks multiple hidden layers to finetune hyperparameters $\Theta$ and extensively smooth data distributions $Pr(\bm{h}_i)$.

As a representative method, Community Discovery based on Deep Sparse Filtering (DSFCD) \cite{xie2018community} is developed on three phases: network representation, community feature mapping and community discovery. The network representation phase performs on the adjacency matrix $\bm{A}$, modularity matrix $\bm{B}$ and two similarity matrices $\bm{S}$ and $\bm{S}'$, respectively. The best representation is selected to input into the DSF for community feature mapping represented on each node $\bm{h}_i$. Meanwhile, $\bm{h}_i$ preserves the node similarity in the original network $\bm{A}$ and latent community membership features. The node similarity is modeled in the loss function:
\begin{equation} 
\small
\mathcal{L} = \sum\nolimits_i \| \bm{h}_{i}\|_1 + \lambda \sum\nolimits_i KL(\bm{h}_i, \bm{h}^*_j),
\end{equation}
where $\|\cdot\|_1$ is the $L_1$ norm penalty to optimize sparseness, $\bm{h}^*_j$ denotes the representation of the most similar node to $v_{i}$ by measuring distances through KL divergence. When the learning process is optimized on the minimized loss, similar nodes are clustered into communities. The DSF architecture is investigated to be significant in experiments on real-world data sets. DSFCD discovers communities in higher accuracy than SF.

\begin{table*}[!t]
\centering \footnotesize
\renewcommand\arraystretch{1}
\setlength{\tabcolsep}{1.5mm}
\caption{Summary of commonly used evaluation metrics in community detection.}
\begin{tabular}{lccl} 
\toprule[1pt]
\textbf{Metrics} & \textbf{Overlap} & \textbf{Ground Truth} & \textbf{Publications} \\ \midrule
\rowcolor{gray!30} 
&  &  & \cite{sperli2019deep,jin2019graph,gcln2020hu,zhang2020commdgi,zhang2019attributed,cui2020adaptive,ipgdn2020liu,ijcai2020-398,gao2019progan,jane2020yang,gao2018deep,jin2018integrative,wang2017mgae,fan2020one2multi,levie2018cayleynets,ijcai2020-486,bo2020structural,wang2019attributed,magcn2020cheng,luo2020deep,wang2021unsupervised,zhang2021spectral,ijcai2021-473,li2021dnc,xu2021graph,xia2021self}, \\
\rowcolor{gray!30} 
\multirow{-2}{*}{ACC} & \multirow{-2}{*}{$\checkmark$} & \multirow{-2}{*}{Yes} & \cite{chen2019variational,shi2020effective,choong2018learning,choong2019optimizing,pan2018adversarially,huang2020community,ye2018deep} \\
& &  & \cite{xin2017deep,cai2020edge,jin2019graph,wang2020evolutionary,wang2021unsupervised,zhang2021spectral,ijcai2021-473,shchur2019overlapping,gcln2020hu,zhang2020commdgi,luo2021detecting,zhang2019attributed,cui2020adaptive,ipgdn2020liu,park2020unsupervised,fu2020magnn,ijcai2020-398,gao2019progan,jane2020yang,jia2019communitygan,yang2016modularity,xie2019high,jin2018integrative,tian2014learning}, \\
\multirow{-2}{*}{NMI} & \multirow{-2}{*}{$\times$} & Yes & \cite{xu2020stacked,10.5555/3015812.3015982,wang2017mgae,li2021dnc,xu2021graph,xia2021self} \\
Overlapping-NMI & $\checkmark$ &  & \cite{ijcai2020-486,bo2020structural,fan2020one2multi,wang2019attributed,magcn2020cheng,luo2020deep,shi2020effective,choong2018learning,choong2019optimizing,pan2018adversarially,huang2020community,ye2018deep,jing2021hdmi,wang2021self} \\
\rowcolor{gray!30} 
Precision & $\checkmark$ & Yes & \cite{xin2017deep,gcln2020hu,ipgdn2020liu,ijcai2020-398,jane2020yang,wang2017mgae,shi2020effective,pan2018adversarially} \\
Recall & $\checkmark$ & Yes & \cite{wang2017mgae} \\
\rowcolor{gray!30} 
F1-Score & $\checkmark$ & Yes &  \cite{cai2020edge,gcln2020hu,zhang2020commdgi,zhang2019attributed,ipgdn2020liu,park2020unsupervised,zhang2020seal,ijcai2020-398,jane2020yang,jia2019communitygan,yang2017graph,wang2017mgae,bo2020structural,fan2020one2multi,wang2019attributed,shi2020effective,pan2018adversarially,de2021deep,luo2021detecting} \\
ARI & $\checkmark$ & Yes & \cite{gcln2020hu,zhang2020commdgi,cui2020adaptive,ipgdn2020liu,fu2020magnn,jane2020yang,wang2017mgae,bo2020structural,fan2020one2multi,wang2019attributed,magcn2020cheng,shi2020effective,pan2018adversarially,ye2018deep,wang2021self,zhang2021spectral,ijcai2021-473,luo2021detecting,li2021dnc,xu2021graph,xia2021self} \\
\rowcolor{gray!30} 
\textit{Q} & $\times$ &  & \cite{xie2019high,wang2020evolutionary,bhatia2018dfuzzy,xu2020stacked,choong2018learning,choong2019optimizing,li2021weighted} \\ 
\rowcolor{gray!30}
Extended-\textit{Q} & $\checkmark$ & \multirow{-2}{*}{No} & \cite{chen2021self} \\
Jaccard & $\checkmark$ & Yes & \cite{xin2017deep,zhang2020seal,yang2017graph} \\
\rowcolor{gray!30} 
CON & $\checkmark$ & No & \cite{choong2018learning,choong2019optimizing,wang2021cane} \\
TPR & $\checkmark$ & No  & \cite{choong2018learning,choong2019optimizing} \\ 
\bottomrule[1pt]
\multicolumn{4}{l}{\makecell[l]{``Overlap'' indicates whether the metric can evaluate the overlapping community that is defined in Section \ref{sec-preliminaries}. \\
``Ground Truth'' indicates whether the actual community structure of the network is required. }}\\
\end{tabular}
\label{tab-metrics}
\end{table*}

\begin{table*}[!t]
\centering \footnotesize
\renewcommand\arraystretch{1}
\setlength{\tabcolsep}{5mm}
\caption{Summary of open-source implementations.}
\begin{tabular}{ccll}
\toprule[1 pt]
\textbf{Category} & \textbf{Tool} & \textbf{Method} & \textbf{URL} \\ \midrule
\rowcolor{gray!30} 
\cellcolor{white!100} & \cellcolor{white!100} & AGE\cite{cui2020adaptive} & \url{https://github.com/thunlp/AGE} \\
\cellcolor{white!100} & \cellcolor{white!100} & LGNN\cite{chen2019supervised} & \url{https://github.com/zhengdao-chen/GNN4CD} \\
\rowcolor{gray!30} 
\cellcolor{white!100} & \cellcolor{white!100} \multirow{-3}{*}{PyTorch} & NOCD\cite{shchur2019overlapping} & \url{https://github.com/shchur/overlapping-community-detection} \\ \cmidrule{2-4}
\cellcolor{white!100} & \cellcolor{white!100} & AGC\cite{zhang2019attributed} & \url{https://github.com/karenlatong/AGC-master} \\
\rowcolor{gray!30} 
\cellcolor{white!100} \multirow{-5}{*}{GCN} & \cellcolor{white!100} \multirow{-2}{*}{TensorFlow} & CayleyNet\cite{levie2018cayleynets} &  \url{https://github.com/amoliu/CayleyNet} \\ \midrule
\cellcolor{white!100} & \cellcolor{white!100} & CP-GNN \cite{luo2021detecting} & \url{https://github.com/RManLuo/CP-GNN} \\
\rowcolor{gray!30} 
\cellcolor{white!100} & \cellcolor{white!100} & DMGI\cite{park2020unsupervised} & \url{https://github.com/pcy1302/DMGI} \\
\cellcolor{white!100} & \cellcolor{white!100} &  MAGNN\cite{fu2020magnn} & \url{https://github.com/cynricfu/MAGNN} \\ 
\rowcolor{gray!30} 
\cellcolor{white!100} & \cellcolor{white!100} & HDMI \cite{jing2021hdmi} & \url{https://github.com/baoyujing/HDMI} \\
\cellcolor{white!100} \multirow{-5}{*}{GAT} & \cellcolor{white!100} \multirow{-5}{*}{PyTorch} &
HeCo \cite{wang2021self} & \url{https://github.com/liun-online/HeCo} \\
\midrule
\rowcolor{gray!30} 
\cellcolor{white!100} & \cellcolor{white!100} PyTorch & SEAL\cite{zhang2020seal} & \url{https://github.com/yzhang1918/kdd2020seal} \\ \cmidrule{2-4}
\cellcolor{white!100} \multirow{-2}{*}{GAN} & \cellcolor{white!100} TensorFlow & CommunityGAN\cite{jia2019communitygan} & \url{https://github.com/SamJia/CommunityGAN} \\ \midrule
\rowcolor{gray!30} 
\cellcolor{white!100} & \cellcolor{white!100} & DANE\cite{gao2018deep} & \url{https://github.com/gaoghc/DANE} \\
\cellcolor{white!100} & \cellcolor{white!100} \multirow{-2}{*}{TensorFlow} & DIME\cite{zhang2017bl} & \url{http://www.ifmlab.org/files/code/Aligned-Autoencoder.zip} \\ \cmidrule{2-4}
\rowcolor{gray!30} 
\cellcolor{white!100} & \cellcolor{white!100} &  & \href{https://github.com/shenxiaocam/Deep-network-embedding-for-graph-representation-learning-in-signed-networks}{https://github.com/shenxiaocam/Deep-network-embedding-for-} \\
\rowcolor{gray!30} 
\cellcolor{white!100} & \cellcolor{white!100} & \multirow{-2}{*}{DNE-SBP\cite{shen2018deep}} & \href{https://github.com/shenxiaocam/Deep-network-embedding-for-graph-representation-learning-in-signed-networks}{graph-representation-learning-in-signed-networks} \\
\cellcolor{white!100} \multirow{-5}{*}{Stacked AE} & \cellcolor{white!100} \multirow{-3}{*}{Matlab} & semi-DRN\cite{yang2016modularity} & \url{http://yangliang.github.io/code/DC.zip} \\ \midrule
\rowcolor{gray!30} 
\cellcolor{white!100} Sparse AE & \cellcolor{white!100} PyTorch & GraphEncoder\cite{tian2014learning} & \url{https://github.com/zepx/graphencoder} \\ \midrule
\cellcolor{white!100} & \cellcolor{white!100} Keras & DNGR\cite{10.5555/3015812.3015982} & \url{https://github.com/MdAsifKhan/DNGR-Keras} \\ \cmidrule{2-4}
\rowcolor{gray!30} 
\cellcolor{white!100} \multirow{-2}{*}{\makecell{Denoising\\AE}} & \cellcolor{white!100} Matlab & MGAE\cite{wang2017mgae} & \url{https://github.com/FakeTibbers/MGAE} \\ \midrule
\cellcolor{white!100} & \cellcolor{white!100} PyTorch & SDCN\cite{bo2020structural} & \url{https://github.com/bdy9527/SDCN} \\ \cmidrule{2-4}
\rowcolor{gray!30} 
\cellcolor{white!100} \multirow{-2}{*}{\makecell{Graph Convolutional\\AE}} & \cellcolor{white!100} TensorFlow & O2MAC\cite{fan2020one2multi} & \url{https://github.com/songzuolong/WWW2020-O2MAC} \\ \midrule
\cellcolor{white!100} Graph Attention & \cellcolor{white!100} & DMGC\cite{luo2020deep} & \url{https://github.com/flyingdoog/DMGC} \\
\rowcolor{gray!30} 
\cellcolor{white!100} AE & \cellcolor{white!100} \multirow{-2}{*}{TensorFlow} & SGCMC\cite{xia2021self} & \url{https://github.com/xdweixia/SGCMC} \\
\midrule
\cellcolor{white!100} Variational & \cellcolor{white!100} & ARGA/ARVGA\cite{pan2018adversarially} & \url{https://github.com/Ruiqi-Hu/ARGA} \\
\rowcolor{gray!30} 
\cellcolor{white!100} AE & \cellcolor{white!100} \multirow{-2}{*}{TensorFlow} & DGLFRM\cite{mehta2019stochastic} & \url{https://github.com/nikhil-dce/SBM-meet-GNN} \\
\bottomrule[1 pt]
\end{tabular}
\label{table-code}
\end{table*}

\section{Published Resources} \label{sec-experimentation} 
From surveyed literature in Sections \ref{sec-model-cn}--\ref{sec-model-filtering}, the popular and available benchmark data sets, evaluation metrics and open-source implementations are summarized in this section with details in APPENDIX \ref{sec-ddds}--\ref{sec-appcode}.

\subsection{Data Sets} \label{sec-datasets}
Both real-world data sets and synthetic data sets are popularly published. Real-world data sets in community detection experiments are collected from real-world applications, which test performances of proposed methods from the real applicable aspect. Synthetic data sets are generated by specific models on manually designed rules, and particular functions can be tested by these data sets. The state-of-the-art popular real-world data sets can be categorized into citation/co-authorship networks, social networks (online and offline), webpage networks and product co-purchasing networks. The typical data sets covering various network shapes (\textit{i.e.}, unattributed, attributed, multi-view, signed, heterogeneous and dynamic) are summarized in TABLE \ref{table-rwdatasets}. The related description is detailed in APPENDIX \ref{sec-ddds} (Real-world Data Sets). Girvan-Newman (GN) networks \cite{girvan2002community} and Lancichinetti–Fortunato–Radicchi (LFR) networks \cite{lancichinetti2008benchmark} are two widely applied synthetic benchmarks, described in APPENDIX \ref{sec-ddds} (Synthetic Benchmark Data Sets).

\subsection{Evaluation Metrics} \label{sec-metrics}
This section summarizes twelve most popular and commonly applied evaluation metrics from the reviewed deep community detection literature (details in APPENDIX \ref{sec-EM}). Evaluations on detection performances vary in community types (\textit{i.e.}, disjoint or overlapping) and ground truth (\textit{i.e.}, available or lacking). We summarize ``Publications'' by metrics in TABLE \ref{tab-metrics} over ``Overlap'' and ``Ground Truth''.

\subsection{Open-source Implementations}\label{sec-code}

The open-source implementations are summarized by tool, category, repository link in TABLE \ref{table-code}. The majority of implementations are under Python 3.x, including PyTorch, TensorFlow and Keras. There are a few implementations that use Matlab. Each implementation project is briefly described and organized by implementation tools and deep learning models in APPENDIX \ref{sec-appcode}.

\section{Practical Applications} \label{sec-application} 

\begin{figure}[!t]
\centering
\includegraphics[width=0.48\textwidth]{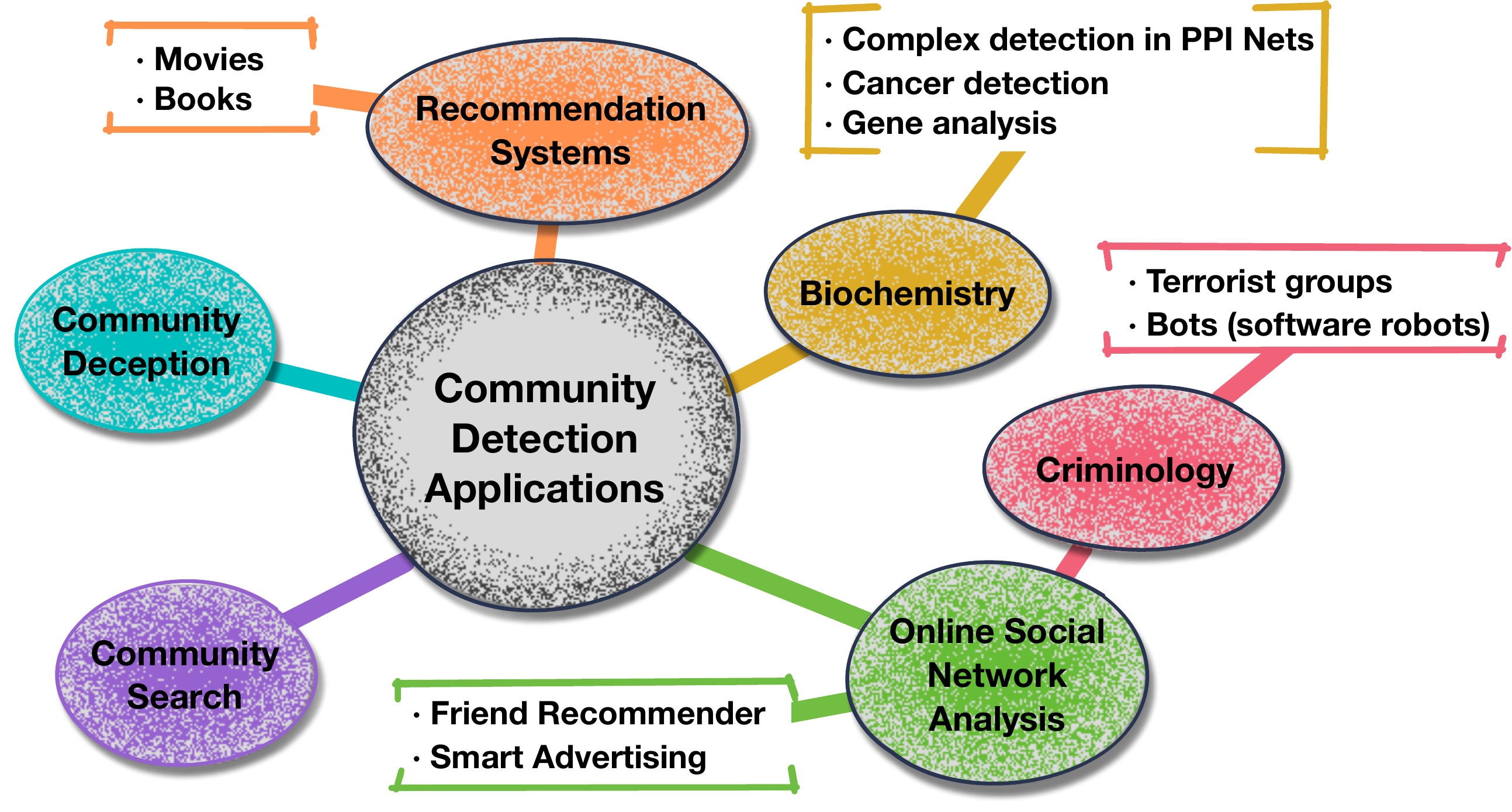}
\caption{\footnotesize Practical applications of community detection.}
\label{fig-application}
\end{figure}

Community detection has many applications across different tasks and domains, as summarized in Fig. \ref{fig-application}. We now detail some typical applications in the following areas.

\vspace{2mm}
\noindent
\textbf{Recommendation Systems.} Community structure plays a vital role for graph-based recommendation systems \cite{AAAI1816261,10.1145_Gao}, as the community members may have similar interests and preferences. By detecting relations between nodes (\textit{i.e.}, users--users, items--tems, users--items), models such as CayleyNets \cite{levie2018cayleynets} and UWMNE/WMCNE-LE \cite{jin2018integrative} produce high-quality recommendations.

\vspace{2mm}
\noindent
\textbf{Biochemistry.} In this field, nodes represent proteins or atoms in compound and molecule graphs, while edges denote their interactions. Community detection can identify complexes of new proteins \cite{xin2017deep,tian2014learning} and chemical compounds which are functional in regional organs (\textit{e.g.}, in brain \cite{garcia2018applications}) or pathogenic factor of a disease (\textit{e.g.}, community-based lung cancer detection \cite{bechtel2005lung}). To various tumor types on genomic data sets, the previous study \cite{haq2016community} shows relevances between communities' survival rates and distributions of tumor types over communities.

\vspace{2mm}
\noindent
\textbf{Online Social Networks.} 
Analyzing online social activities can identify online communities and correlate them in the real world. The practice on online social networks \cite{girvan2002community} such as Twitter, LinkedIn and Facebook reveals similar interests among online users that individual preferences can be automatically provided. Meanwhile, community detection can be used for online privacy \cite{remy2017tracking} and to identify \textbf{criminals} based on online social behaviors \cite{chen2012community}, who support and diffuse criminal ideas and even who may practice terrorism activities \cite{waskiewicz2012friend}.

\vspace{2mm}
\noindent
\textbf{Community Deception.} To hide from the community detection, community deception \cite{fionda2016community} covers a group of users in social networks such as Facebook. Deception is either harmful to virtual communities or harmless that provides justifiable benefits. From community-based structural entropy, Residual Entropy Minimization (REM) effectively nullify community detection algorithms \cite{liu2019rem}. A systematic evaluation \cite{fionda2017community} on community detection robustness to deception is carried out in large networks.

\vspace{2mm}
\noindent
\textbf{Community Search.} Community search aims to queue up a series of nodes dependent on communities \cite{li2017most}. For example, a user (node) search on interests (communities) after another user (node). To this end, communities are formed temporally based on the user's interest. There are several practices applied in this scenario. Local community search \cite{cui2014local} assumes one query node at a time and expands the search space around it. The strategy is attempted repeatedly until the community finds all its members. Attributed Truss Communities (ATC) \cite{huang2017attribute} interconnects communities on query nodes with similar query node attributes.

\section{Future Directions} \label{sec-future}

Although deep learning has brought community detection into a prospering era, several open issues need further investigation. In this section, we recommend twelve future directions.

\subsection{An Unknown Number of Communities}
For community detection in real-world scenario, the majority of data are unlabeled due to the high-cost acquisition, therefore the number of communities $K$ is often unknown. Unsupervised detection methods in deep learning provides an effective way to handle such unknown scenario, but they generally need to specify $K$ as the prior knowledge. This phenomenon brings us a catch-22: approaches require prior knowledge which does not exist. Therefore, there is an urgent demand to handle the unknown community number issue. 

\vspace{2mm}
\noindent 
\textbf{Opportunities:} Analysis of network topological structure offers a potential solution to tackle this challenge, and some research efforts have been made \cite{bhatia2018dfuzzy}. Typically, these methods perform random walks to get preliminary communities and refine the detection results by modularity. Nevertheless, when they come to disconnected networks in practice, random walks cannot involve every node and degrade the detection performance. This open issue, therefore, calls for a complete solution and further research.

\subsection{Community Embedding}
Node embedding methods traditionally preserve nodes' neighborhood information, in which structural information of communities are ignored \cite{perozzi2014deepwalk}. To this end, designing community-aware learning processes to characterize community information can improve the accuracy of community detection \cite{cavallari2017learning}.

\vspace{2mm}
\noindent
\textbf{Opportunities:} To date, few works integrate the community embedding into a deep learning model, so more efforts are desired for this promising area. In general, as community embedding that learns representations on each community increases computational cost, future work needs to develop fast computation-aimed algorithms. Furthermore, the optimization mechanism to hyperparameters in the community embedding needs a new design so that deep community detection can meet the expectation.

\subsection{Hierarchical Networks}
Networks such as the Web often show a treelike hierarchical organization of communities in different scales, in which small communities in lower levels group together to form larger ones in higher levels. Hence, community detection is required to detect communities from low to high levels.

\vspace{2mm}
\noindent
\textbf{Opportunities:} Traditional methods generally follow one of three work lines: (1) estimating the hierarchy directly, (2) recursively merging communities in a bottom-up fashion, and (3) recursively splitting communities in a top-down fashion. Their performance is limited by either a large number of parameters or the requirements for network density \cite{li2020hierarchical}. Network embedding shows the efficiency to these issues \cite{du2018galaxy,long2019hierarchical}. However, preserving the community hierarchy into the embeddings remains unsolved \cite{long2019hierarchical}. With the advancement of high-order relationship representations, we believe deep learning models can facilitate the development of hierarchical community detection.

\subsection{Multi-layer Networks}
Realistic systems always possess the nature of multiple layers. To exhibit this property, the multi-layer network is extracted. It is composed of multiple interdependent graphs, called layers, where each layer represents a different type of interactions among entities, and the cross-layer links reflect dependencies among entities\cite{8047481}. In the multi-layer social networks, for example, nodes can be individuals, the intralayer connections of each layer can represent friendship, kinship, or schoolmates, etc., while the interlayer connections align the same individuals \cite{kim2015community}. Thus, community detection in such networks can leverage more information to benefit the results. 

\vspace{2mm}
\noindent
\textbf{Opportunities:} In contrast with the prosperity of community detection works in single-layer networks, the development of research on multi-layer networks is still in its infancy \cite{huang2021a}. A straightforward process is fusing information of multi-layer networks into one layer, followed by monolayer community detection methods. In deep learning architecture, a similar solution is to flatten the multi-layer information in low dimensional representations with the following general open issues: (1) differences among the interaction types, (2) varying levels of sparsity in layers, (3) possible connections across layers, and (4) the scalability scheme of layers. 

\subsection{Heterogeneous Networks}
For an accurate depiction of reality, networks are required to involve heterogeneous information \cite{cao2021knowledgepreserving} that characterizes relationships between different types of entities, such as the role-play relation between actors and movies. Since community detection methods designed for homogeneous networks cannot be directly employed, due to lack of capacity to model the complex structural and semantic information, community detection research on heterogeneous networks is challenging. 

\vspace{2mm}
\noindent
\textbf{Opportunities:} Metapath is a promising research effort to deal with diverse semantic information, which describes a composite relation between the node types involved. It allows deep models to represent nodes by aggregating information from the first-order structure with different metapaths and measuring node similarity to cluster communities \cite{zhang2017bl,wang2019heterogeneous,fu2020magnn}. However, the selection of most meaningful metapaths remains an open problem. The future efforts should focus on a flexible schema for metapath selection and novel models that can exploit various types of relationships.

\subsection{Network Heterophily}
Network heterophily \cite{Zhu2020BeyondHI} can be interpreted as a phenomenon that connected nodes belong to different communities or characterized by dissimilar features. For example, fraudsters intentionally make a connection with normal users to hide from being discovered. For community detection, boundary nodes connected across communities comply with this property. It is significant to capture network heterophily providing valuable information on community division.

\vspace{2mm}
\noindent
\textbf{Opportunities:} Since most methods heavily rely on homophily, they assume connected nodes share more similarities and are more likely from the same community. Deep learning methods that exploit network heterophily are expected for better community detection performance.

\subsection{Topologically Incomplete Networks}
Relationships in real-world scenarios are not always available, which leads to incomplete network topology and isolated subgraphs \cite{xin2017deep}. For example, protein-protein interaction (PPI) networks are usually incomplete since monitoring all protein-protein interactions are expensive \cite{jeong2001lethality}. Deriving meaningful knowledge of communities from limited topology information has been crucial to this case.

\vspace{2mm}
\noindent
\textbf{Opportunities:} The requirement of complete network topology reduces the applicability of community detection methods, especially those based on neighborhood aggregations. To this end, deep learning methods should be further developed with an information recovery mechanism to achieve accurate community detection on TINs.

\subsection{Cross-domain Networks} 
In the real world, there are many similar networks such as Facebook and Twitter. Each network has its independent domain. Borrowing knowledge from an information-rich domain benefits the network learning in the target domain \cite{XUE2019135}. Therefore, community detection is encouraged to develop its own deep learning models to improve detection results \cite{xie2019high}.

\vspace{2mm}
\noindent
\textbf{Opportunities:} Through community adaptations, the latent representation transferring from the source to the target domain faces the following challenges: (1) a lack of explicit community structure, (2) node label shortages, (3) missing a community's ground truth, (4) poor representation performance caused by the inferior network structure, and (5) deep learning with small-scale networks. The future methods aim to measure cross-domain coefficiency, distribution shifts, and control computational costs.
	
\subsection{Attributed Multi-view Networks}
Real-world networks are complex and contain distinctively featured data \cite{6729567}. Attributed multi-view networks provide a new way to describe relational information in multiple informative views, each containing a type of attributes \cite{7023376}. Communities are initially detected on each single-view network and deeply encoded over multiple views in an expensive way. Hence, cheap community detection methods are aimed to exploit the multi-view complementarity \cite{li2019flexible}.

\vspace{2mm}
\noindent
\textbf{Opportunities:} A straightforward workflow is to combine representations learned separately from each view, but it may cause noises and redundancy, which worsens community detection results. To develop an integrated framework, deep learning tries to extract the consistency information among multiple views by learning a common clustering embedding for community detection \cite{magcn2020cheng}. Since multi-view node attributes still require better integration scheme in the learning process, future works on networks with multi-view attributes need to develop a suitable integration scheme in the learning process. Meanwhile, more works are encouraged to study the under-explored issue of global representation for community detection over multi-views to avoid sub-optimization.

\subsection{Signed Networks}
It has been increasingly noticed that not all occurring connections get entities close. Generally, friendship indicates the positive sentiment (\textit{e.g.}, like and support), while foes express the negative attitude (\textit{e.g.}, dislike). These distinctions are reflected on network edges, called signed edges and signed networks \cite{10.1145_Xu}. As impacts of positive and negative ties are different, existing community detection methods working on unsigned networks are not applicable to signed networks. 

\vspace{2mm}
\noindent
\textbf{Opportunities:} To conduct community detection in signed network, the main challenges lie in adapting negative ties. Deep learning techniques should be exploited to represent both ties. Negative ties offer distinguishing community knowledge in learning signed network embeddings \cite{shen2018deep}. Future works are expected to cope with signed edges and further identify positive/negative information automatically on edges.

\subsection{Dynamic Networks}
Real networks may evolve over dynamic structures and temporal semantic features. The addition/removal of nodes/edges changes the community structure. Accordingly, deep learning models should rapidly capture the network evolvement to explore community changes. Based on our literature review, only one study touches this topic, designing an evolutionary AE to discover smoothly changing community structure \cite{wang2020evolutionary}.

\vspace{2mm}
\noindent
\textbf{Opportunities:} Both deep learning and community detection need to handle the shifting distributions and evolving data scales. Repeated training over snapshots is expensive and slow. Technical challenges in detecting communities from a dynamic network mainly come from the dynamics control from both spatial and temporal parts. Future directions include: (1) the analysis of spatial changes on communities, (2) the recognition of dynamic network patterns, and (3) tracks of network dynamics and community detection robustness.

\subsection{Large-scale Networks}
Large-scale networks may contain massive number of nodes, edges and communities. Their inherent scale characteristics, such as scale-free or named as power-law degree distribution \cite{watts1998collective, barabasi2003scale} in social networks, can influence the performance of deep community detection. Scalability is another crucial issue in large-scale networks \cite{li2017most}. A suggested direction on this topic is to develop a flexible deep learning approach aiming at collaborative computing with high robustness.

\vspace{2mm}
\noindent
\textbf{Opportunities:} Common dimension reduction strategies in deep learning, such as matrix low-rank approximation, cannot cope with the high-dimensional relationships in large-scale networks. Current solutions of distributed computing are too expensive. Consequently, there is a crying need for novel deep learning frameworks, models and algorithms exceeding current benchmarks in terms of precision and speed.
	
\section{Conclusions} \label{sec-conclusion}
This survey provides a comprehensive overview of the state-of-the-art community detection methods. During the recent decade, deep learning model-based methods are notably developed. This trend completely changes the community detection framework, resulting in a large number of publications in multiple fields in high-impact international conferences and peer-view journals. According to our survey, these methods significantly increase the effectiveness, efficiency, robustness and applicability of community detection. New techniques are much more flexible in use than traditional methods, and a larger volume of data can be leveraged in a rough preprocess. Selectively included into six categories, a taxonomy is newly designed. In each category, community detection is aimed by the deep learning model, \textit{i.e.}, encoding representations and optimizing clustering results. We discussed the contributions of each deep learning model to the community detection task. Furthermore, we summarized and provided handy resources, \textit{i.e.}, data sets, evaluation metrics, open-source implementations, based on the reviewed literature. We also offered an insight into a range of community detection applications. To stimulate further research in this area, we identified twelve open directions. 

\bibliographystyle{IEEEtran}
\bibliography{ref}

\newpage
\vspace{-1cm}

\begin{IEEEbiography}[{\includegraphics[width=1in,height=1.25in,clip,keepaspectratio]{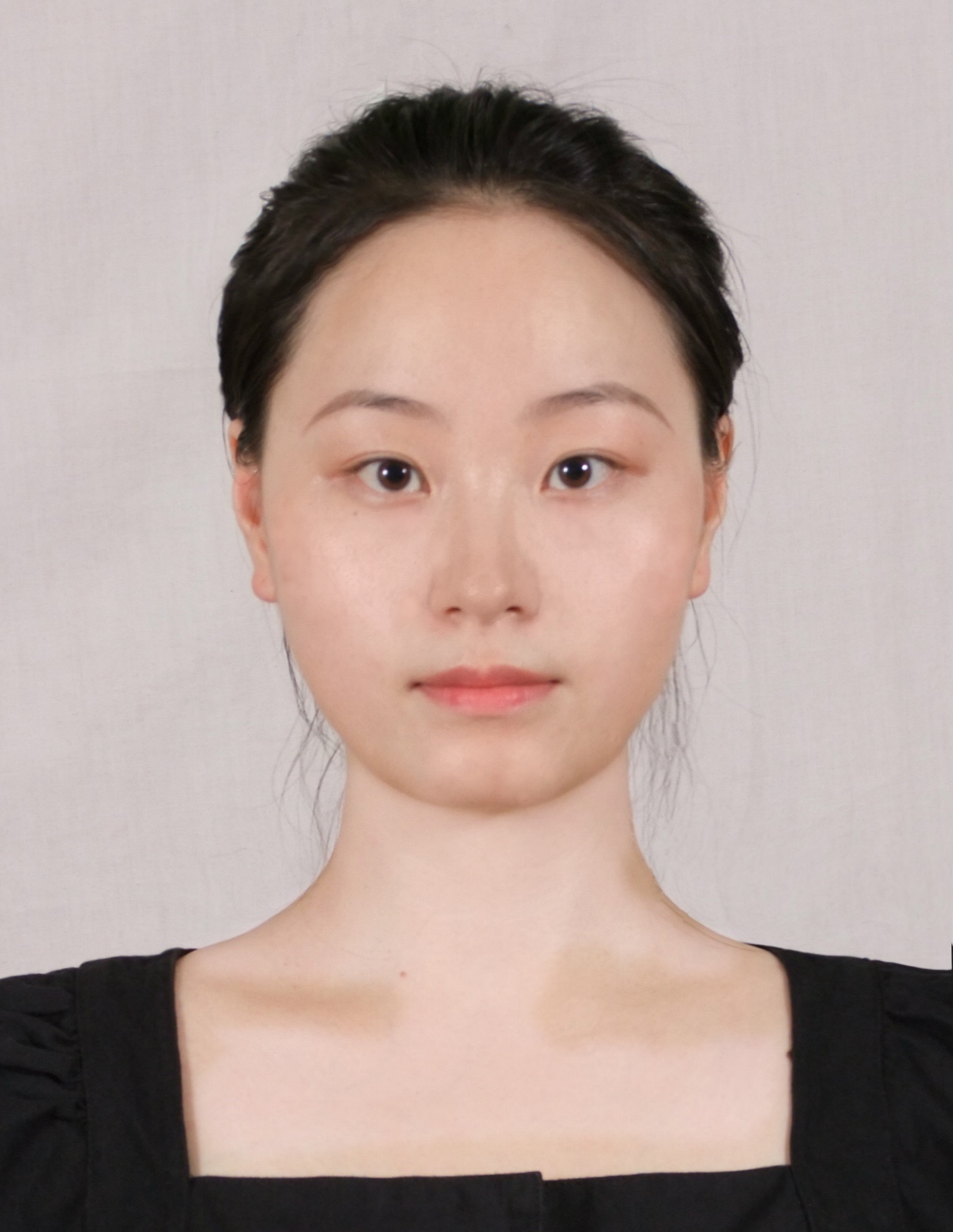}}]{Xing Su} received her M.Eng. degree in computer technology from Lanzhou University, China in 2020. She is currently a Ph.D. candidate in School of Computing at Macquarie University, Australia. Her current research interests include community detection, deep learning and social network analysis.
\end{IEEEbiography}

\vspace{-1cm}

\begin{IEEEbiography}[{\includegraphics[width=1in,height=1.25in,clip,keepaspectratio]{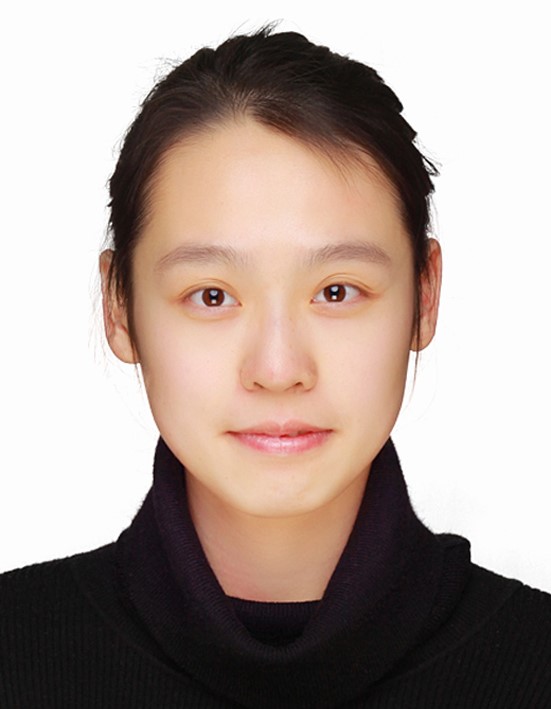}}]{Shan Xue} received her Ph.D. in Computer Science from the Center for Artificial Intelligence, University of Technology Sydney, Australia in 2019 and a Ph.D. in Information Management from the School of Management, Shanghai University, Shanghai, China in 2018. She is a postdoctoral research fellow of School of Computing, Faculty of Science and Engineering, Macquarie University, Sydney, Australia as well as a researcher of CSIRO Data61, Sydney, Australia. Her current research interests include artificial intelligence, machine learning and knowledge mining.
\end{IEEEbiography}

\vspace{-1cm}

\begin{IEEEbiography}[{\includegraphics[width=1in,height=1.25in,clip,keepaspectratio]{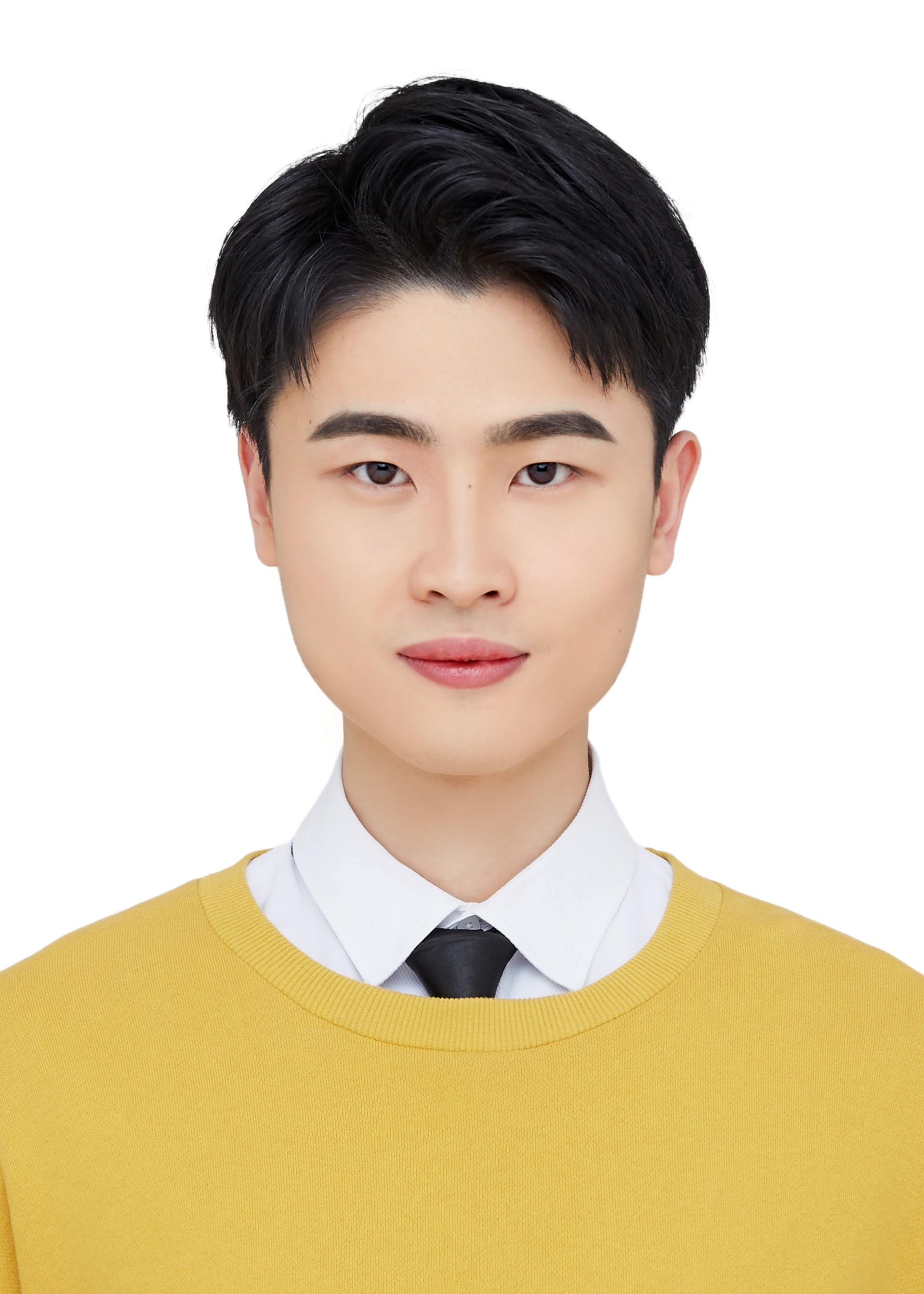}}]{Fanzhen Liu} received the M.Res. (Master of Research) degree from Macquarie University, Sydney, Australia. He is currently pursuing the Ph.D. degree in computer science with Macquarie University, Sydney, Australia. His current research interests include graph mining and machine learning.
\end{IEEEbiography}

\vspace{-1cm}

\begin{IEEEbiography}[{\includegraphics[width=1in,height=1.25in,clip,keepaspectratio]{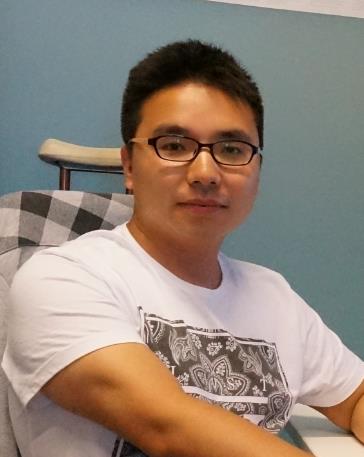}}]{Jia Wu} is an ARC DECRA Fellow at the School of Computing, Macquarie University, Sydney, Australia. He received the PhD from the University of Technology Sydney, Australia. His current research interests include data mining and machine learning. He has published 100+ refereed research papers in the above areas such as  TPAMI, TKDE, TNNLS, TMM, NIPS, KDD, ICDM, IJCAI, AAAI and WWW. Dr Wu was the recipient of SDM'18 Best Paper Award in Data Science Track and IJCNN'17 Best Student Paper Award. He currently  serves  as an Associate Editor of ACM \textit{Transactions on Knowledge Discovery from Data} (TKDD). He is a Senior Member of IEEE.
\end{IEEEbiography}

\vspace{-1cm}

\begin{IEEEbiography}[{\includegraphics[width=1in,height=1.25in,clip,keepaspectratio]{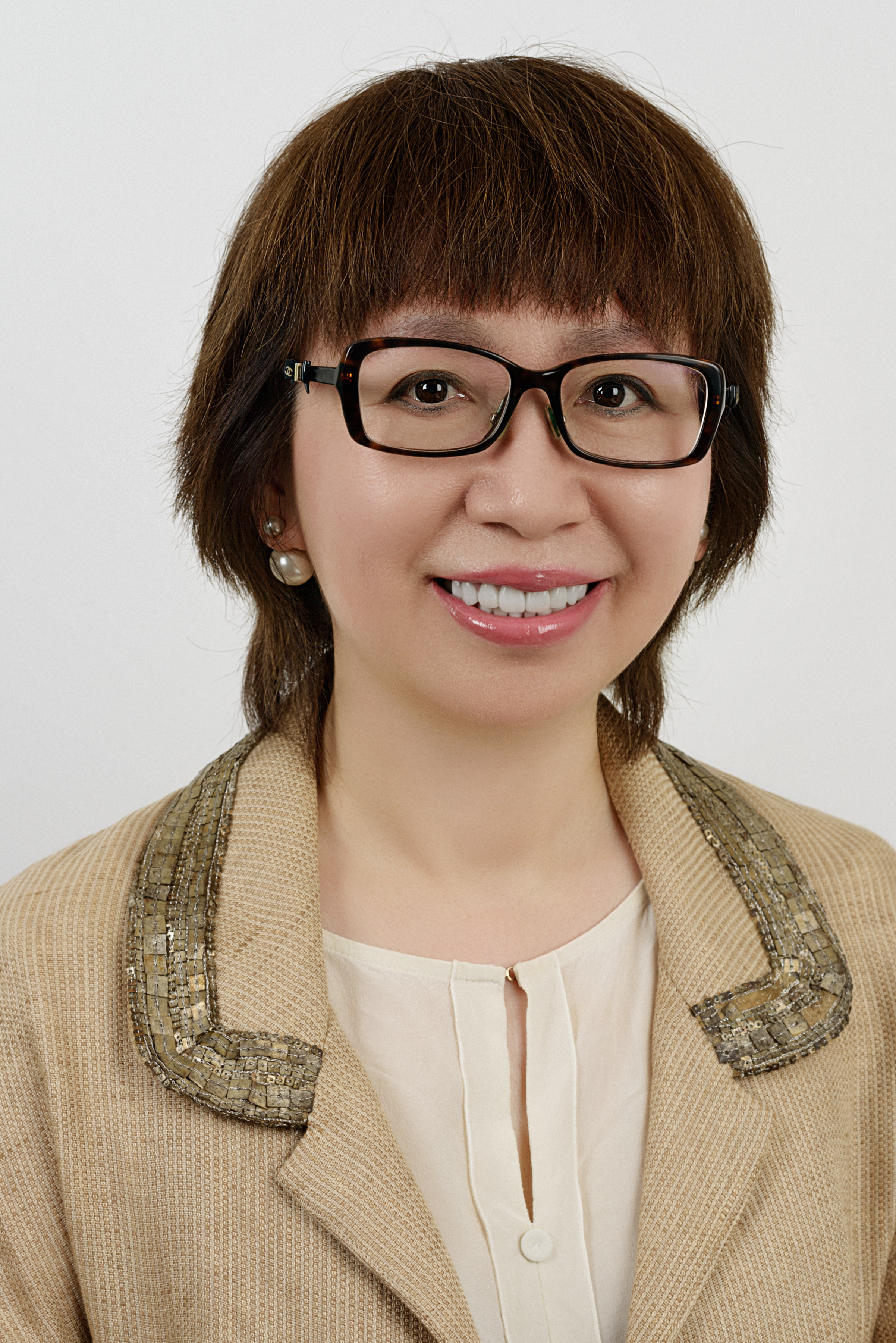}}]{Jian Yang} is a full professor at the School of Computing, Macquarie University. She received her PhD in Data Integration from the Australian National University in 1995. Her main research interests are: business process management; data science; social networks. Prof. Yang has published more than 200 journal and conference papers in international journals and conferences such as IEEE Transactions, Information Systems, Data and Knowledge Engineering, VLDB, ICDE, ICDM, CIKM, etc. She is currently serving as an Executive Committee for the Computing Research and Education Association of Australasia.
\end{IEEEbiography}

\vspace{-1cm}

\begin{IEEEbiography}[{\includegraphics[width=1in,height=1.25in,clip,keepaspectratio]{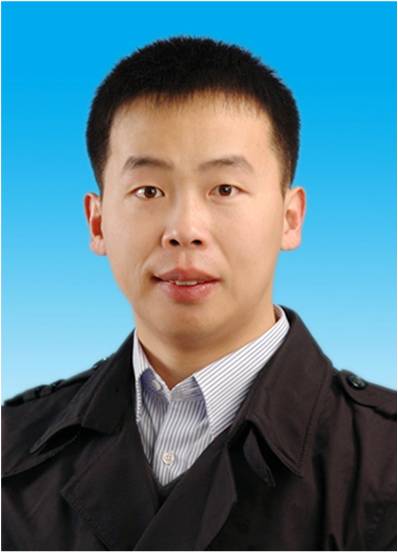}}]{Chuan Zhou} obtained Ph.D. degree from Chinese Academy of Sciences in 2013. He won the outstanding doctoral dissertation of Chinese Academy of Sciences in 2014, the best paper award of ICCS-14, and the best student paper award of IJCNN-17. Currently, he is an Associate Professor at the Academy of Mathematics and Systems Science, Chinese Academy of Sciences. His research interests include social network analysis and graph mining. To date, he has published more than 80 papers, including TKDE, ICDM, AAAI, IJCAI, WWW, etc.
\end{IEEEbiography}

\vspace{-1cm}

\begin{IEEEbiography}[{\includegraphics[width=1in,height=1.25in,clip,keepaspectratio]{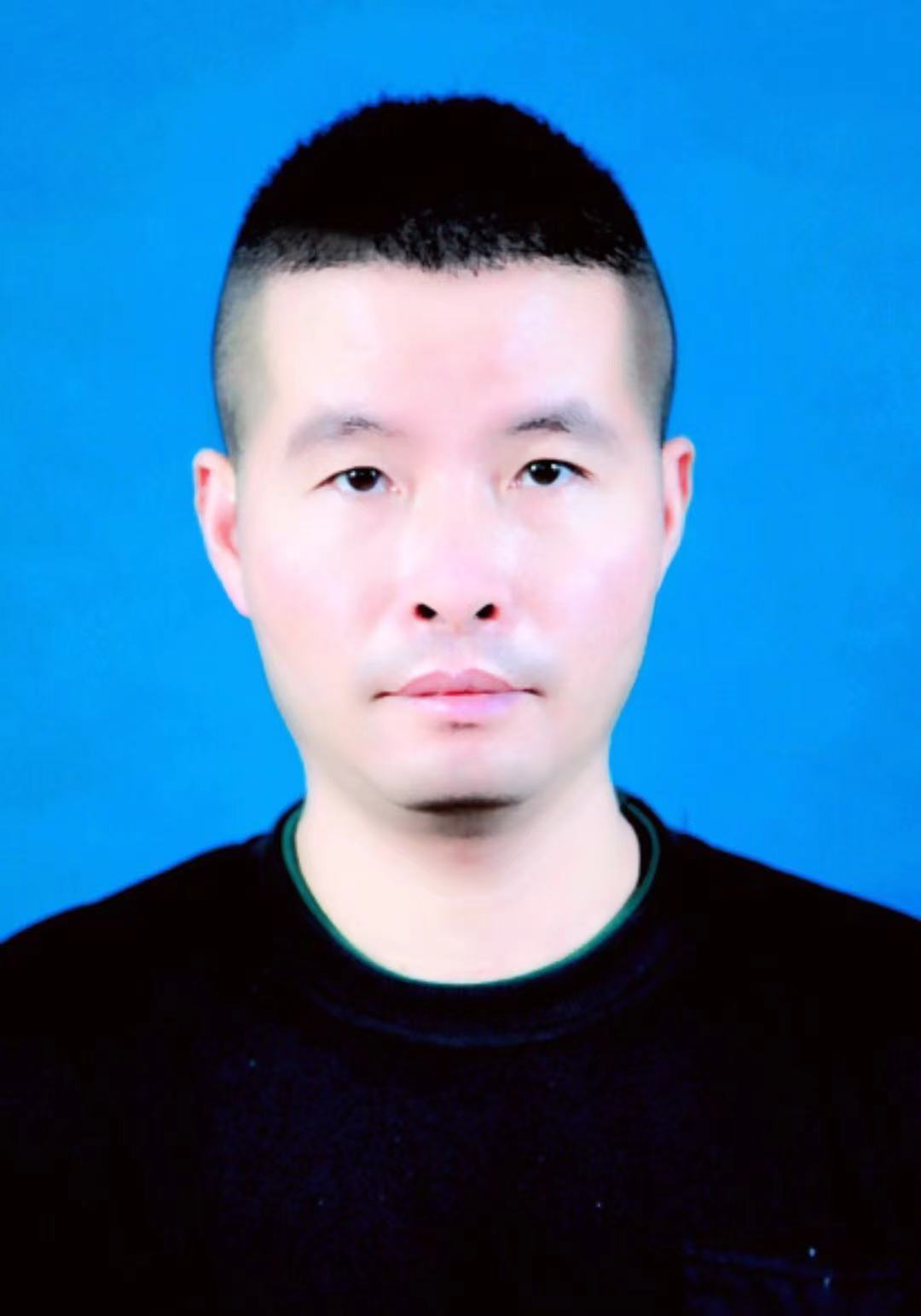}}]{Wenbin Hu} is currently a professor with the School of Computer Science, Wuhan University. His current research interests include intelligent and complex network, data mining and social network. He has more than 80 publications appeared in several top conferences such as SIGKDD, IJCAI, AAAI and ICDM, and journals such as IEEE TKDE, IEEE TMC, IEEE TITS, and ACM TKDD.
\end{IEEEbiography}

\vspace{-1cm}

\begin{IEEEbiography}[{\includegraphics[width=1in,height=1.25in,clip,keepaspectratio]{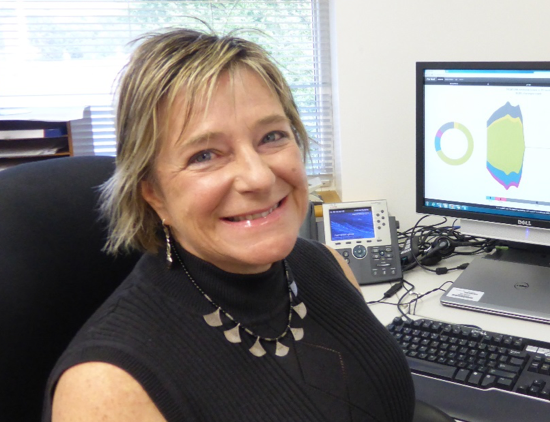}}]{Cecile Paris } is a pioneer in Natural Language Processing and User Modelling, and, more generally, in Artificial Intelligence and human-machine communication. With a Bachelor Degree from the University of Berkeley (California) and a PhD from Columbia University (New York), she has over to 25 years of experience in research and research management, and over 300 refereed publications. She is a Chief Research Scientist at CSIRO Data61, a Fellow of the Academy for Technology, Science and Engineering (ATSE) and of the Royal Society of NSW, and an Honorary Professor at Macquarie University. 
\end{IEEEbiography}

\vspace{-1cm}

\begin{IEEEbiography}[{\includegraphics[width=1in,height=1.25in,clip,keepaspectratio]{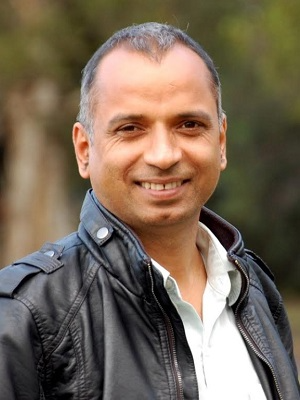}}]{Surya Nepal} is a Senior Principal Research Scientist at CSIRO Data61 and the Deputy Research Director of Cybersecurity Cooperative Research Centre (CRC). He has been with CSIRO since 2000 and currently leads the distributed systems security group comprising 30 staff and 50 PhD students. His main research focus is on distributed systems, with a specific focus on security, privacy and trust. He has more than 250 peer-reviewed publications to his credit. He is a member of the editorial boards of IEEE TSC, ACM TOIT, and IEEE DSC. Dr Nepal also holds an honorary professor position at Macquarie University and a conjoint faculty position at UNSW.
\end{IEEEbiography}

\vspace{-1cm}

\begin{IEEEbiography}[{\includegraphics[width=1in,height=1.25in,clip,keepaspectratio]{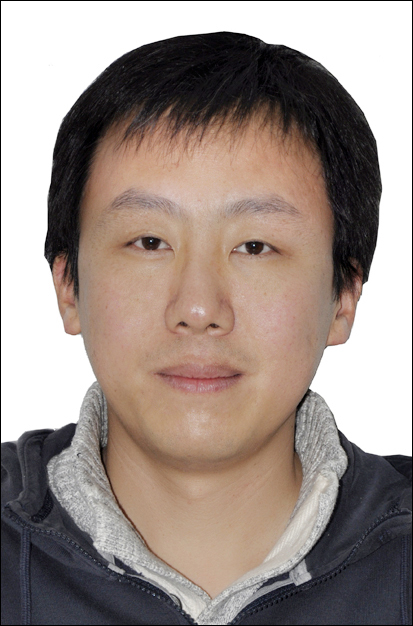}}]{Di Jin} received the Ph.D. degree in
computer science from Jilin University, Changchun,
China, in 2012. He was a Postdoctoral Research Fellow with the
School of Design, Engineering, and Computing,
Bournemouth University, Poole, U.K., from 2013 to
2014. He is currently an Associate Professor with
the College of Intelligence and Computing, Tianjin
University, Tianjin, China. He has published more
than 50 papers in international journals and conferences in the areas of community detection, social network analysis, and machine learning.
\end{IEEEbiography}

\vspace{-1cm}

\begin{IEEEbiography}[{\includegraphics[width=1in,height=1.25in,clip,keepaspectratio]{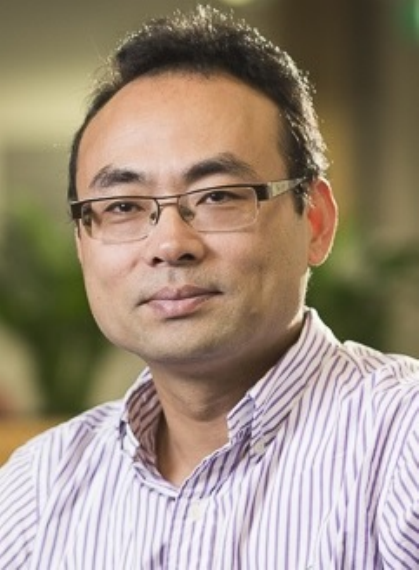}}]{Quan Z. Sheng} is a full Professor and Head of School of Computing at Macquarie University, Sydney, Australia. His research interests include big data analytics, service oriented computing, and Internet of Things. Michael holds a PhD degree in computer science from the University of New South Wales. He has more than 400 publications. Prof Michael Sheng is the recipient of AMiner Most Influential Scholar Award in IoT in 2019, ARC Future Fellowship in 2014, Chris Wallace Award for Outstanding Research Contribution in 2012, and Microsoft Fellowship in 2003. He is ranked by Microsoft Academic as one of the Most Impactful Authors in Services Computing (ranked Top 5 all time).
\end{IEEEbiography}

\vspace{-1cm}

\begin{IEEEbiography}[{\includegraphics[width=1in,height=1.25in,clip,keepaspectratio]{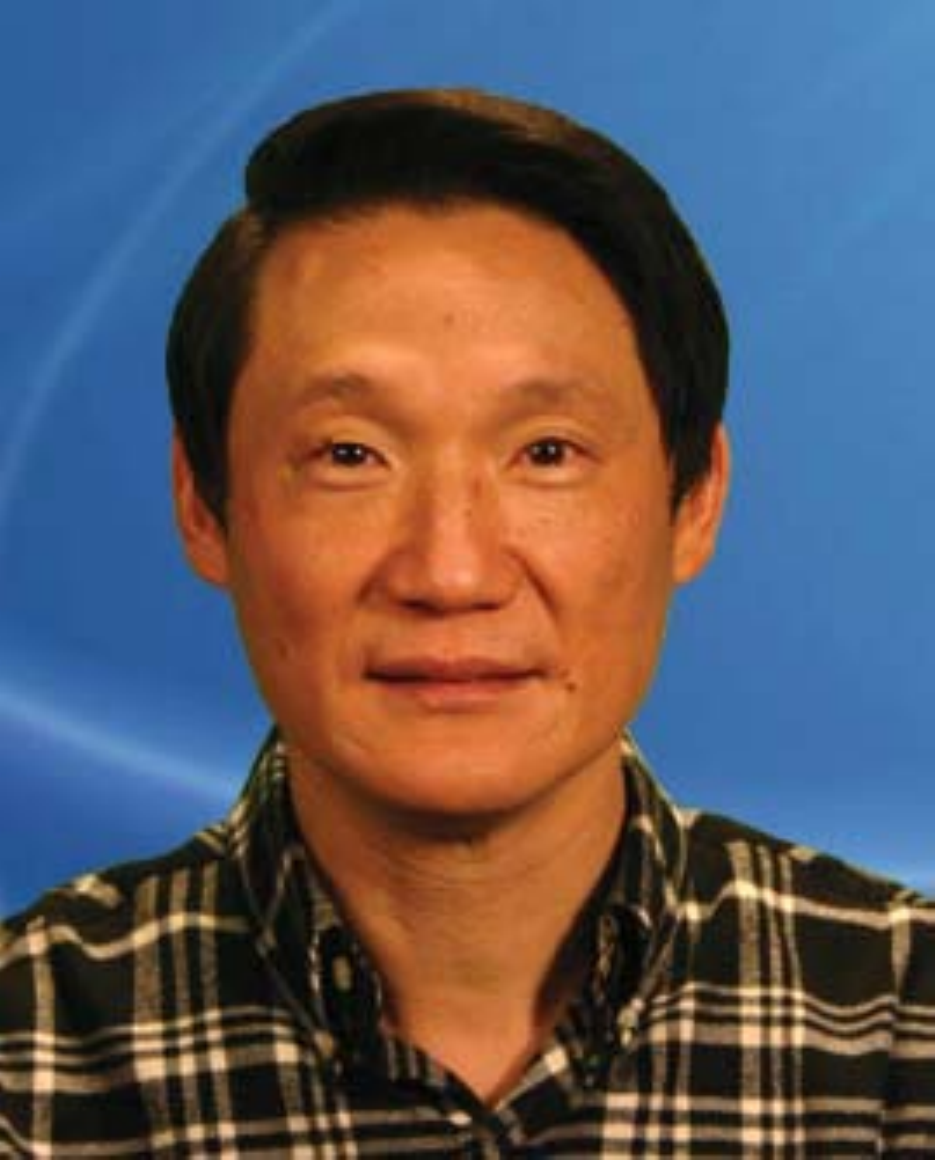}}]{Philip S. Yu}
(F'93) is a distinguished professor of computer science with the University of Illinois at Chicago and also holds the Wexler chair in information technology. His research interest is on big data, including data mining, data stream, database, and privacy. He has published more than 990 papers in refereed journals and conferences. He holds or has applied for more than 300 US patents. He was the editor-in-chief of the IEEE TKDE (2001-2004). He has received several IBM honors including two IBM Outstanding Innovation Awards. Prof. Yu is a fellow of the ACM and the IEEE. \end{IEEEbiography}

\clearpage 
\appendices
\setcounter{page}{1}
\twocolumn

\section{Summarized Techniques of Deep Learning-based Community Detection Methods}\label{sec-stdlcd}

TABLE \ref{table-cnn}--\ref{table-autoencoder} summarize the reviewed literature in sections \ref{sec-model-cnn} (CNN-based Community Detection), \ref{sec-model-gcn} (GCN-based Community Detection), \ref{sec-model-gat} (GAT-based Community Detecion), \ref{sec-model-gan} (GAN-based Community Detection) and \ref{sec-model-autoencoder} (AE-based Community Detection), respectively. Regarding the ``Input'', different methods require different inputs, such as network topology $\bm{A}$, node attributes $\bm{X}$ and other preprocessed data. Regarding the availability of ground truth labels, methods vary in supervised, semi-supervised and unsupervised ``Learning''. Meanwhile, methods that are able to output overlapping communities are marked by ``$\checkmark$'' in the ``Overlap'' column. Methods that can detect overlapping communities are also suitable to disjoint community detection. To other special requirements summarized in the tables, please refer to the corresponding sections.

\begin{table*}[h]
\centering 
\footnotesize
\renewcommand\arraystretch{1.1}
\setlength{\tabcolsep}{4.5mm}
\caption{Summary of CNN-based community detection methods.}
\rowcolors{2}{white}{gray!30}
\begin{tabular}{lcccccc}
\toprule[1 pt]
\textbf{Method} & \textbf{Input} & \textbf{Learning} &  \textbf{Preprocess} & \textbf{Co-technique} & \textbf{Overlap} & \textbf{Network} \\ 
\midrule
Xin \textit{et al.} \cite{xin2017deep} & $\bm{A}$ & Supervise & Node to image & -- & $\times$ & TINs \\ 
SparseConv \cite{sperli2019deep} & $\bm{A}$ & Supervise & Node to image & Sparse matrix & $\times$ & Sparse network \\
SparseConv2D \cite{de2021deep} & $\bm{A}$ & Semi-supervise & Node to image & Sparse matrix & $\times$ & Sparse network \\ 
ComNet-R \cite{cai2020edge} & $\bm{A}$ & Supervise & Edge to image & Local modularity & $\times$ & Large-scale network \\
\bottomrule[1 pt]
\end{tabular}
\label{table-cnn}
\end{table*}

\begin{table*}[h]
\centering 
\footnotesize
\renewcommand\arraystretch{1.1}
\setlength{\tabcolsep}{2.5mm}
\caption{Summary of GCN-based community detection methods.}
\rowcolors{2}{white}{gray!30}
\begin{tabular}{lcccccc}
\toprule[1 pt]
\textbf{Method} & \textbf{Input} & \textbf{Learning} & \textbf{Convolution} & \textbf{Clustering} & \textbf{Co-technique} & \textbf{Overlap} \\ 
\midrule
LGNN\cite{chen2019supervised} & $\bm{A},\bm{X}$ & Supervise & First-order + Line graph & -- & Edge features & $\checkmark$ \\ 
MRFasGCN\cite{jin2019graph} & $\bm{A},\bm{X}$ & Semi-supervise & First-order + Mean Field Approximate & -- & eMRF & $\times$  \\
SGCN\cite{wang2021unsupervised} & $\bm{A},\bm{X}$ & Unsupervise & First-order & -- & Label sampling & $\times$ \\
NOCD\cite{shchur2019overlapping} & $\bm{A},\bm{X}$ & Unsupervise & First-order & -- & Bernoulli–Poisson & $\checkmark$ \\  
GCLN\cite{gcln2020hu} & $\bm{A},\bm{X}$ & Unsupervise & First-order & $k$-means & U-Net architecture & $\times$ \\  
IPGDN\cite{ipgdn2020liu} & $\bm{A},\bm{X}$ & Unsupervise & \makecell[c]{First-order + Disentangled\\ representation} & $k$-means & HSIC as regularizer & $\times$ \\
AGC\cite{zhang2019attributed} & $\bm{A}, \bm{X}$ & Unsupervise & $k$-order + Laplacian smoothing filter & Spectral Clustering & -- & $\times$ \\  
AGE\cite{cui2020adaptive} & $\bm{A},\bm{X}$ & Unsupervise & Laplacian smoothing filter & Spectral Clustering & Adaptive learning & $\times$ \\
CayleyNet\cite{levie2018cayleynets} & $\bm{A},\bm{X}$ & Semi-supervise & Laplacian smoothing filter & -- & Cayley polynomial & $\times$ \\ 
SENet\cite{zhang2021spectral} & $\bm{A},\bm{X}$ & Unsupervise & Third-order + Spectral clustering loss & $k$-means & Kernel matrix learning & $\times$ \\
CommDGI\cite{zhang2020commdgi} & $\bm{A},\bm{X}$ & Unsupervise & First-order + Sampling & -- & Joint optimization & $\times$ \\
Zhao \textit{et al.}\cite{ijcai2021-473} & $\bm{A},\bm{X}$ & Unsupervise & First-order + Sampling & -- & Joint optimization & $\times$ \\
\bottomrule[1 pt]
\end{tabular}
\label{table-gcn}
\end{table*}

\begin{table*}[h]
\centering \footnotesize
\renewcommand\arraystretch{1.1}
\setlength{\tabcolsep}{2.1mm}
\caption{Summary of GAT-based community detection methods.}
\rowcolors{2}{white}{gray!30}
\begin{tabular}{lcccccccc}
\toprule[1 pt]
\textbf{Method} & \textbf{Input} & \textbf{Metapath} & \textbf{Learning} & \textbf{Attention Mechanism} & \textbf{Co-technique} & \textbf{Clustering} & \textbf{Overlap} & \textbf{Network} \\ \midrule
DMGI\cite{park2020unsupervised} & $V,E^{(r)},\bm{X}$ & $\times$ & Unsupervise & \cite{bahdanau2014neural} & Contrastive learning & $k$-means & $\times$ & Multiplex \\
HDMI\cite{jing2021hdmi} & $V,E^{(r)},\bm{X}$ & $\times$ & Unsupervise & \cite{you2016image} & MI & $k$-means & $\times$ & Multiplex \\
MAGNN\cite{fu2020magnn} & $\mathcal{V},\mathcal{E},\mathcal{X}$ & $\checkmark$ & Unsupervise & \cite{vaswani2017attention} & -- & $k$-means & $\times$ & Heterogeneous\\
HeCo\cite{wang2021self} & $\mathcal{V},\mathcal{E}$ & $\checkmark$ & Unsupervise & \cite{bahdanau2014neural} & Contrastive learning & $k$-means & $\times$ & Heterogeneous \\ 
CP-GNN\cite{luo2021detecting} & $\mathcal{V},\mathcal{E}$ & $\times$ & Unsupervise & \cite{vaswani2017attention} & -- & $k$-means & $\times$ & Heterogeneous \\ 
\bottomrule[1 pt]
\end{tabular}
\label{table-gat}
\end{table*}

\begin{table*}[h]
\centering \footnotesize
\renewcommand\arraystretch{1.1}
\setlength{\tabcolsep}{3.5mm}
\caption{Summary of GAN-based community detection methods.} 
\begin{tabular}{lccccccc}
\toprule[1 pt]
\textbf{Method} & \textbf{Input} & \textbf{Learning} & \textbf{Generator} & \textbf{Discriminator} & \textbf{Generated Samples} & \textbf{Clustering} & \textbf{Overlap} \\ 
\midrule
\rowcolor{gray!30}
SEAL\cite{zhang2020seal} & $\bm{A},\bm{X}$ & Semi-supervise & iGPN & GINs & Communities & -- & $\checkmark$ \\
DR-GCN\cite{ijcai2020-398} & $\bm{A},\bm{X}$ & Semi-supervise & MLP & MLP & Embeddings & $k$-means & $\times$ \\
\rowcolor{gray!30}
&  &  &  &  & Topology, attributes, &  &  \\
\rowcolor{gray!30}
\multirow{-2}{*}{JANE\cite{jane2020yang}} & \multirow{-2}{*}{$\bm{A},\bm{X}$} & \multirow{-2}{*}{Unsupervise} & \multirow{-2}{*}{Various} & \multirow{-2}{*}{MLP} & embeddings & \multirow{-2}{*}{--} & \multirow{-2}{*}{$\times$} \\
ProGAN\cite{gao2019progan} & $\bm{A},\bm{X}$ & Unsupervise & MLP & MLP & Triplets & $k$-means & $\times$ \\
\rowcolor{gray!30}
CommunityGAN\cite{jia2019communitygan} & $\bm{A}$ &  Unsupervise & AGM & AGM & Motifs & -- & $\checkmark$ \\
CANE\cite{wang2021cane} & $\bm{A}$ & Unsupervise & Softmax & MLP & Node pairs & $k$-means & $\times$ \\
\rowcolor{gray!30}
ACNE\cite{chen2021self} & $\bm{A}$ & Unsupervise & Softmax & MLP & Nodes, Communities & -- & $\checkmark$ \\
\bottomrule[1 pt]
\end{tabular}
\label{table-gan}
\end{table*}

\begin{table*}[h] 
\centering \footnotesize
\renewcommand\arraystretch{1.1}
\setlength{\tabcolsep}{1mm}
\caption{Summary of AE-based community detection methods.} 
\begin{tabular}{clcccccc}
\toprule[1 pt]
\textbf{Category} & \textbf{Method} & \textbf{Input} & \textbf{Learning} & \textbf{Encoder} & \textbf{Decoder} & \textbf{Loss} & \textbf{Overlap} \\ \midrule
\rowcolor{gray!30}
\cellcolor{white} & semi-DNR\cite{yang2016modularity} & $\bm{B}$ & Semi-supervise & MLP & MLP & reconstruction+pairwise & $\times$ \\
\cellcolor{white} & DNE-SBP\cite{shen2018deep} & $\bm{A}(+,-)$ & Semi-supervise & MLP & MLP & reconstruction+regularization+pairwise & $\times$ \\
\rowcolor{gray!30}
\cellcolor{white} & UWMNE/WMCNE-LE\cite{jin2018integrative} & $\bm{B},\bm{X}$ & Unsupervise & MLP & MLP & reconstruction+pairwise & $\times$ \\
\cellcolor{white} & sE-Autoencoder\cite{wang2020evolutionary} & $\{\bm{A}_t\}$ & Semi-supervise & MLP & MLP & reconstruction+regularization+pairwise & $\times$ \\
\rowcolor{gray!30}
\cellcolor{white} & DANE\cite{gao2018deep} & $\bm{A}, \bm{X}$ & Unsupervise & MLP & MLP & reconstruction+proximity & $\times$ \\
\cellcolor{white} & Transfer-CDDTA\cite{xie2019high} & $\bm{S}_s,\bm{S}_t$ & Unsupervise & MLP & MLP & reconstruction+regularization+proximity & $\times$ \\
\rowcolor{gray!30}
\cellcolor{white} &  & $\mathcal{V}$, $\mathcal{E}$, &  &  &  & reconstruction+regularization &  \\
\rowcolor{gray!30}
\cellcolor{white} \multirow{-10}{*}{\makecell[c]{Stacked\\AE}} & \multirow{-2}{*}{DIME \cite{zhang2017bl}} & $\mathcal{X}, \{\mathcal{A}_{ij}\}$ & \multirow{-2}{*}{Unsupervise} & \multirow{-2}{*}{MLP} & \multirow{-2}{*}{MLP} & +information fusion  & \multirow{-2}{*}{$\times$} \\ 
\midrule  
\cellcolor{white} & GraphEncoder\cite{tian2014learning} & $\bm{A},\bm{D}, \bm{S}$ & Unsupervise & MLP & MLP & reconstruction+regularization+sparsity & $\times$ \\
\rowcolor{gray!30}
\cellcolor{white} & WCD\cite{li2021weighted} & $\bm{S}$ & Unsupervise & MLP & MLP & reconstruction+sparsity & $\times$ \\
\cellcolor{white} & DFuzzy\cite{bhatia2018dfuzzy} & $\bm{A}$ & Unsupervise & MLP & MLP & reconstruction+sparsity & $\checkmark$ \\
\rowcolor{gray!30}
\cellcolor{white} \multirow{-5}{*}{\makecell[c]{Sparse\\AE}} & CDMEC\cite{xu2020stacked} & $\bm{S}_s,\bm{S}_t$ & Unsupervise & MLP & MLP & reconstruction+sparsity & $\times$ \\ \midrule
\cellcolor{white} & DNGR\cite{10.5555/3015812.3015982} & $\bm{A}$ & Unsupervise & MLP & MLP & reconstruction & $\times$ \\
\rowcolor{gray!30}
\cellcolor{white} & DNC\cite{li2021dnc} & $\bm{A}$ & Unsupervise & MLP & MLP & reconstruction+clustering & $\times$ \\
\cellcolor{white} & GRACE\cite{yang2017graph} & $\bm{A},\bm{X}$ & Unsupervise & MLP & MLP & reconstruction+clustering & $\times$ \\
\rowcolor{gray!30}
\cellcolor{white} \multirow{-4}{*}{\makecell[c]{Denoising\\AE}} & MGAE\cite{wang2017mgae} & $\bm{A},\bm{X}$ & Unsupervise & GCN & GCN & reconstruction+regularization & $\times$ \\ \midrule  
\cellcolor{white} & GUCD\cite{ijcai2020-486} & $\bm{A},\bm{X}$ & Unsupervise & MRFasGCN & MLP & reconstruction+pairwise & $\times$ \\
\rowcolor{gray!30}
\cellcolor{white} & SDCN\cite{bo2020structural} & $\bm{A},\bm{X}$ & Unsupervise & GCN+DNN & DNN & reconstruction+clustering & $\times$ \\
\cellcolor{white} \multirow{-3}{*}{\makecell[c]{Graph\\Convolutional\\ AE}} & O2MAC\cite{fan2020one2multi} & $\{\bm{A}\},\bm{X}$ & Unsupervise & GCN & Inner Product & reconstruction+clustering & $\times$ \\ \midrule
\rowcolor{gray!30}
\cellcolor{white} & DAEGC\cite{wang2019attributed} & $\bm{A},\bm{X}$ & Unsupervise & GAT & Inner Product & reconstruction+clustering & $\times$ \\
\cellcolor{white} &  &  &  &  & Inner Product & reconstruction+regularization & \\
\cellcolor{white} & \multirow{-2}{*}{GEC-CSD\cite{xu2021graph}} & \multirow{-2}{*}{$\bm{A},\bm{X}$} & \multirow{-2}{*}{Unsupervise} & \multirow{-2}{*}{GAT} &  +GAT & +clustering+adversarial & \multirow{-2}{*}{$\times$} \\
\rowcolor{gray!30}
\cellcolor{white} &  &  &  &  & Inner Product & reconstruction+clustering & \\
\rowcolor{gray!30}
\cellcolor{white} & \multirow{-2}{*}{MAGCN\cite{magcn2020cheng}} & \multirow{-2}{*}{$\bm{A},\{\bm{X}\}$} & \multirow{-2}{*}{Unsupervise} & \multirow{-2}{*}{GAT+MLP} & +GCN & +consistency & \multirow{-2}{*}{$\times$} \\
\cellcolor{white} &  &  &  &  & & reconstruction+regularization & \\
\cellcolor{white} & \multirow{-2}{*}{SGCMC\cite{xia2021self}} & \multirow{-2}{*}{$\bm{A},\{\bm{X}\}$} & \multirow{-2}{*}{Unsupervise} & \multirow{-2}{*}{GAT} & \multirow{-2}{*}{GAT} & +clustering+consistency & \multirow{-2}{*}{$\times$} \\
\rowcolor{gray!30}
\cellcolor{white} &  &  &  &  &  & reconstruction+regularization & \\
\rowcolor{gray!30}
\cellcolor{white} \multirow{-10}{*}{\makecell[c]{Graph\\Attention\\AE}} & \multirow{-2}{*}{DMGC\cite{luo2020deep}} & \multirow{-2}{*}{$\{A\}$} & \multirow{-2}{*}{Unsupervise} & \multirow{-2}{*}{MLP} & \multirow{-2}{*}{MLP} & +proximity+clustering & \multirow{-2}{*}{$\times$} \\ \midrule
\cellcolor{white} & TGA/TVGA\cite{shi2020effective} & $\bm{A}$,$\bm{X}$ & Unsupervise & GCN & Triad & reconstruction & $\times$ \\
\rowcolor{gray!30}
\cellcolor{white} & VGECLE\cite{chen2019variational} & $\bm{A}$ & Semi-supervise & DNN & DNN & reconstruction+pairwise & $\times$ \\
\cellcolor{white} & DGLFRM\cite{mehta2019stochastic} & $\bm{A},\bm{X}$ & Semi-supervise & GCN & DNN & reconstruction+regularization & $\checkmark$ \\
\rowcolor{gray!30}
\cellcolor{white} & LGVG\cite{Sarkar2020LGVG} & $\bm{A},\bm{X}$ & Semi-supervise & GCN & DNN & reconstruction+regularization & $\checkmark$ \\
\cellcolor{white} & VGAECD\cite{choong2018learning} & $\bm{A},\bm{X}$ & Unsupervise & GCN & Inner Product & reconstruction+clustering & $\times$ \\
\rowcolor{gray!30}
\cellcolor{white} & VGAECD-OPT\cite{choong2019optimizing} & $\bm{A},\bm{X}$ & Unsupervise & GCN & Inner Product & reconstruction+clustering & $\times$ \\
\cellcolor{white} \multirow{-8}{*}{\makecell[c]{Variational \\AE}} & ARGA/ARVGA\cite{pan2018adversarially} & $\bm{A},\bm{X}$ & Unsupervise & GCN & Inner Product & reconstruction & $\times$ \\ 
\bottomrule[1 pt]
\end{tabular}
\label{table-autoencoder}
\end{table*}

\section{Detailed description: Data Sets}\label{sec-ddds}

\subsection{Real-world Data Sets}

\textbf{Citation/Co-authorship Networks.}
\textit{Citeseer}, \textit{Cora} and \textit{Pubmed}\footnote{\url{https://linqs.soe.ucsc.edu/data}} are the most popular group of paper citation networks used in community detection experiments as attributed networks, where nodes represent publications and links mean citations. The nodes are described by binary word vectors. Topics are class labels. 
\textit{DBLP}\footnote{\url{http://snap.stanford.edu/data/com-DBLP.html}} is a co-authorship network from computer science bibliography website. A ground truth community labels that the authors who published papers on one journal or conference. In complicated scenarios, the heterogeneous \textit{DBLP} can be extracted with considering diverse information such as co-authorship, citation.
\textit{ACM}\footnote{\url{http://dl.acm.org/}} is a paper network with two views, one view describes whether two papers are written by the same author, the other view depicts if the subjects of two papers are the same. The nodes (\textit{i.e.} papers) are described by the bag-of-words of the keywords. The heterogeneous \textit{ACM} can be constructed as well.

\textbf{Online Social Networks.}
In most of online social network data sets, nodes represent users, edges depict relationship between them such as friendships or followings. In the \textit{YouTube}\footnote{\url{http://snap.stanford.edu/data/com-Youtube.html}} dataset, social network is formed by users having friendship, each user-created group for video sharing is a community. \textit{LiveJournal}\footnote{\url{http://snap.stanford.edu/data/soc-LiveJournal1.html}} is a friendship social network collected from a online blog community where users state friendship to each other and define groups that can be ground truth communities. \textit{PolBlogs}\footnote{\url{http://www-personal.umich.edu/~mejn/netdata/}\label{newman}} is a social blog network of weblogs on US politics, where nodes are blogs and hyperlinks between them extract to edges. 
Some data sets also collect node attributes. In the \textit{Facebook}\footnote{\url{http://snap.stanford.edu/data/ego-Facebook.html}} and \textit{Twitter}\footnote{\url{http://snap.stanford.edu/data/ego-Twitter.html}} data sets, for example, attributes are described by user profiles while nodes and edges represent users and their connections, respectively. The heterogeneous \textit{Twitter} extracts different types information from users, tweets, and locations. \textit{Flickr} provides an online platform for users to share photos, the network is formed by users following each other, attributes are image tags, and community labels represent different interest groups. \textit{Blogcatalog} is another blogger community, where attributes are extracted from users' blogs, community labels denote various topics. \textit{Gplus}\footnote{\url{http://snap.stanford.edu/data/ego-Gplus.html}} is the ego network constructed by users sharing their \textit{circles} in Google+, in which links exist if users follow each other and their profiles are extracted as attributes. Furthermore, there are three signed online social networks, \textit{Slashdot} is extracted from the technology news site Slashdot where users form the relationships of friends or foes which represented by positive or negative links. \textit{Epinions} is a \textit{who trust whom} network constructed from the Epinions site\footnote{\url{http://www.epinions.com/}}, where the positive and negative edges reflects that users trust or distrust each other. In the signed \textit{Wiki} data set, the network describes election of Wikipedia admins among users, edges indicate that the users vote \textit{for} (positive) or \textit{against} (negative) to the other. Considering heterogeneous information, \textit{Last.fm} as a music website\footnote{\url{https://www.last.fm/}} constructs the network of users, artists and artist tags by tracking user listening information from various sources. \textit{Foursquare} is another heterogeneous social network, which is formed by various kinds of location-related services. 

\textbf{Traditional Social Networks.}
The data sets of \textit{Karate}, \textit{Dolphin} and \textit{Football} are the most widely-used social networks in community detection. \textit{Karate}\textsuperscript{\ref{newman}} describes friendships between 34 members of Zachary’s karate club. In \textit{Dolphin}\textsuperscript{\ref{newman}} social network, nodes and edges represent dolphins and tighter connection between two dolphins, respectively. In \textit{Football}\textsuperscript{\ref{newman}} data set, nodes represent different football teams while edges indicate the matches between them. Another traditional social network data sets include: the \textit{Friendship6} and \textit{Friendship7} data sets that construct high school friendship networks, and \textit{Email}\footnote{\url{https://snap.stanford.edu/data/email-Eu-core.html}} data set describing email communication of the EU research institution. Considering the evolution of networks, \textit{Cellphone Calls}\footnote{\url{http://www.cs.umd.edu/hcil/VASTchallenge08/download/Download.htm}} consists of a set of cell phone call records over ten days, where nodes are cell phones and edges represent phone calls. \textit{Rados}\footnote{\url{http://networkrepository.com/ia-radoslaw-email.php}} and \textit{Enron Email}\footnote{\url{http://www.cs.cmu.edu/\%7Eenron/}} are popular email data sets from employees' communication over a period of time, where senders or recipients are represented by nodes and edges exist if two individuals emailed. Another three dynamic netowrks are \textit{High School}, \textit{Hospital} 
and \textit{Hypertext}\footnote{\url{http://www.sociopatterns.org/datasets}}. \textit{High School} forms the temporal friendship network of contacts between students in different classes of a high school within one week. \textit{Hospital} constructs the time-varing network of contacts between patients, patients and health-care workers and among the workers in a hospital during a few days. \textit{Hypertext} extracts a social network of conference attendees during the ACM Hypertext 2009 conference.

\begin{table*}[!t]
\centering \footnotesize
\renewcommand\arraystretch{1}
\setlength{\tabcolsep}{1mm}
\caption{Summary of real-world benchmark data sets.} 
\begin{tabular}{clrrrrl}
\toprule[1 pt]
\textbf{Network} & \textbf{Data Set} & \textbf{Number of} & \textbf{Number of}  & \textbf{Number of} & \textbf{Given Number of} &  \\
\textbf{Category} & \textbf{Name} & \textbf{Nodes} & \textbf{Edges}  & \textbf{Attributes} & \textbf{Communities} & \textbf{Publications} \\ \midrule
\rowcolor{gray!30} 
\cellcolor{white!100} & Karate \cite{zachary1977information} & 34 & 78 & 0 & 2 & \cite{cai2020edge,yang2016modularity,xie2019high,xu2020stacked,choong2018learning,xie2018community,wang2021unsupervised} \\
\cellcolor{white!100} & Dolphins \cite{lusseau2003emergent} &  62 & 159 &  0 & 2 & \cite{cai2020edge,yang2016modularity,xie2019high,xu2020stacked,xie2018community,wang2021unsupervised} \\
\rowcolor{gray!30} 
\cellcolor{white!100} & Friendship6 \cite{xie2013overlapping} & 69 & 220 & 0 & 6 & \cite{yang2016modularity,xie2019high,xu2020stacked} \\
\cellcolor{white!100} & Friendship7 \cite{xie2013overlapping} & 69 & 220 & 0 & 7 & \cite{yang2016modularity,xie2019high,xu2020stacked} \\
\rowcolor{gray!30} 
\cellcolor{white!100} & Polbooks \cite{newman2006modularity} &  105 & 441 &  0 & 3 & \cite{yang2016modularity,xie2019high,xu2020stacked,wang2021unsupervised} \\
\cellcolor{white!100} &  &  &  &  &  & \cite{xin2017deep,cai2020edge,yang2016modularity,xie2019high,xu2020stacked,huang2020community,li2021weighted,wang2021unsupervised}, \\
\cellcolor{white!100} & \multirow{-2}{*}{Football \cite{girvan2002community}} & \multirow{-2}{*}{115} & \multirow{-2}{*}{613} & \multirow{-2}{*}{0} & \multirow{-2}{*}{12} & \cite{xie2018community} \\
\rowcolor{gray!30} 
\cellcolor{white!100} & Email\cite{yin2017local} & 1,005 & 25,571 & 0 & 42 & \cite{ye2018deep,huang2020community} \\
\cellcolor{white!100} & Polblogs \cite{adamic2005political} & 1,490 & 16,718 & 0 & 2 & \cite{yang2016modularity,xie2019high,xu2020stacked,choong2018learning,huang2020community,xie2018community} \\
\rowcolor{gray!30}
\cellcolor{white!100} & Amazon \cite{yang2015defining} & 334,863 & 925,872 &  0 & 75,149 & \cite{cai2020edge,chen2019supervised,zhang2020seal,jia2019communitygan,xie2019high,bhatia2018dfuzzy,xu2020stacked,sun2017non} \\
\cellcolor{white!100} & DBLP \cite{yang2015defining} & 317,080 & 1,049,866 &  0 & 13,477 & \cite{cai2020edge,chen2019supervised,zhang2020seal,jia2019communitygan,xie2019high,bhatia2018dfuzzy,bo2020structural,sun2017non} \\
\rowcolor{gray!30}
\cellcolor{white!100} & YouTube \cite{yang2015defining} & 1,134,890 & 2,987,624 & 0 & 8,385 & \cite{xin2017deep,cai2020edge,chen2019supervised,zhang2020seal,jia2019communitygan,bhatia2018dfuzzy,de2021deep} \\
\cellcolor{white!100} \multirow{-11}{*}{\makecell[c]{Unattributed\\Network}} & LiveJournal \cite{yang2015network} & 3,997,962 & 34,681,189 &  0 & 287,512 & \cite{xin2017deep,bhatia2018dfuzzy} \\ \midrule 
\rowcolor{gray!30} 
\cellcolor{white!100} & Texas \cite{sen2008collective} & 187 & 328 &  1,703 & 5 & \cite{jin2019graph,jin2018integrative,ijcai2020-486,wang2021unsupervised} \\
\cellcolor{white!100} & Cornell \cite{sen2008collective}  & 195 & 304 & 1,703 & 5 & \cite{jin2019graph,jin2018integrative,ijcai2020-486,wang2021unsupervised} \\
\rowcolor{gray!30}
\cellcolor{white!100} & Washington \cite{sen2008collective} & 230 & 446 &  1,703 & 5 & \cite{jin2019graph,jin2018integrative,ijcai2020-486,wang2021unsupervised} \\
\cellcolor{white!100} & Wisconsin \cite{sen2008collective} & 265 & 530 &  1,703 & 5 & \cite{jin2019graph,jin2018integrative,ijcai2020-486,wang2021unsupervised} \\
\rowcolor{gray!30}
\cellcolor{white!100} & Wiki \cite{yang2015network} & 2,405 & 17,981 &  4,973 & 19 & \cite{zhang2019attributed,cui2020adaptive,gao2018deep,10.5555/3015812.3015982,wang2017mgae,ye2018deep,chen2021self,xia2021self} \\
\cellcolor{white!100} &  &  &  &  &  & \cite{jin2019graph,yang2016modularity,gcln2020hu,ipgdn2020liu,yang2017graph,zhang2019attributed,cui2020adaptive,jane2020yang,gao2019progan,wang2021unsupervised},\\
\cellcolor{white!100} &  &  &  &  &  & \cite{zhang2021spectral,zhang2020commdgi,ijcai2021-473,gao2018deep,ijcai2020-398,xie2019high,xu2020stacked,10.5555/3015812.3015982,wang2017mgae,chen2021self},\\
\cellcolor{white!100} & \multirow{-3}{*}{Cora \cite{sen2008collective}} & \multirow{-3}{*}{2,708} & \multirow{-3}{*}{5,429} & \multirow{-3}{*}{1,433} & \multirow{-3}{*}{7} &  \cite{ijcai2020-486,wang2019attributed,magcn2020cheng,shi2020effective,mehta2019stochastic,Sarkar2020LGVG,choong2018learning,chen2019variational,pan2018adversarially,ye2018deep,xie2018community,xu2021graph,xia2021self} \\
\rowcolor{gray!30} 
\cellcolor{white!100} &  &  &  &  &  & \cite{jin2019graph,yang2017graph,gcln2020hu,ipgdn2020liu,zhang2019attributed,cui2020adaptive,jane2020yang,gao2019progan,wang2021unsupervised}, \\
\rowcolor{gray!30} 
\cellcolor{white!100} &  &  &  &  &  & \cite{zhang2020commdgi,zhang2021spectral,ijcai2021-473,gao2018deep,ijcai2020-398,xie2019high,xu2020stacked,10.5555/3015812.3015982,wang2017mgae,ijcai2020-486,bo2020structural,li2021dnc,mehta2019stochastic}, \\
\rowcolor{gray!30} 
\cellcolor{white!100} & \multirow{-3}{*}{Citeseer \cite{sen2008collective}} & \multirow{-3}{*}{3,312} & \multirow{-3}{*}{4,715} & \multirow{-3}{*}{3,703} & \multirow{-3}{*}{6} & \cite{xu2021graph,magcn2020cheng,shi2020effective,Sarkar2020LGVG,pan2018adversarially,ye2018deep,xie2018community,xia2021self} \\
\cellcolor{white!100} & UAI2010 \cite{sen2008collective} & 3,363 & 45,006 &  4,972 & 19 & \cite{jin2019graph,jin2018integrative,ijcai2020-486} \\
\rowcolor{gray!30} 
\cellcolor{white!100} & Facebook \cite{leskovec2012learning} & 4,039 & 88,234 & 1,283 & 193 & \cite{shchur2019overlapping,zhang2020seal,bhatia2018dfuzzy,yang2017graph} \\
\cellcolor{white!100} & Blogcatalog \cite{huang2017accelerated} & 5,196 & 171,743 & 8,189 & 6 & \cite{gao2019progan,chen2019variational,zhang2021spectral,li2021dnc} \\
\rowcolor{gray!30}
\cellcolor{white!100} & Flickr \cite{huang2017accelerated} & 7,564 & 239,365 & 12,047 & 9 & \cite{gao2019progan,chen2019variational,li2021dnc} \\ 
\cellcolor{white!100} & DBLP\cite{wu2020comprehensive} & 17,725 & 52,890 & 6,974 & 4 & \cite{ijcai2020-398,chen2021self} \\
\rowcolor{gray!30}
\cellcolor{white!100} &  &  &  &  &  & \cite{yang2017graph,wang2021unsupervised,zhang2020commdgi,gcln2020hu,ipgdn2020liu,zhang2019attributed,cui2020adaptive,jane2020yang,zhang2021spectral,ijcai2021-473},\\
\rowcolor{gray!30}
\cellcolor{white!100} &  &  &  &  &  & \cite{jin2018integrative,gao2018deep,10.5555/3015812.3015982,ijcai2020-486,wang2019attributed,magcn2020cheng,mehta2019stochastic,xu2021graph,choong2018learning},\\
\rowcolor{gray!30}
\cellcolor{white!100} & \multirow{-3}{*}{PubMed \cite{namata2012query}} & \multirow{-3}{*}{19,717} & \multirow{-3}{*}{44,338} & \multirow{-3}{*}{500} & \multirow{-2}{*}{3} & \cite{Sarkar2020LGVG,pan2018adversarially,ye2018deep} \\ 
\cellcolor{white!100} & Twitter \cite{leskovec2012learning} & 81,306 & 1,768,149 & 216,839 & 4,065 & \cite{zhang2020seal,bhatia2018dfuzzy,yang2017graph,ijcai2020-486} \\
\rowcolor{gray!30}
\cellcolor{white!100} \multirow{-21}{*}{\makecell[c]{Attributed \\Network}} & GPlus \cite{leskovec2012learning} & 107,614 & 13,673,453 & 15,907 & 468 & \cite{bhatia2018dfuzzy,yang2017graph} \\ \midrule 
& & & \textbf{View1} ~~~~\textbf{View2} & & & \\ \cmidrule{4-4}
\rowcolor{gray!30} 
\cellcolor{white!100} & ACM \cite{wang2019heterogeneous} & 3,025 & 29,281  2,210,761 & 1,830 & 3 & \cite{park2020unsupervised,fan2020one2multi,jing2021hdmi} \\ 
\cellcolor{white!100} \multirow{-2}{*}{\makecell[c]{Multi-view \\Network}} & IMDB \cite{wang2019heterogeneous} & 4,780 & 98,010 ~~~ 21,018 & 1,232 & 3 & \cite{park2020unsupervised,fu2020magnn,fan2020one2multi,jing2021hdmi} \\ \midrule 
&  &  & \textbf{Edges} (+) &\textbf{Edges} (-) &  &  \\ \cmidrule{4-5}
\rowcolor{gray!30}
\cellcolor{white!100} & Epinions \cite{massa2005controversial} & 7,000 & 404,006 & 47,143 & -- & \cite{shen2018deep} \\
\cellcolor{white!100} &  Slashdot \cite{kunegis2009slashdot} & 7,000 & 181,354 & 56,675 & -- & \cite{shen2018deep} \\
\rowcolor{gray!30}
\cellcolor{white!100} \multirow{-3}{*}{\makecell[c]{Signed \\Network}} & Wiki \cite{leskovec2010governance} & 7,118 & 81,318 & 22,357 & -- & \cite{shen2018deep} \\ \midrule
&  &  &  & \textbf{Metapath} &  &  \\ \cmidrule{5-5}
\rowcolor{gray!30}
\cellcolor{white!100} &  ACM\cite{wang2019heterogeneous} & 8,916 & 12,769 & \checkmark & -- & \cite{luo2021detecting,wang2021self} \\
\cellcolor{white!100} &  IMDB\cite{wang2019heterogeneous} & 11,616 & 17,106 & \checkmark & -- & \cite{fu2020magnn,luo2021detecting} \\
\rowcolor{gray!30}
\cellcolor{white!100} & Last.fm\cite{cantador2011second} & 20,612 & 128,804 & \checkmark & -- & \cite{fu2020magnn} \\
\cellcolor{white!100} & DBLP\cite{gao2009graph} & 26,128 & 119,783 & \checkmark & -- & \cite{fu2020magnn,luo2021detecting,wang2021self} \\
\rowcolor{gray!30}
\cellcolor{white!100} & Foursquare\cite{zhang2017bl} & 93,069 & 174,484 & $\times$ & -- & \cite{zhang2017bl} \\
\cellcolor{white!100} \multirow{-7}{*}{\makecell[c]{Heterogeneous \\Network}} &  Twitter\cite{zhang2017bl} &  9,793,112 & 10,271,142 & $\times$ & -- & \cite{zhang2017bl} \\ \midrule
&  &  &  & \textbf{Time steps} &  &  \\ \cmidrule{5-5}
\rowcolor{gray!30}
\cellcolor{white!100} & Hospital \cite{vanhems2013estimating} & 75 & 32,424 & 9 & -- & \cite{wang2020evolutionary} \\
\cellcolor{white!100} &  Hypertext \cite{isella2011s} & 113 & 20,818 & 5 & -- & \cite{wang2020evolutionary} \\
\rowcolor{gray!30}
\cellcolor{white!100} & Enron Mail \cite{klimt2004introducing} & 151 & 33,124 & 12 & -- & \cite{wang2020evolutionary} \\
\cellcolor{white!100} & Rados \cite{rossi2015network} & 167 & 82,927 & 10 & -- & \cite{wang2020evolutionary} \\
\rowcolor{gray!30}
\cellcolor{white!100} & High School \cite{mastrandrea2015contact} & 327 & 188,508 & 9 & -- & \cite{wang2020evolutionary} \\
\cellcolor{white!100}\multirow{-6}{*}{\makecell[c]{Dynamic \\Network}} & Cellphone Calls \cite{wang2020evolutionary} & 400 & 9,834 & 10 & -- & \cite{wang2020evolutionary}  \\
\bottomrule[1 pt]
\end{tabular}
\label{table-rwdatasets}
\end{table*}

\textbf{Webpage Networks.} The webpage networks connect between one type of world wide web resources. \textit{IMDB}\footnote{\url{https://www.imdb.com/}} is a website of movies, containing information like casts, storyline and user reviews. Heterogeneous \textit{IMDB} is constructed by representing nodes through movies, actors and directors. Multi-view \textit{IMDB} exploits co-actor relationship and co-director relationship. The bag-of-words of plots are extracted as attributes, community can be labeled by movie's genre. \textit{Wiki}\footnote{\url{https://linqs.soe.ucsc.edu/data}} is a webpage network where a node represents the certain webpage, edge means that hyperlink connects any two webpages, and attributes from weighted word vectors. Another wiki data set is \textit{UAI2010} which extracts related article information from English-Wikipedia pages. \textit{WebKB}\footnote{\url{https://linqs-data.soe.ucsc.edu/public/lbc/WebKB.tgz}} is the webpage data set which contains 4 sub-datasets extracted from four different universities, \textit{i.e.}, \textit{Cornell}, \textit{Texas}, \textit{Washington} and \textit{Wisconsin}.

\textbf{Product Co-purchasing Networks.}
The \textit{Amazon}\footnote{\url{http://snap.stanford.edu/data/\#amazon}} data set is provided by Amazon website to analyze the co-purchased products where nodes represent products and edges link any two products which are co-purchased frequently. The ground truth communities are labeled by product categories. \textit{Polbooks}\textsuperscript{\ref{newman}} describes the purchase of political books during the time of US presidential election in 2004. Nodes donate books, edges link any two frequently co-purchased books by the same buyers, and the community labels reflect buyers' political attitude.

\subsection{Synthetic Benchmark Data Sets} 

\textbf{Girvan-Newman (GN) Networks}\cite{girvan2002community}:
The classic GN benchmark network contains 128 nodes that partitioned into 4 communities with 32 nodes in each community. Every node has a fixed average degree $k_{in}$ and connects a pre-defined number of nodes in another community $k_{out}$. For example, $k_{in}+k_{out}=16$. A parameter $\mu$ controls the ratio of neighbors in other communities for each node.

\textbf{Lancichinetti–Fortunato–Radicchi (LFR) Networks} \cite{lancichinetti2008benchmark}: The LFR benchmark data set simulates the degree of nodes in a real-world network and the scale-free nature of the network for more flexible networks. The community validation is more challenging, and the results are more convincing. The LFR generation tool controls synthetic network topology via a set of parameters, including network size $n$, the average $k$ and maximum $Maxk$ degree, the minimum $Minc$ and maximum $Maxc$ community size and the mixing parameter $\mu$. The node degree and community size are governed by power laws distributions with exponents of $\gamma$ and $\beta$, respectively.
LFR is the most common simulation benchmark in community detection research, the extended LFR benchmark can generate overlapping communities in directed and weighted networks \cite{lancichinetti2009benchmarks}.

\section{Detailed DESCRIPTION: Evaluation Metrics} \label{sec-EM}

\noindent
\textbf{Normalized Mutual Information (NMI)}: The metric provides a standard evaluation on community detection outputs $\mathcal{C}$ and the ground truth $\mathcal{C^{*}}$. It reports the mutual parts in a normalized score:
\begin{equation}
\small
\operatorname{NMI}(\mathcal{C}, \mathcal{C^{*}})=\frac{-2 \sum_{i=1}^{K} \sum_{j=1}^{K^{*}} n_{i j} \log \frac{n_{i j} n}{n_{i} n_{j}}}{\sum_{i=1}^{K} n_{i} \log \frac{n_{i}}{n}+\sum_{j=1}^{K^{*}} n_{j} \log \frac{n_{j}}{n}},
\end{equation}
where $K$ and $K^{*}$ are the number of detected and ground truth communities, respectively, and $n$ is the number of nodes. $n_{ij}$ denotes the number of nodes appearing in both the $i$-th detected community $C_{i}$ and the $j$-th ground truth community $C_{j}^{*}$. $n_{i}$ and $n_{j}$ sum up the number of nodes in $C_{i}$ and $C_{j}^{*}$, respectively. NMI can evaluate overlapping communities by its variant: Normalized Mutual Information for Overlapping Community (Overlapping-NMI)\cite{mcdaid2011normalized}.

\vspace{2mm}
\noindent	
\textbf{Accuracy (ACC)}: ACC evaluates the community membership on each node as follows:
\begin{equation}
\small
\operatorname{ACC}\left(\mathcal{C}, \mathcal{C^{*}} \right) = \frac{1}{n}\sum_{i=1}^{n}\delta(y_{i},y_{i}^{*}), 
\end{equation}
where $y_{i}$ and $y_{i}^{*}$ denote the $i$-th node's community label in community detection results and the ground truth, respectively. The Kronecker delta function $\delta = 1$ if $y_{i}=y_{i}^{*}$, otherwise $0$.

\vspace{2mm}
\noindent
\textbf{Precision}: In the community detection evaluation, precision describes the percentage of clustered nodes in each detected community existing in its ground truth:
\begin{equation}
\small
\operatorname{Precision}(C_{k}, C^*_{k})= \frac{|C_k \cap C^*_k|} {|C_k|}, 
\end{equation}
where $C_k$ and $C_k^*$ denote the $k$-th detected community and groud truth community, respectively.

\vspace{2mm}
\noindent
\textbf{Recall}: Accordingly, recall measures the percentage of ground truth nodes of each community discovered in the detected community by counting the $k$-th community as below: 
\begin{equation}
\small
\operatorname{Recall}(C_{k}, C^*_{k}) = \frac{|C_k \cap C^*_k|} {|C_k^*|}. 
\end{equation}

\vspace{2mm}
\noindent
\textbf{F1-score}: The F1-score balances Precision and Recall: 
\begin{equation}
\small
\begin{aligned}
\operatorname{F1-score}&\left(C_{k}, C^*_{k}\right)=\\
& 2 \cdot \frac{\operatorname{Precision}\left(C_{k}, C^*_{k}\right) \cdot \operatorname{Recall}\left(C_{k}, C^*_{k}\right)}{\operatorname{Precision}\left(C_{k}, C^*_{k}\right)+\operatorname{Recall}\left(C_{k}, C^*_{k}\right)}
\end{aligned}.
\end{equation}
Over the whole set of communities, the F1-score evaluates the true/false positive/negative node clustering results in detail:
\begin{equation}
\small
\begin{aligned}
\operatorname{F1-score}&\left(\mathcal{C}, \mathcal{C}^*\right)= \\
&\sum\nolimits_{C_{k}\in \mathcal{C}} \frac{|C_{k}|}{\sum_{C_{k}\in \mathcal{C}} |C_{k}|} \max\limits_{C_{k}^{*}\in \mathcal{C}^{*}} \operatorname{F1-score}\left(C_{k}, C_{k}^{*} \right)
\end{aligned}. 
\end{equation}

\vspace{2mm}
\noindent
\textbf{Adjusted Rand Index (ARI)}: ARI is a variation of \textit{Rand Index (RI)} that measures the percentage of correctly divided nodes based on true/false positive/negative:
\begin{equation}
\small
\begin{aligned}
\operatorname{ARI}&(\mathcal{C}, \mathcal{C}^*) = \\ &\frac{\sum_{ij}\tbinom{n_{ij}}{2}-\left[\sum_i \tbinom{n_{i}}{2}\sum_{j} \tbinom{n_{j}}{2}\right]/\tbinom{n}{2}}{\frac{1}{2}\left[\sum_{i} \tbinom{n_{i}}{2}+\sum_{j} \tbinom{n_{j}}{2}\right]-\left[\sum_{i} \tbinom{n_{i}}{2}\sum_{j} \tbinom{n_{j}}{2}\right]/\tbinom{n}{2}}
\end{aligned}, 
\end{equation}
where $n$ is the number of nodes, 
$n_{i}$ is the number of nodes in the $i$-th detected community $C_{i}$ while $n_{j}$ denotes the number of nodes in the $j$-th ground truth community $C_{j}^{*}$. $n_{ij}$ indicates the number of nodes simultaneously appearing in $C_{i}$ and $C_{j}^{*}$.

\vspace{2mm}
\noindent	
\textbf{Modularity (\textit{Q})}: \textit{Q} is widely used to evaluate the strength of detected communities without ground truth. It compares with a null model which is a random graph with an equivalent degree distribution as the original graph:  
\begin{equation}
\small
Q = \sum_{k} (e_{kk} - a_{k}^{2}), 
\label{eq-modularity}
\end{equation}
where $e_{kk}$ is the ratio of edges linking nodes within the $k$-th community $C_{k}$ to the total edges in network, $a_{k}$ is the proportion of edges that associate with nodes in community $C_{k}$ to the number of total edges. The extended modularity \cite{shen2009detect} evaluates overlapping community detection performances.  

\vspace{2mm}
\noindent
\textbf{Jaccard}: Jaccard measures whether the $k$-th detected community $C_{k}$ is similar to its ground truth $C_{k}^{*}$ as follows: 
\begin{equation}
\small
\operatorname{JC}(C_{k}, C_{k}^{*})=\frac{|C_{k}\cap C_{k}^{*}|}{|C_{k} \cup C_{k}^{*}|}. 
\end{equation}
For the whole set of communities, Jaccard is computed as: 
\begin{equation}
\small
\begin{aligned}
\operatorname{JC}\left(\mathcal{C}, \mathcal{C}^*\right)= &\sum_{C_{k}\in \mathcal{C}}\frac{\max_{C_{k}^{*}\in \mathcal{C}^{*}}\operatorname{JC}(C_{k}, C_{k}^{*})}{2|\mathcal{C}|}\\
&+\sum_{C_{k}^{*}\in \mathcal{C}^{*}} \frac{\max_{C_{k}\in \mathcal{C}}\operatorname{JC}(C_{k}, C_{k}^{*})}{2|\mathcal{C^*}|}
\end{aligned}. 
\end{equation}

\vspace{2mm}
\noindent
\textbf{Conductance (CON)}: CON measures the separability of the $k$-th detected community through the fraction of total edge numbers of $C_{k}$ linking to outside communities:
\begin{equation}
\small
\operatorname{CON}(C_{k})=\frac{m_{k}^{out}}{2 m_{k}^{in} + m_{k}^{out}},
\end{equation}
where $m_{k}^{in}$ indicates the number of internal edges in community $C_{k}$, and $m_{k}^{out}$ is the number of external edges on the cross-community boundary of $C_{k}$.

\vspace{2mm}
\noindent
\textbf{Triangle Participation Ratio (TPR)}: TPR indicates the community density through measuring the fraction of triads within a detected community $C_{k}$. It is defined as follows:
\begin{equation}
\small
\begin{array}{r}
\operatorname{TPR}(C_{k})=\mid\left\{v_{i}:v_{i}\in C_{k},\left\{\left(v_{j}, v_{w}\right):v_{j}, v_{w} \in C_{k},  \right.\right. \\
\left.\left.\left(v_{i}, v_{j}\right),\left(v_{j}, v_{w}\right),\left(v_{i}, v_{w}\right) \in E\right\} \neq \emptyset\right\}| / |C_{k}|
\end{array}.
\end{equation}

\section{Detailed description: Open-source Implementations }\label{sec-appcode}

\textbf{Under PyTorch and GCN:} The implementation of LGNN\cite{chen2019supervised} for community detection in graphs runs 5-class distortive SBMs-based GNN and GCN-based LGNN, respectively. The main algorithm and other utilities of NOCD\cite{shchur2019overlapping} implement overlapping community detection. AGE\cite{cui2020adaptive} implements adaptive graph encoder for attributed graph embedding and tests node clustering experiment results. \textbf{Under PyTorch and GAT:} DMGI\cite{park2020unsupervised} and HDMI\cite{jing2021hdmi} examine graph clustering performance on unsupervised attributed multiplex network embedding. The implementation of MAGNN\cite{fu2020magnn} provides instructions to detect node clusters in heterogeneous graphs via metapath aggregated GNN. HeCo\cite{wang2021self} testifies the model's performance of node clustering on heterogeneous networks with providing four preprocessed data sets. CP-GNN\cite{luo2021detecting} implements experiments of community detection on four real-world heterogeneous graphs. SEAL \cite{zhang2020seal} runs algorithms on Python and Shell validating on 50 communities. \textbf{Under PyTorch and AEs:}  GraphEncoder\cite{tian2014learning} implements for graph clustering on deep representations. Under the DNN architecture, SDCN\cite{bo2020structural} cluster the graph on AE embeddings.

Under \textbf{TensorFlow} and GCN:  AGC\cite{zhang2019attributed} clusters attributed graph via adaptive graph convolution and intra-cluster distance computation. CayleyNet\cite{levie2018cayleynets} presents an implementation of community detection for a variety of different networks, including a sparse filter and rational spectral filters with Jacobi. There is an implementation on the CommunityGAN \cite{jia2019communitygan} method under TensorFlow and GAN. DIME\cite{zhang2017bl} and DANE\cite{gao2018deep} are stacked AE-based implementations applying TensorFlow. On heterogeneous networks: Foursquare and Twitter, DIME\cite{zhang2017bl} examines community detection results and hyperparameters. For graph convolutional AE-based community detection tasks, TensorFlow is employed by O2MAC\cite{fan2020one2multi} and SGCMC\cite{xia2021self} over multi-view graphs, while DMGC\cite{luo2020deep} via attentive cross-graph association for deep multi-graph clustering. Moreover, DGLFRM\cite{mehta2019stochastic} and ARGA/ARVGA\cite{pan2018adversarially} apply TensorFlow to VAE-based community detection. Other Python packages, such as \textbf{Keras}, are employed in our summarized implementations, \textit{i.e.}, DNGR\cite{10.5555/3015812.3015982}.

Despite the popularity of Python, \textbf{Matlab} is another popular tool to implement deep community detection methods. semi-DRN\cite{yang2016modularity} implements community detection via stacked AE. DNE-SBP \cite{shen2018deep} inputs signed adjacency matrix to detect $k$ network communities upon node representations. MGAE\cite{wang2017mgae} makes use of marginalized graph autoencoder to cluster graph.

\clearpage
\onecolumn
\section{Abbreviations}\label{sec-abbr}

\begin{table*}[h]
\centering \footnotesize
\renewcommand\arraystretch{1.2}
\caption{Abbreviations in this survey.}
\begin{tabular}{llll}
\toprule[1pt]
\textbf{Abbr.} & \textbf{Full Name} & \textbf{Ref.} & \textbf{Paper Title} \\ \midrule
\rowcolor{gray!30}
ACC & Accuracy & & \\
& Network Embedding and overlapping Community & & Self-Training Enhanced: Network embedding and overlapping  \\ 
\multirow{-2}{*}{ACNE\cite{chen2021self}} & detection with Adversarial learning & \multirow{-2}{*}{\cite{chen2021self}} & community detection with adversarial learning \\
\rowcolor{gray!30}
AE & AutoEncoder &  &  \\
AGC & Adaptive Graph Convolution & \cite{zhang2019attributed} & Attributed graph clustering via adaptive graph convolution \\
\rowcolor{gray!30}
AGE & Adaptive Graph Encoder & \cite{cui2020adaptive} & Adaptive graph encoder for attributed graph embedding \\
ARGA & Adversarial Regularized Graph AutoEncoder & \cite{pan2018adversarially} & Adversarially regularized graph autoencoder for graph embedding \\ 
\rowcolor{gray!30}
ARI & Adjusted Rand Index &  &  \\
& Adversarially Regularized Variational Graph & & \\
\multirow{-2}{*}{ARVGA} & AutoEncoder & \multirow{-2}{*}{\cite{pan2018adversarially}} & \multirow{-2}{*}{Adversarially regularized graph autoencoder for graph embedding} \\
\rowcolor{gray!30}
ATC & Attributed Truss Communities & \cite{huang2017attribute} & Attribute-driven community search \\ \midrule
BP & Bernoulli--Poisson & & \\ \midrule
\rowcolor{gray!30}
& & & CANE: community-aware network embedding via adversarial \\
\rowcolor{gray!30}
\multirow{-2}{*}{CANE} & \multirow{-2}{*}{Community-Aware Network Embedding} & \multirow{-2}{*}{\cite{wang2021cane}} & training \\
& Graph Convolutional Neural Networks with & & CayleyNets: Graph convolutional neural networks with complex \\ 
\multirow{-2}{*}{CayleyNets} & Cayley Polynomials & \multirow{-2}{*}{\cite{levie2018cayleynets}} & rational spectral filters \\ 
\rowcolor{gray!30}
& Community Detection Algorithm based on  &  & Community detection in complex networks using structural \\
\rowcolor{gray!30}
\multirow{-2}{*}{CDASS} & Structural Similarity & \multirow{-2}{*}{\cite{zarandi2018community}} & similarity \\
& (Stacked Autoencoder-Based) Community &  & Stacked autoencoder-based community detection method via  \\
\multirow{-2}{*}{CDMEC} & Detection Method via Ensemble Clustering & \multirow{-2}{*}{\cite{xu2020stacked}} & an ensemble clustering framework \\
\rowcolor{gray!30}
CNN & Convolutional Neural Network &  &  \\
CommDGI & Community Deep Graph Infomax & \cite{zhang2020commdgi} & CommDGI: Community detection oriented deep graph infomax \\
\rowcolor{gray!30}
& Community Detection with Generative & & CommunityGAN: Community detection with generative \\
\rowcolor{gray!30}
\multirow{-2}{*}{CommunityGAN} & Adversarial Nets & \multirow{-2}{*}{\cite{jia2019communitygan}} & adversarial nets \\
&  &  & Edge classification based on Convolutional Neural Networks  \\
\multirow{-2}{*}{ComNet-R} & \multirow{-2}{*}{Community Network Local Modularity R} & \multirow{-2}{*}{\cite{cai2020edge}} & for community detection in complex network \\
\rowcolor{gray!30}
CON & Conductance &  &  \\ 
& & & Detecting communities from heterogeneous graphs: A Context  \\
\multirow{-2}{*}{CP-GNN} & \multirow{-2}{*}{Context Path-based Graph Neural Network} & \multirow{-2}{*}{\cite{luo2021detecting}} & path-based graph neural network model \\
\midrule
\rowcolor{gray!30}
DAEGC & Deep Attentional Embedded Graph Clustering & \cite{wang2019attributed} & Attributed graph clustering: A deep attentional embedding approach \\ 
DANE & Deep Attributed Network Embedding & \cite{gao2018deep} & Deep attributed network embedding \\
\rowcolor{gray!30}
& Deep Autoencoder-like Nonnegative & & Deep autoencoder-like nonnegative matrix factorization for \\
\rowcolor{gray!30}
\multirow{-2}{*}{DANMF} & Matrix Fatorization & \multirow{-2}{*}{\cite{ye2018deep}} & community detection \\
& Density-Based Spatial Clustering of &  & A density-based algorithm for discovering clusters in large \\
\multirow{-2}{*}{DBSCAN} & Applications with Noise & \multirow{-2}{*}{\cite{ester1996density}} & spatial databases with noise \\
\rowcolor{gray!30}
DCSBM & Degree-Corrected Stochastic Block Model & \cite{karrer2011stochastic} & Stochastic blockmodels and community structure in networks \\
& Deep Learning-based Fuzzy &  & DFuzzy: A deep learning-based fuzzy clustering model for large \\
\multirow{-2}{*}{DFuzzy} & Clustering Model & \multirow{-2}{*}{\cite{bhatia2018dfuzzy}} & graphs \\
\rowcolor{gray!30}
& Deep Generative Latent Feature Relational &  &  \\ 
\rowcolor{gray!30}
\multirow{-2}{*}{DGLFRM} & Model & \multirow{-2}{*}{\cite{mehta2019stochastic}} & \multirow{-2}{*}{Stochastic blockmodels meet graph neural networks} \\ 
&  &  & BL-MNE: Emerging heterogeneous social network embedding \\
\multirow{-2}{*}{DIME} & \multirow{-2}{*}{Deep Aligned Autoencoder-based Embedding} & \multirow{-2}{*}{\cite{zhang2017bl}} & through broad learning with aligned autoencoder \\
\rowcolor{gray!30}
DMGC & Deep Multi-Graph Clustering & \cite{luo2020deep} & Deep multi-graph clustering via attentive cross-graph association \\
& Deep Graph Infomax for attributed &  &  \\ 
\multirow{-2}{*}{DMGI} & Multiplex network embedding & \multirow{-2}{*}{\cite{park2020unsupervised}} & \multirow{-2}{*}{Unsupervised attributed multiplex network embedding} \\
\rowcolor{gray!30}
& Deep Neural network-based Clustering-oriented & & DNC: A deep neural network-based clustering-oriented network \\
\rowcolor{gray!30}
\multirow{-2}{*}{DNC} & network embedding & \multirow{-2}{*}{\cite{li2021dnc}} & embedding algorithm \\
& Deep Network Embedding with &  & Deep network embedding for graph representation learning \\
\multirow{-2}{*}{DNE-SBP} & Structural Balance Preservation & \multirow{-2}{*}{\cite{shen2018deep}} & in signed networks \\
\rowcolor{gray!30}
DNGR & Deep Neural Networks for Graph Representation & \cite{10.5555/3015812.3015982} & Deep neural networks for learning graph representations \\
DNMF & Deep Nonnegative Matrix Factorization & & \\
\bottomrule[1pt]
\end{tabular} \label{tab-abbr1}
\end{table*}

\begin{table*}[!t]
\centering \footnotesize
\renewcommand\arraystretch{1.2}
\caption{Abbreviations in this survey (continue-1).}
\begin{tabular}{llll}
\toprule[1pt]
\textbf{Abbr.} & \textbf{Full Name} & \textbf{Ref.} & \textbf{Paper Title} \\ \midrule
\rowcolor{gray!30}
DNN & Deep Neural Network & & \\
& Dual-Regularized Graph Convolutional &  &  \\
\multirow{-2}{*}{DR-GCN} & Networks & \multirow{-2}{*}{\cite{ijcai2020-398}} & \multirow{-2}{*}{Multi-class imbalanced graph convolutional network learning} \\
\rowcolor{gray!30}
DSF & Deep Sparse Filtering & & \\
& Community Discovery based on Deep &  &  \\
\multirow{-2}{*}{DSFCD} & Sparse Filtering & \multirow{-2}{*}{\cite{xie2018community}} & \multirow{-2}{*}{Community discovery in networks with deep sparse filtering} \\ \midrule
\rowcolor{gray!30}
ELBO & Evidence Lower Bound & &  \\
EM & Expectation-Maximization algorithm  & & \\
\rowcolor{gray!30}
&  &  & Graph convolutional networks meet Markov random fields: \\ 
\rowcolor{gray!30}
\multirow{-2}{*}{eMRF} & \multirow{-2}{*}{extended Network Markov Random Fields} & \multirow{-2}{*}{\cite{jin2019graph}} & Semi-supervised community detection in attribute networks \\ \midrule
Fast\textit{Q} & Fast Modularity &  &  \\ \midrule
\rowcolor{gray!30}
GAN & Generative Adversarial Network &  &  \\
GAT & Graph Attention Network &  &  \\
\rowcolor{gray!30}
GCLN & Graph Convolutional Ladder-shape Networks & \cite{gcln2020hu} & Going deep: Graph convolutional ladder-shape networks \\
GCN & Graph Convolutional Network &  &  \\
\rowcolor{gray!30}
& Graph embedding clustering with & & Graph embedding clustering: Graph attention auto-encoder with  \\
\rowcolor{gray!30}
\multirow{-2}{*}{GEC-CSD} & cluster-specificity distribution & \multirow{-2}{*}{\cite{xu2021graph}} & cluster-specificity distribution \\
GIN & Graph Isomorphism Network & &  \\
\rowcolor{gray!30}
GMM & Gaussian Mixture Model  & &  \\
GN & Girvan-Newman networks &  &  \\
\rowcolor{gray!30}
GRACE & Graph Clustering with Dynamic Embedding & \cite{yang2017graph} & Graph clustering with dynamic embedding \\
GraphEncoder & Autoencoder-based Graph Clustering Model & \cite{tian2014learning} & Learning deep representations for graph clustering \\
\rowcolor{gray!30}
& GCN-based approach for Unsupervised &  & Community-centric graph convolutional network for unsupervised \\
\rowcolor{gray!30}
\multirow{-2}{*}{GUCD} & Community Detection & \multirow{-2}{*}{\cite{ijcai2020-486}} &  community detection \\ \midrule
HDMI & High-order Deep Multiplex Infomax & \cite{jing2021hdmi} & HDMI: High-order deep multiplex infomax \\
\rowcolor{gray!30}
& Co-contrastive learning mechanism for & & Self-supervised heterogeneous graph neural network with co-contrastive \\
\rowcolor{gray!30}
\multirow{-2}{*}{HeCo} & Heterogeneous graph neural networks & \multirow{-2}{*}{\cite{wang2021self}} &  learning \\
HIN & Heterogeneous Information Network  & & \\
\rowcolor{gray!30}
HSIC & Hilbert-Schmidt Independence Criterion & \cite{gretton2005measuring} & Measuring statistical dependence with Hilbert-Schmidt norms \\ \midrule
InfoMap & Information Mapping & \cite{rosvall2008maps} & Maps of random walks on complex networks reveal community structure \\
\rowcolor{gray!30}
& Graph Pointer Network with incremental &  & SEAL: Learning heuristics for community detection with generative \\
\rowcolor{gray!30}
\multirow{-2}{*}{iGPN} & updates & \multirow{-2}{*}{\cite{zhang2020seal}} & adversarial networks \\
& Independence Promoted Graph Disentangled &  &  \\ 
\multirow{-2}{*}{IPGDN} & Network & \multirow{-2}{*}{\cite{ipgdn2020liu}} & \multirow{-2}{*}{Independence promoted graph disentangled networks} \\ \midrule
\rowcolor{gray!30}
JANE & Jointly Adversarial Network Embedding & \cite{jane2020yang} & JANE: Jointly adversarial network embedding \\ \midrule
KL & Kullback-Leibler divergence & &  \\ \midrule
\rowcolor{gray!30}
& Locating Structural Centers for Community &  & Locating structural centers: A density-based clustering method for \\
\rowcolor{gray!30}
\multirow{-2}{*}{LCCD} & Detection & \multirow{-2}{*}{\cite{wang2017locating}} & community detection \\
LDA & Latent Dirichlet Allocation & & \\
\rowcolor{gray!30}
LFR & Lancichinetti–Fortunato–Radicchi networks &  &  \\ 
LGNN & Line Graph Neural Network & \cite{chen2019supervised} & Supervised community detection with line graph neural networks \\
\rowcolor{gray!30}
& Ladder Gamma Variational Autoencoder for &  &  \\
\rowcolor{gray!30}
\multirow{-2}{*}{LGVG} & Graphs & \multirow{-2}{*}{\cite{Sarkar2020LGVG}} & \multirow{-2}{*}{Graph representation learning via ladder gamma variational autoencoders} \\
LPA & Label Propagation Algorithm & & \\
\midrule
\rowcolor{gray!30}
&  &  & A multi-agent genetic algorithm for community detection in complex \\
\rowcolor{gray!30}
\multirow{-2}{*}{MAGA-Net} & \multirow{-2}{*}{Multi-Agent Genetic Algorithm} & \multirow{-2}{*}{\cite{li2016multi}} &  networks \\
& Multi-View Attribute Graph Convolution &  &  \\
\multirow{-2}{*}{MAGCN} & Networks & \multirow{-2}{*}{\cite{magcn2020cheng}} & \multirow{-2}{*}{Multi-view attribute graph convolution networks for clustering} \\
\rowcolor{gray!30}
&  &  & MAGNN: Metapath aggregated graph neural network for heterogeneous \\
\rowcolor{gray!30}
\multirow{-2}{*}{MAGNN} & \multirow{-2}{*}{Metapath Aggregated Graph Neural Network} & \multirow{-2}{*}{\cite{fu2020magnn}} &  graph embedding \\
& Modularized Deep NonNegative Matrix &  & Community detection based on modularized deep nonnegative matrix \\
\multirow{-2}{*}{MDNMF} & Factorization & \multirow{-2}{*}{\cite{huang2020community}} &  factorization \\
\rowcolor{gray!30}
MGAE & Marginalized Graph AutoEncoder & \cite{wang2017mgae} & MGAE: Marginalized graph autoencoder for graph clustering \\
\bottomrule[1pt]
\end{tabular}\label{tab-abbr2}
\end{table*}

\begin{table*}[!t]
\centering \footnotesize
\renewcommand\arraystretch{1.2}
\caption{Abbreviations in this survey (continue-2).}
\begin{tabular}{llll}
\toprule[1pt]
\textbf{Abbr.} & \textbf{Full Name} & \textbf{Ref.} & \textbf{Paper Title} \\ \midrule
\rowcolor{gray!30}
MI & Mutual Information &  &  \\
MMB & Mixed Membership Stochastic Blockmodel & \cite{airoldi2008mixed} & Mixed membership stochastic blockmodels \\ 
\rowcolor{gray!30}
& Markov Random Fields as a convolutional &  & Graph convolutional networks meet Markov random fields: \\
\rowcolor{gray!30}
\multirow{-2}{*}{MRFasGCN} & layer in Graph Convolutional Networks & \multirow{-2}{*}{\cite{jin2019graph}} & Semi-supervised community detection in attribute networks \\  \midrule
NMF & Nonnegative Matrix Factorization &  &  \\
\rowcolor{gray!30}
NMI & Normalized Mutual Information&  &  \\
NOCD & Neural Overlapping Community Detection & \cite{shchur2019overlapping} & Overlapping community detection with graph neural networks \\ \midrule
\rowcolor{gray!30}
One2Multi & One-view to Multi-view &\cite{fan2020one2multi}  & One2multi graph autoencoder for multi-view graph clustering \\
& One2Multi Graph Autoencoder for Multi-view &  & \\ 
\multirow{-2}{*}{O2MAC} & Graph Clustering & \multirow{-2}{*}{\cite{fan2020one2multi}} & \multirow{-2}{*}{One2multi graph autoencoder for multi-view graph clustering} \\ \midrule
\rowcolor{gray!30}
PPI & Protein-Protein Interaction & &  \\ 
&  &  & Progan: Network embedding via proximity generative \\
\multirow{-2}{*}{ProGAN} & \multirow{-2}{*}{Proximity Generative Adversarial Network} & \multirow{-2}{*}{\cite{gao2019progan}} & adversarial network \\ \midrule
\rowcolor{gray!30}
\textit{Q} & Modularity &  &  \\ \midrule
SBM & Stochastic Block Model &  &  \\
\rowcolor{gray!30}
SCAN & Structural Clustering Algorithm for Networks & \cite{xu2007scan} & SCAN: A structural clustering algorithm for networks \\
SDCN & Structural Deep Clustering Network & \cite{bo2020structural} & Structural deep clustering network \\
\rowcolor{gray!30}
& Seed Expansion with generative Adversarial &  & SEAL: Learning heuristics for community detection with \\
\rowcolor{gray!30}
\multirow{-2}{*}{SEAL} & Learning & \multirow{-2}{*}{\cite{zhang2020seal}} & generative adversarial networks \\
& Semi-supervised Nonlinear Reconstruction &  &  \\ 
\multirow{-2}{*}{semi-DRN} & Algorithm with Deep Nerual Network & \multirow{-2}{*}{\cite{yang2016modularity}} & \multirow{-2}{*}{Modularity based community detection with deep learning} \\ 
\rowcolor{gray!30}
sE-Autoencoder & Semi-supervised Evolutionary Autoencoder & \cite{wang2020evolutionary} & An evolutionary autoencoder for dynamic community detection \\
SENet & Spectral Embedding Network & \cite{zhang2021spectral} & Spectral embedding network for attributed graph clustering \\
\rowcolor{gray!30}
SF & Sparse Filtering &  &  \\
SGC &  Simplification of GCN & \cite{wu2019simplify} & Simplifying graph convolutional networks \\
\rowcolor{gray!30}
& Self-supervised Graph Convolutional network &  &  \\ 
\rowcolor{gray!30}
\multirow{-2}{*}{SGCMC} & for Multi-view Clustering & \multirow{-2}{*}{\cite{xia2021self}} & \multirow{-2}{*}{Self-supervised graph convolutional network for multi-view clustering} \\ 
& A framework of local label sampling model & & Unsupervised learning for community detection in attributed \\ 
\multirow{-2}{*}{SGCN} & and GCN model & \multirow{-2}{*}{\cite{wang2021unsupervised}} & networks based on graph convolutional network \\
\rowcolor{gray!30}
SGVB & Stochastic Gradient Variational Bayes & & \\
SparseConv & Sparse Matrix Convolution & \cite{sperli2019deep} & A deep learning based community detection approach \\ \midrule
\rowcolor{gray!30}
REM & Residual Entropy Minimization & & \\ \midrule
TGA/TVGA & Triad (Variational) Graph Autoencoder & \cite{shi2020effective} & Effective decoding in graph auto-encoder using triadic closure \\
\rowcolor{gray!30}
TIN & Topologically Incomplete Networks&  &  \\
& Transfer Learning-inspired Community & & High-performance community detection in social networks using \\
\multirow{-2}{*}{Transfer-CDDTA} & Detection with Deep Transitive Autoencoder & \multirow{-2}{*}{\cite{xie2019high}} &  a deep transitive autoencoder \\
\rowcolor{gray!30}
TPR & Triangle Participation Ratio & & \\ \midrule
& Unified Weight-free Multi-component & &  \\ 
\multirow{-2}{*}{UWMNE} & Network Embedding & \multirow{-2}{*}{\cite{jin2018integrative}} & \multirow{-2}{*}{Integrative network embedding via deep joint reconstruction} \\ \midrule
\rowcolor{gray!30}
VAE & Variational AutoEncoder & \cite{kingma2013auto} & Auto-encoding variational bayes \\
VGAE & Variational Graph AutoEncoder & \cite{kipf2016variational} & Variational graph auto-encoders \\
\rowcolor{gray!30}
& Variational Graph Autoencoder for &  &  \\ 
\rowcolor{gray!30}
\multirow{-2}{*}{VGAECD} & Community Detection & \multirow{-2}{*}{\cite{choong2018learning}} & \multirow{-2}{*}{Learning community structure with variational autoencoder} \\ 
& Optimizing Variational Graph Autoencoder &  &  \\
\multirow{-2}{*}{VGAECD-OPT} & for Community Detection & \multirow{-2}{*}{\cite{choong2019optimizing}} & \multirow{-2}{*}{Optimizing variational graph autoencoder for community detection} \\
\rowcolor{gray!30}
& Variational Graph Embedding and Clustering &  &  \\
\rowcolor{gray!30}
\multirow{-2}{*}{VGECLE} & with Laplacian Eigenmaps & \multirow{-2}{*}{\cite{chen2019variational}} & \multirow{-2}{*}{Variational graph embedding and clustering with Laplacian eigenmaps} \\ \midrule
WalkTrap & WalkTrap  & \cite{pons2005computing} & Computing communities in large networks using random walks \\
\rowcolor{gray!30}
& & & A weighted network community detection algorithm based on deep \\
\rowcolor{gray!30}
\multirow{-2}{*}{WCD} & \multirow{-2}{*}{Weighted Community Detection} & \multirow{-2}{*}{\cite{li2021weighted}} & learning \\
& Weight-free Multi-Component Network &  &  \\
\multirow{-2}{*}{WMCNE-LE} & Embedding with Local Enhancement & \multirow{-2}{*}{\cite{jin2018integrative}} & \multirow{-2}{*}{Integrative network embedding via deep joint reconstruction} \\
\bottomrule[1pt]
\end{tabular}\label{tab-abbr3}
\end{table*}

\end{document}